%% file: jvla_SED_v1.tex
\documentclass[twocolumn]{aastex62}

\hypersetup{filecolor=cyan, hypertexnames=true}
\usepackage{color}
\usepackage{captcont}
\usepackage{comment}
\usepackage{amsmath}
\usepackage{lineno}

\newcommand\asec{{$^{\prime\prime}$}}

\graphicspath{{./}{figures/}}

\submitjournal{ApJ}

\shorttitle{Radio Spectra of WISE-NVSS Quasars}
\shortauthors{Patil et al.}

\begin{document}

\title{Radio Spectra of Luminous,  Heavily Obscured WISE-NVSS Selected Quasars}

\correspondingauthor{Pallavi Patil}
\email{ppatil@nrao.edu}

\author[0000-0002-9471-8499]{Pallavi Patil} \altaffiliation{Jansky Fellow of the National Radio Astronomy Observatory}
\affiliation{National Radio Astronomy Observatory, 1003 Lopezville Rd, Socorro, NM 87801, USA}

\author{Mark Whittle}
\affil{Department of Astronomy, University of Virginia, 530 McCormick Road, Charlottesville, VA 22903, USA}

\author[0000-0003-1991-370X]{Kristina Nyland}
\affiliation{U.S. Naval Research Laboratory, 4555 Overlook Ave. SW, Washington, DC 20375, USA}

\author{Carol Lonsdale}
\affiliation{National Radio Astronomy Observatory, 520 Edgemont Road, Charlottesville, VA 22903, USA}

\author{Mark Lacy}
\affiliation{National Radio Astronomy Observatory, 520 Edgemont Road, Charlottesville, VA 22903, USA}

\author{Amy E Kimball}
\affiliation{National Radio Astronomy Observatory, 1003 Lopezville Rd, Socorro, NM 87801, USA}

\author{Colin Lonsdale}
\affiliation{Massachusetts Institute of Technology, Haystack Observatory, Westford, MA 01886, USA}

\author[0000-0002-5187-7107]{Wendy Peters}
\affiliation{U.S. Naval Research Laboratory, 4555 Overlook Ave. SW, Washington, DC 20375, USA}

\author[0000-0001-6812-7938]{Tracy E. Clarke}
\affiliation{U.S. Naval Research Laboratory, 4555 Overlook Ave. SW, Washington, DC 20375, USA}

\author{Andreas Efstathiou}
\affiliation{School of Sciences, European University Cyprus, Engomi, 1516 Nicosia, Cyprus}

\author{Simona Giacintucci}
\affiliation{U.S. Naval Research Laboratory, 4555 Overlook Ave. SW, Washington, DC 20375, USA}

\author{Minjin Kim}
\affiliation{Department of Astronomy and Atmospheric Sciences, Kyungpook National University, Daegu 41566, Korea}

\author[0000-0002-3249-8224]{Lauranne Lanz}
\affiliation{Department of Physics, The College of New Jersey, 2000 Pennington Road, Ewing, NJ 08628, USA}

\author{Dipanjan Mukherjee}
\affiliation{Inter-University Centre for Astronomy and Astrophysics, Post Bag 4, Ganeshkhind, Pune - 411007, India.}

\author{Emil Polisensky}
\affiliation{U.S. Naval Research Laboratory, 4555 Overlook Ave. SW, Washington, DC 20375, USA}



\begin{abstract}

We present radio spectra spanning $0.1 - 10$ GHz for the sample of heavily obscured luminous quasars with extremely red mid-infrared-optical colors and compact radio emission. The spectra are constructed from targeted 10 GHz observations and archival radio survey data, which together yield $6-11$ flux density measurements for each object. 
Our primary result is that most (62\%) of the sample have peaked or curved radio spectra and many (37\%) could be classified as Gigahertz Peaked Spectrum (GPS) sources. This indicates compact emission regions likely arising from recently triggered radio jets. Assuming synchrotron self-absorption (SSA) generates the peaks, we infer compact source sizes ($3 - 100$ pc) with strong magnetic fields ($6 - 100$ mG) and young ages ($30 - 10^4$ years). Conversely, free-free absorption (FFA) could also create peaks due to the high column densities associated with the deeply embedded nature of the sample. However, we find no correlations between the existence or frequency of the peaks and any parameters of the MIR emission. The high-frequency spectral indices are steep ($\alpha \approx -1$) and correlate, weakly, with the ratio of MIR photon energy density to magnetic energy density, suggesting that the spectral steepening could arise from inverse Compton scattering off the intense MIR photon field. This study provides a foundation for combining multi-frequency and mixed-resolution radio survey data for understanding the impact of young radio jets on the ISM and star formation rates of their host galaxies. \href{https://github.com/paloween/Radio_Spectral_Fitting}{\faGithub}

\end{abstract}



\section{Introduction}\label{sec:c3_intro}

Cosmological simulations predict that gas-rich mergers are likely to trigger an intense starburst phase followed by a rapidly accreting supermassive black hole (SMBH), whose early stages are expected to be heavily obscured \citep[e.g.,][]{dimatteo+05, hopkins+10, alexander+12}.  On the observational side, the connection between ultraluminous infrared galaxies (ULIRGs) and quasars is  broadly consistent with the merger-driven, dust enshrouded  quasar growth scenario \citep[e.g.,][]{sanders+88, lonsdale+06, petric+11}. In the early stages, these heavily obscured quasars will be faint in the optical and X-rays, but bright in the mid- and far-infrared (MIR, FIR) and sub-mm due to reprocessed emission from dust and, in some cases, also bright in the radio due to synchrotron emitting relativistic particles arising from jets. Infrared (IR) satellites such as the Infrared Astronomical Satellite (IRAS; \citealt{neugebauer+84}), the Spitzer Space Telescope \citep{werner+04}, and the WideField Infrared Survey Explorer (WISE; \citealt{wright+10}) as well as ground-based sub-mm instruments (e.g., SCUBA \citealt{holland+99}, ALMA) have led to the discovery of ultraluminous, MIR-bright galaxies, and Submillimeter Galaxies (SMGs; e.g.  \citealt{blain+02}), 
at all redshifts.  
Different selection techniques are employed to identify these heavily obscured active galactic nuclei (AGN), but they all have luminous hosts harboring a recently triggered, high-accretion rate SMBH at the peak of its fueling, and many are found at the peak of galaxy mass assembly at $z\sim2$ \cite[e.g.,][]{tsai+15,fan+16,  zappacosta+18}.  
In recent years, WISE has opened the MIR sky and found some of the most luminous and heavily obscured AGN, including Hot Dust-Obscured Galaxies \citep[``Hot DOGs'', e.g.,][]{wu+12, eisenhardt+12,bridge+13}
and their  radio-bright counterparts \citep{lonsdale+15}.


This paper focuses on the sample selected by \citet[][hereafter Paper I]{lonsdale+15} that combined MIR and radio properties to find heavily obscured but luminous ($L_{\rm Bol} \sim 10^{11.7}-10^{14.2} L_\odot$)  AGN at redshifts $z\sim1-2$ with compact radio sources. The unique selection method identified 167 radio AGN with very red colors in WISE bands 1 (3.6$\mu$m), 2 (4.5$\mu$m), and 3 (12$\mu$m) with faint or no optical counterparts. An important feature of the selection is the requirement of compact and bright radio emission in the NRAO VLA Sky Survey (NVSS; \citealt{condon+98}) and/or the Faint Images of the Radio Sky at Twenty-one centimeters (FIRST; \citealt{becker+95}) survey. The overall aim was to select young radio AGN that are still enshrouded by dust following a recent gas rich merger, allowing the early stages of black hole accretion and radio source expansion to be studied in detail. 

This sample has been the subject of multi-wavelength follow-up  observations, including the Atacama Large Millimeter Array (ALMA; Paper I), the Karl G. Jansky Very Large Array (VLA; \citealt{patil+20}, Paper II), the Very Long Baseline Array (VLBA; Lonsdale et al.\ in prep), and OIR imaging and spectra using the Large Binocular Telescope (LBT, Whittle et al.\ in prep), Gemini \citep{kim+13}, and the Very Large Telescope (VLT, \citealt{ferris+21}). 

Paper I showed that the rest-frame MIR-submillimeter spectral energy distributions (SEDs) are AGN dominated with a possible contribution from a starburst, and are therefore similar to the radio-blind samples of Hot DOGs. The sources also have high bolometric luminosity similar to those of ULIRGs and HyperLIRGs ($L_{IR} > 10^{12-13} L_\odot$). 
They are typically found in over-dense environments, suggesting some of our sources are likely to be tracers of unvirialized protocluster regions \citep{silva+15, jones+15, penny+19}, consistent with both observations  \citep[e.g.,][]{miley+08, dannerbauer+14} and simulations \citep[e.g.,][]{chiang+17} of $z>2$ radio-loud quasars. 
The black hole masses have been estimated from MIR-submillimeter SED modeling to be in the range log(M$_{BH}$/M$_\odot$) =  $7.7-10.2$ (Paper I), and from [OIII] line luminosities as a proxy for bolometric luminosity to be in the range log(M$_{BH}$/M$_\odot$) = $7.9-9.4$ \citep{kim+13,ferris+21}. 
\citet{kim+13} and \citet{ferris+21} have also measured broad [OIII] lines (FWHM $\sim 1000-2000$ km s$^{-1}$)  
suggesting strong AGN or jet-induced outflows. 

Paper II presented 10 GHz  sub-arcsecond-resolution VLA images of 93\% (155) of the sample. While 57\% are unresolved (median upper limit $0.1^{\prime\prime}$, $<0.8$ kpc at $z\sim2$) the remainder are still compact (median $1.0^{\prime\prime}$, $8$ kpc at $z\sim2$). 
The radio characteristics of many sources are consistent with powerful ($26.8< \log (L_{1.4\;\textrm{\scriptsize GHz}}/\textrm{W Hz}^{-1} ) < 28 $), sub-galactic, and high pressure ($\log (P_{\textrm{\scriptsize10 GHz}}/ \textrm {dyne cm}^{-3}) >-7$) radio sources typical of those seen in compact and young radio AGN \citep[e.g.,][]{ readhead+96, orienti+14}. The radio sources in our sample are therefore similar to other well-known classes of young and compact radio sources: the Compact Steep Spectrum (CSS) sources, Gigahertz-Peaked Spectrum (GPS)  sources, and High-Frequency Peakers (HFP) \citep[see reviews by][]{odea+98, odea+20}.

While Paper II focused on radio morphology, in this paper we study the radio spectra from $\sim 0.1 - 10$ GHz by combining our 10 GHz observations with archival survey data. Such spectra can yield important additional information about the radio source, its environment, and its evolutionary stage. For example, compact radio sources often have a peak in their radio spectra at frequencies near or below $\sim 1$~GHz which is thought to arise from absorption at lower frequencies \citep[e.g.,][]{dekool+89, tingay+03}. The source of the absorption is often unclear but could arise from high synchrotron optical depth within the radio source \citep[i.e., synchrotron self-absorption; e.g.,][]{ken+66} or free-free absorption from surrounding ionized gas \citep[e.g.,][]{bicknell+97}. Establishing either of these processes would yield important information about the radio source properties and/or the near nuclear environment. In addition, a possible correlation between peak frequency and source size has been interpreted within an evolutionary framework \citep[e.g.,][]{odea+98, orienti+14}, and so measuring spectral shapes can also shed light on the age of our radio sources. In a different context, many studies of compact radio sources have focused on outflows generated by radio jets, but these tend to be limited to objects at lower redshift \citep[e.g.,][]{odea+20}. Our sample also provides an opportunity to pursue this important process at higher-redshift, near cosmic noon.





Here is an outline of our paper. In Section~\ref{sec:sed_sample}, we summarize our sample selection. In Section~\ref{sec:radio_data} we describe our 10~GHz observations and the archival radio data. In Section~\ref{sec:spectral_fitting}, we present our radio spectral fitting technique and classify the various spectral shapes. In Section~\ref{sec:shape_morph}, we present the relative frequency of spectral shape classes and their relation to the radio source morphology. In Sections~\ref{sec:steep_alpha},~\ref{sec:mir_radio}, and~\ref{sec:alma_emission}, we discuss the key results from the radio spectral measurements. In Sections~\ref{sec:lin_vs_to} and~\ref{sec:peaked}, we analyze a subset of sources with peaked spectra, and in Section~\ref{sec:nonpeak}, we briefly discuss the properties of non-peaked sources. Finally, in Section~\ref{sec:sed_conclusion}, we summarize our main results. 
Throughout, we adopt a $\Lambda$CDM cosmology with $H_{0}$ = 67.7  km s$^{-1}$ Mpc$^{-1}$, $\Omega_{\Lambda}$ = 0.691, and $\Omega_{\textrm{\small{M}}}$ = 0.307 \citep{planck+15}.

\section{Sample Selection}\label{sec:sed_sample}

Here we briefly review the sample selection (see  Paper I for a detailed description). The primary sample was selected by cross-matching the WISE All Sky catalog (WISE; \citealt{wright+10}) with the NRAO VLA Sky Survey (NVSS; \citealt{condon+98}). When available, higher resolution radio images from the Faint Images of the Radio Sky at Twenty-one centimeters (FIRST; \citealt{becker+95}) were also evaluated, and their more reliable astrometry used for the cross-matching. We required WISE sources to be unresolved with S/N $>7$ in W4($22\mu$m) or W3($12\mu$m) bands. The corresponding radio sources were selected to be unresolved at the angular resolution of NVSS ($45\arcsec$). Although sources were not strictly required to be unresolved in FIRST ($5\arcsec$), the majority (45/51) have compact morphology (Papers I and II).  
We required the radio source flux density to satisfy the criterion: $\log(f_{22 \mu \textrm{m}}/f_{21\,\textrm{cm}}) < 0$. We further required the sources to have very red colors in the three WISE bands $3.6 (W1), 4.5 (W2)$, and $12(W3) \mu$m, with color cut:  $(W1-W2) + 1.25(W2-W3) > 7$. 

Sources that satisfied the above criteria were then visually inspected in the Sloan Digital Sky Survey (SDSS; \citealt{york+00}) or Digitized Sky Survey (DSS) and required to be optically faint or undetected, as a means to reject low$-z$ sources. As a result, our final sample comprised 167 sources. 

A spectroscopic followup secured redshifts of 71/80 sources in the range $0.4-2.8$ with a median of 1.53 \citep[Paper I,][]{ ferris+21}. Overall, their submillimeter and MIR properties indicate that these are MIR-bright heavily obscured quasars, with high IR luminosities ($L_{\rm IR} > 10^{11.7} L_{\odot}$).

\section{Radio Observations}\label{sec:radio_data}

\begin{figure}[ht!]
    \centering
    \includegraphics[width=\linewidth, clip=true, trim = 0.4cm 0.2cm 0.2cm 0.2cm ]{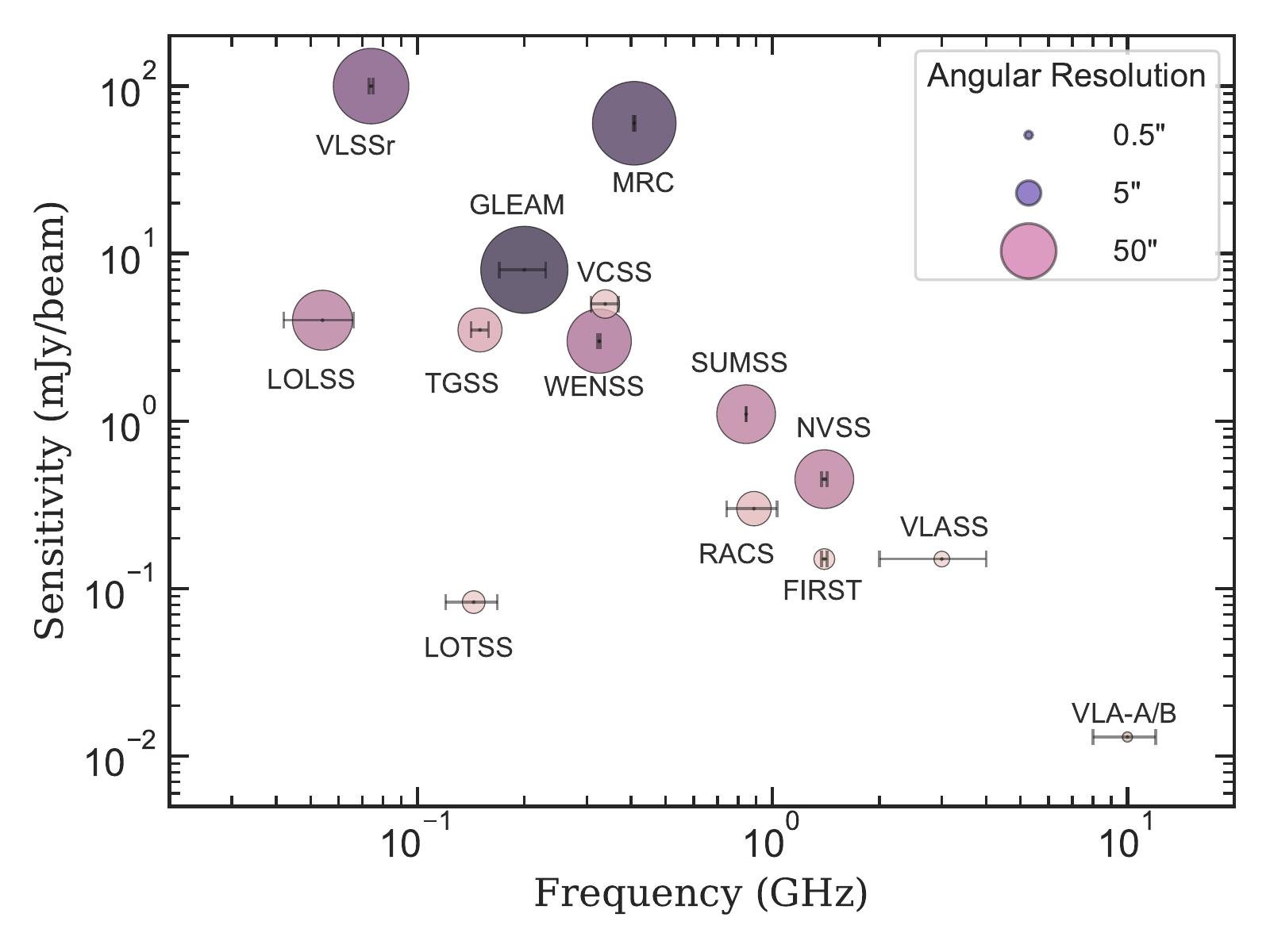}
    \caption{Comparison of metrics of surveys used in the spectral fitting. The nominal 1$-\sigma$ sensitivity is plotted as a function of frequency. The circle size is proportional to the resolution of the survey. The length of the horizontal bar corresponds to the bandwidth of the observation. The legend box shows three marker sizes for the resolutions of $0.5\arcsec$, $5\arcsec$, and $50\arcsec$, respectively.\label{fig:res_nu}}
\end{figure}

\begin{figure*}[ht!]
    \centering
    \includegraphics[width=\linewidth]{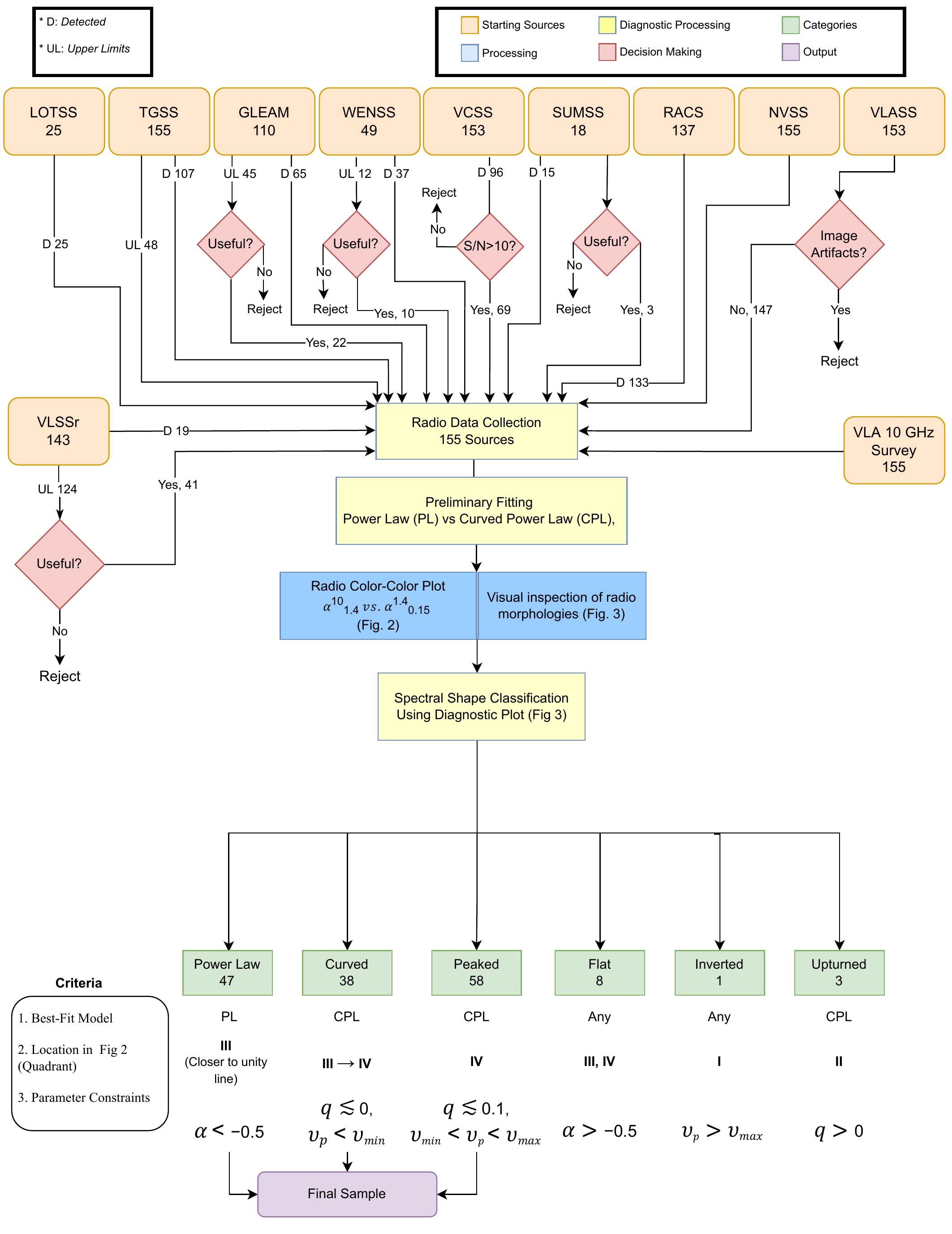}
    \caption{This flowchart summarizes the overall process of radio spectral data gathering and analysis. The survey name is followed by its central frequency in GHz, and below is the number of our sources within the footprint of the survey. Symbols ($\alpha$, $q$, $\nu_p$, $\nu_{min}$, $\nu_{max}$) are defined in Section\ref{sec:sedparam}.  ``Useful'' refers to upper limits whose value is sufficiently low to have  an impact on the fit. }
    \label{fig:flowchart}
\end{figure*}

To construct the radio spectra of our sample, we combined targeted VLA observations from Paper II with public archival radio surveys spanning a frequency range of 0.1$-$ 10 GHz. We found 12 radio surveys that have at least one successful detection. Table~\ref{tab:surveys} lists these surveys and their characteristics, and Figure~\ref{fig:res_nu} compares the resolution and sensitivity of the different surveys used. Figure~\ref{fig:flowchart} summarizes our approach to generating the spectral fits and their classification. A more detailed discussion is given in this section and the next.

\subsection{VLA 10 GHz Imaging}
We obtained 10~GHz (X-band) images of the entire sample using the Karl G. Jansky Very Large Array (VLA). Paper II presents these data and initial results. Briefly, the 10~GHz VLA images were obtained with A and B arrays, with resolution of $\sim0.2^{\prime\prime}$ and $\sim0.6^{\prime\prime}$, respectively. Overall, 118 sources have good quality data in A-array, and 147 sources have good quality data in B-array.  In total, 155 sources from the parent sample of 167 sources have followup 10 GHz VLA images.

The A-array observations were taken in October and December 2012, while the B-array data were taken in June, July, and August 2012. An identical WIDAR correlator setup was used for both arrays that provided two 1$-$GHz bands spanning $8-12$ GHz.  The imaging was performed in snapshot mode with typical on-source integration time of $\sim4-5$ minutes. 

The data were reduced in a standard manner using  the Common Astronomy Software Applications (CASA; \citealt{mcmullin+07}) package version 4.7.0. This involved manual data editing, then pipeline calibration, followed by a few rounds of self-calibration. The calibrated $uv-$data were then imaged using the CASA task CLEAN. Paper II gives further details of the data reduction and analysis.

\subsection{Archival Radio Surveys}

Radio flux densities and their uncertainties from the following surveys are given in an online table. With the exception of WENSS (see below) the flux density values are taken directly from the published catalogs with no further corrections applied for different flux density scales as discussed by \citet[][PB17]{perley+17}. The corrections are typically $<1-$few\% and do not significantly affect our spectral fits. We note that our 10 GHz flux densities used the PB17 scale. However, we added 5\% to the uncertainties of all published flux densities to allow for potential differences between the flux density scales of the different surveys. This additional uncertainty is inherent when combining the measurements from many different surveys using different instruments.


\subsubsection{NVSS}

The NVSS observed the entire northern sky ($\delta > -40$ deg) at 1.4~GHz. This survey was conducted in the D configuration of the VLA, which resulted in an angular resolution of 45\asec. As NVSS was used for our primary selection, our entire sample is detected in the NVSS and---from our selection criteria---all sources are unresolved ($<45\arcsec$) with 1.4~GHz emission brighter than 7~mJy. 

\subsubsection{FIRST}

FIRST is a 1.4 GHz VLA survey that covered 10,575 deg$^2$ of the total sky at a resolution of about 5$^{\prime\prime}$ and sensitivity limit of $\sim 0.15$ mJy beam$^{-1}$. Note that in this work we use updated FIRST flux densities from the more recent catalog of 
\citet{helfand+15}. In total, 51 of our sources have flux densities taken from the FIRST survey.

\subsubsection{TGSS-ADR1}

The Tata Institute of Fundamental Research Giant Meterwave Radio Telescope Sky Survey Alternative Data Release (TGSS-ADR1; \citealt{intema+16}) is a 150 MHz survey covering about 90\% of the total sky (36,900 deg$^2$). TGSS has a resolution of about $25^{\prime\prime}$ and rms noise of $3-5$ mJy beam$^{-1}$. Our entire sample falls within the TGSS footprint and yields 87 source detections and 68  non-detections with a median $3\sigma$ upper limit of 10.8 mJy.


 \subsubsection{VLASS}

The VLA Sky Survey (VLASS; \citealt{lacy+20}) is an ongoing 3 GHz (S-band: $2-4$ GHz) continuum survey covering 33,885 deg$^2$, which is similar to the footprint of NVSS. The planned survey will take place over 3 epochs, with the first recently completed. While the complete source catalog is still in preparation, quick-look images are currently available for public access. The typical $1\sigma$ sensitivity of a single epoch image is $\sim 120\,\mu$Jy beam$^{-1}$ with an angular resolution of 2.5\asec. 
The VLASS flux densities used in our analysis are estimates based on the epoch 1 quick-look source catalog (Lacy et al. private communication) generated using the Python Blob Detector and Source Finder (PyBDSF; \citealt{mohan+15}) with standard input parameters.\footnote{The VLASS Quicklook Epoch 1 catalog is now available \citep{gordon+20}. We find that both VLASS catalogs provide the same number of  cross matches for our sample.} We use the peak flux density from the source catalog for unresolved sources. When sources have multi-components or significantly extended single components, we measure the flux densities using the CASA VIEWER, and add to the uncertainties given by PyBDSF or the CASA VIEWER an additional systematic uncertainty of 20\% due to antenna pointing issues \citep{lacy+memo}.  
All of our sources except two are detected in the VLASS quick-look images. However, due to the preliminary nature of the quick-look imaging\footnote{For details on the limitations of the VLASS quick-look images, see \citet{lacy+20}.}, 
we visually inspected all of the VLASS image cutouts for our sample, finding six sources with severe quick-look image artifacts for which reasonably accurate (within 20\%; see \citealt{lacy+memo}) flux density estimates cannot be obtained.

\subsubsection{VCSS} 
The VLA Low-band Ionosphere and Transient Experiment (VLITE; \citealt{clarke+16, polisensky+16}) is a commensal instrument on the VLA which records and correlates data from a 64 MHz sub-band centered at 340 MHz for up to 18 antennas during nearly all regular VLA operations. Unfortunately, our VLA X-band data were taken prior to the start of VLITE operation in November 2014, so we do not have a complete set of targeted data for our sample.  However, VLITE was operational during the VLASS survey, yielding images and a source catalog: VCSS (VLITE Commensal Sky Survey; \citealt{peters+21}). Quality assurance checks are ongoing for epoch 1, and final data products are not yet publicly available.  For the work presented in this paper we have obtained preliminary images and source fits from VCSS.  

The VCSS images have a typical angular resolution of $15\arcsec$ and an rms of 3 mJy bm$^{-1}$, with some variation due to piggybacking a survey which was optimized for a much higher observing frequency.  Because these are still preliminary images, we restrict the sample to sources with an undistorted peak above 10$\sigma$ significance.  Images are available for 153/155 sources in the sample, of which 96 are reliably detected ($>10\sigma$).  The sources are fit with PyBDSF, and the fitted values are increased by 7.5\% to correct a known bias in the survey images. We also add a $20\%$ flux density  uncertainty to the PyBDSF fitting errors to reflect the preliminary nature of the measurements.

\subsubsection{RACS}
The Rapid ASKAP Continuum Survey (RACS; \citealt{mcconnell+20}) is the first all-sky survey conducted with the Australian Square Kilometer Array Pathfinder (ASKAP; \citealt{johnston+07, hotan+21}), centered at a frequency of 887.5 MHz. The first epoch observations use 36 12-m antennas to cover the entire sky south of +41$^o$ with sensitivity $25-40 \mu$Jy and $ ^{-1}$ and $\sim15\arcsec$ beam. Within the RACS footprint, 137 of our sources were observed with 133 sources detected above $10\sigma$. 

\subsection{LOTSS}
The Low Frequency Aarray (LOFAR) Two-meter Sky Survey (LOTSS) is an ongoing all sky-survey covering northern sky (declination$>34^{\circ}$ ) centered at 144 MHz \citep{shimwell+19, shimwell+22}. The second data release\footnote{See \url{https://lofar-surveys.org/dr2_release.html} for more details.} covers 27\% of the sky with a point-source sensitivity of 83 $\mu$Jybeam$^{-1}$.  A total of 25 sources fall within LOTSS-DR2 footprint, all of which are detected in the source catalog. 

\subsubsection{Other surveys}
In addition to these large surveys with good overlap for our sample, we also searched radio surveys with less complete sky coverage, shallower depths, and/or lower angular resolution. These include:
the Galactic and Extra-galactic Murchinson Widefield Array (GLEAM; \citealt{hurley+17}) survey,  
the Green-Bank 6-cm Radio Source Catalog (GB6; \citealt{gregory+96}), the Sydney University Molonglo Sky Survey (SUMSS; \citealt{mauch+03}),  the Texas Survey of Radio Sources (TEXAS; \citealt{douglas+96}), the Westerbork Northern Sky Survey (WENSS; \citealt{rengelink+97}), the VLA Low-Frequency Sky Survey (VLSSr; \citealt{lane+14}), LOFAR LBA Sky Survey (LOLSS; \citealt{gasperin+21}), and Molonglo Reference Catalog of Radio Source (MRC; \citealt{large+81}).
For WENSS, we decreased all flux density values by 19\% to convert to the Baars scale~\citep{hardcastle+16}. 
Due to the relatively low angular resolution ($\sim1^{\prime}$)  of GLEAM, SUMSS, WENSS, and VLSSr, some sources with catalog detections suffer from source blending.  We discuss this further in Section~\ref{sec:resolution}.  The number of sources with available data from each survey, as well as the number of detected sources, is summarized in Table~\ref{tab:surveys}. Finally, we found no useful flux density measurements for our sources in the Australia Telescope 20 GHz Catalog (AT20G; \citealt{murphy+10}).
The sensitivity limits of AT20G is too high for our sources. 

\input{Survey_tab}


\subsection{Multi-resolution Concerns}\label{sec:resolution}
The radio observations come from a  range of telescopes with different resolutions. This can affect the flux density measurements and the form of the radio spectra. The low-frequency  ($<$1~GHz) surveys typically have a larger synthesized beam and are thus more sensitive to diffuse emission and are also more likely to suffer from source confusion.  In contrast, higher-resolution observations, typically at higher frequencies, 
may resolve out extended low surface brightness emission. When combined, these effects can cause artificial steepening of the radio spectrum, since the high-frequency observations are missing emission.

To assess these possible resolution effects, we visually inspected our sample in all the available surveys. 
We find that the majority of our sample (83/155) show no complex structure in any of the primary surveys, which are TGSS, LOTSS, VCSS, RACS, NVSS, FIRST, VLASS, and our VLA 10 GHz observations. The remaining surveys with lower angular resolution than NVSS were compared with NVSS image cutouts to look for possible source blending.
We find 2/31 WENSS sources and  1/9 VLSSr sources show source confusion and these observations are therefore excluded from our analysis. As a further check, we inspected optical images of our sample in the Panoramic Survey Telescope and Rapid Response System (Pan-STARRS; \citealt{chambers+16}) for potential false identifications or confusion with nearby sources, especially for radio surveys with $\sim$ arcminute resolution. We find 7/39 GLEAM sources may be affected in this way, and their GLEAM flux densities were not included in our analysis. 

The recent LOTSS survey is particularly helpful in detecting fainter extended emission from older radio sources because of its high surface brightness sensitivity ($\sim80~\mu$Jybeam$^{-1}$) and relatively high (6\arcsec) resolution at low frequencies \citep[e.g.,][]{brienza+17, mahatma+18, jurlin+20}. Of the 25 sources in the LOTSS DR2, all but one are consistent with a compact source. Source J1238+52 is a distorted triple, with emission components $\sim 7\arcsec$ N and $\sim 10\arcsec$ SW of the nuclear source. The extended components aren't detected in any other survey, suggesting they have steep spectral index ($\alpha < -1$). Since we could isolate the central component and measure its flux, we have kept it in our sample for analysis. 

Additionally, 38 sources are extended in high-frequency observations. Of these, 19 appear to have missing emission at higher frequency and we exclude these sources from the analysis. The remaining 19 have single-power-law or upturned spectra and so we keep them in our analysis.


In summary, the majority (96/155) of the sources included in our spectral analysis are unresolved at all frequencies. 

\subsection{Faint Detections and Upper Limits}
Radio surveys typically publish source catalogs using a S/N threshold of $\sim 5-7\sigma$ in order to reduce the inclusion of false detections and image artifacts. While this is appropriate for a blind survey, we have prior source positions and this allows us to measure flux densities  that are below the formal catalog threshold. By inspecting the survey's images at our source locations, we find 6 VLSSr sources, 2 WENSS sources, 16 GLEAM sources, 3 SUMSS sources, and 24 TGSS sources below the corresponding catalog source  detection limit. 
We use CASA task IMFIT to obtain flux density  measurements of these faint sources which typically have S/N ratios between $\sim 3.9-5$. Conversely, we found  45 GLEAM sources, 12 WENSS sources, 124 VLSSr sources, and 48 TGSS sources are undetected. For these  we use $5\,\sigma$ as an upper limit on the flux density, where $\sigma$ is the rms noise level in the image. Due to relatively poor sensitivity, the upper limits for fainter sources are not usually useful in constraining the spectral shape. After visual inspection, we excluded these upper limits from our spectral fitting in 83/124 VLSSr, 2/12 WENSS, and 23/45 GLEAM non-detections.


\section{Radio Spectral Fitting}\label{sec:spectral_fitting}
The radio spectrum of a source contains information about the source's physical condition. The observed radio spectrum is usually thought of as a combination of emission processes, energy losses, and absorption processes. In the case of an AGN, the emission is thought to be synchrotron emission with a power-law energy distribution of relativistic electrons generating a power-law radio spectrum, increasing to lower frequency. Over time, the most energetic electrons can radiate their energy causing a break to steeper spectra at high frequencies. Conversely, at lower frequencies, the radiation can be absorbed by the relativistic electrons (synchrotron self-absorption) and/or by thermal electrons (free-free absorption), causing a spectral turnover with characteristic inverted slope at low-frequencies. Thus, identifying and measuring these features in a radio spectrum can help ascertain a number of important physical properties of the radio source.

\subsection{Fitting Procedure}

We developed a suite of Python tools to perform all of the spectral modelling presented here. We have made these tools available to the community on Github\footnote{\faGithub\,\href{https://github.com/paloween/Radio_Spectral_Fitting}{Radio\_Spectral\_Fitting}}. Given the sparse spectral sampling, we choose not to fit idealized physical models to the spectra, such as synchrotron self absorption (SSA) or free-free absorption (FFA).  
Instead, we use simple functions to characterize the overall form of the spectrum in $\log \nu-\log S_\nu$, and then use these fits to help guide our physical interpretation. Our overall approach is to first fit a power law, and if significant deviations are found then fit a parabola \citep[see e.g., ][]{callingham+17}. Hence, the underlying synchrotron emission mechanism can be captured by the power law, while any curvature on the high or low frequency side, including a turnover, can be captured by the parabola in log space.  
 
The power law is given by:  

\begin{equation}\label{eqn:pl}
    S_\nu = S_o\,\nu^{\alpha}
\end{equation}
where $S_\nu$ is the flux density in mJy at frequency $\nu$ GHz, $S_o$ is the flux density in mJy at 1 GHz, and $\alpha$ is the spectral index.

The parabola is given by: 

 \begin{equation}\label{eqn:cpl}
    S_\nu = S_o\,\nu^\alpha\,e^{q(ln\nu)^2}
\end{equation}
which has peak frequency in GHz, $\nu_{\rm peak}$ and flux density  at the peak in mJy, $S_{\rm peak}$, given by  

\begin{equation}\label{eqn:cpl_nup}
\nu_{peak} = e^{-\alpha/2q} 
\end{equation}
and
\begin{equation}\label{eqn:cpl_Sp}
S_{peak} = S_o e^{-\alpha^2/4q} .
\end{equation}
This function describes a parabola in $\log S_\nu$ vs.\ $\log \nu$ (a Gaussian in $S_\nu$ vs.\ $\log \nu$), where $q$ characterizes the width of the peak, or the degree of curvature, with full width at half the peak flux density (for negative $q$), $S_{\rm peak}/2$ of $\log \textrm{FWHM} = 0.72/\sqrt{-q}$. The function becomes a power-law of index $\alpha$ as $q\to0$. Significant spectral curvature is usually defined as $|q| \ge 0.2$ \citep{duffy+12}.

We perform the data fitting using a $\chi^2-$minimization routine\footnote{We used the curve\_fit function available in a Python module called SciPy \citep{2020SciPy}.} that uses the Lavenberg-Marquardt algorithm. This method is useful for solving non-linear equations but can sometimes find local rather than global minima for the best-fit solution. To guard against this possibility, we visually inspected the best-fit solutions; those with poor-fits were fitted again after adjusting the range of initial parameters.  
We chose not to include the 10 GHz in-band spectral indices ($\alpha_{IB}$) in our fit because this high-frequency index is astrophysically important and we prefer to keep two independent measurements of it for our analysis. In section~\ref{sec:alpha_distrib}, we confirm the overall agreement between $\alpha_{IB}$ and the slope of our fit at 10 GHz. In just three cases, the spectral sampling was so sparse we chose to include $\alpha_{IB}$ in the fit (See Section 4.4 in Paper II for further discussion on $\alpha_{IB}$).


\subsection{Spectral Shape Classification}\label{sec:sedclass}
We inspected all radio spectra to assess the best-fit model and checked the reduced $\chi^2$ and $q$ to decide which model fits best.  We divide our sample into six broad categories:


\begin{figure}[ht]
\centering
\includegraphics[width=\linewidth]{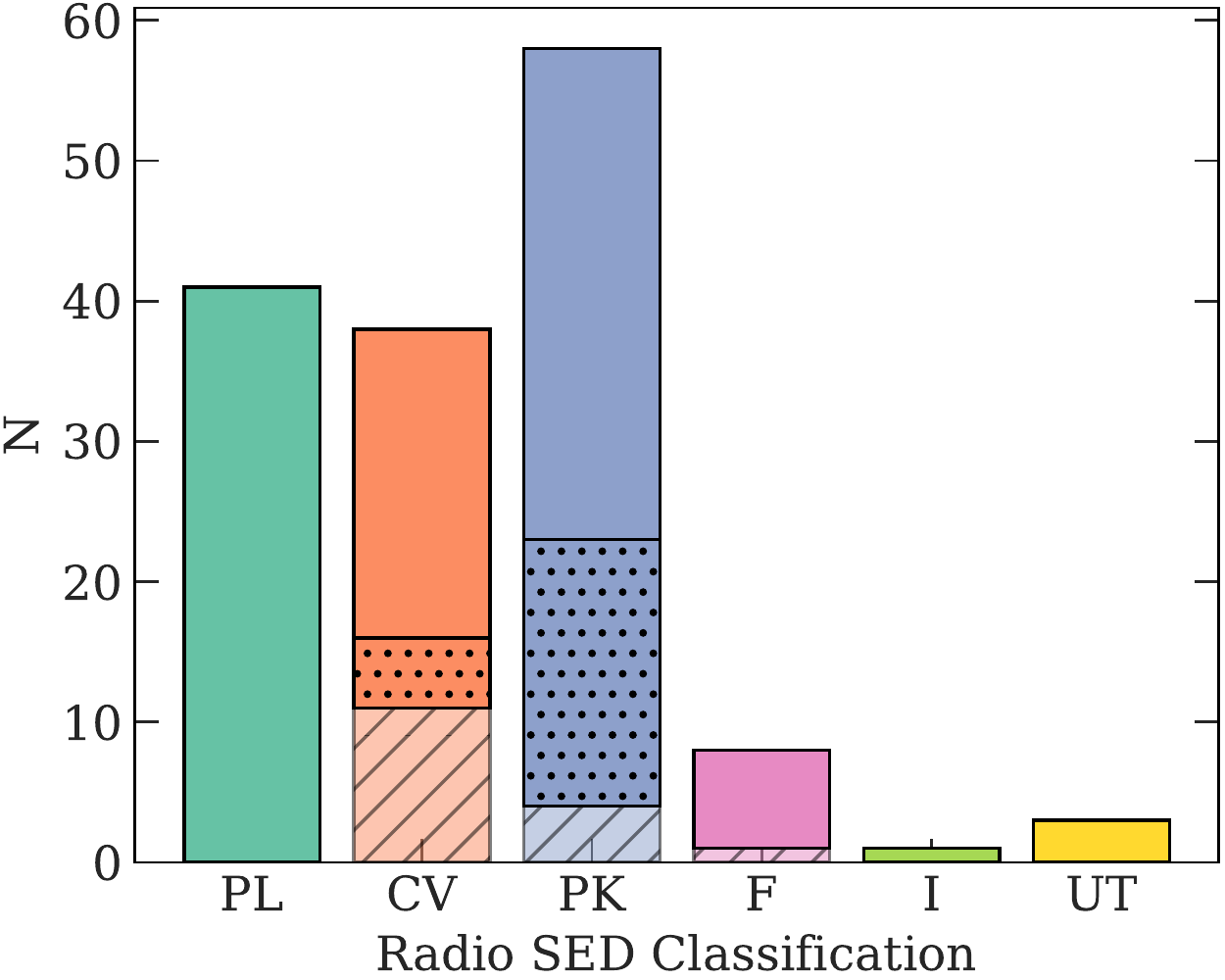}
\caption{The distribution of radio spectral shape classes in our sample of 155 sources. The dotted portion in the middle gives the number of sources with relatively poor quality data. The hatched portion at the bottom indicates the number of sources rejected from the final sample (Section~\ref{sec:final}) due to various issues with spectra (Section~\ref{sec:resolution}). The $x-$axis labels are the radio spectral shape classes defined in Section~\ref{sec:sedclass}. The codes are PL: Power-law, CV: Curved, PK: Peaked, F: Flat, I: Inverted, and UT: Upturned.\label{fig:spshape}}
\end{figure}

\begin{itemize}
    \item \textbf{Power Law (PL):} This is a standard power law given by Equation~\ref{eqn:pl} spanning the full spectral range with relatively steep spectral index $\alpha<-0.7$ consistent with optically thin synchrotron emission.  Either the reduced $\chi^2$ is lower for the power-law model or the $q$ value in the parabolic fit is consistent with zero within its uncertainty.  
    
     \item \textbf{Peaked  (PK):} A turnover is detected  within the observed spectral range. 
     Thus, if $\nu_u$ and $\nu_l$ are the 1$\sigma$ upper and lower limits on the best-fit $\nu_{\rm peak}$, then $\nu_u<\nu_{max}$ and $\nu_l>\nu_{min}$, where $\nu_{max}$ and $\nu_{min}$ are the lowest and highest frequencies of the observations in an individual spectrum.

    \item \textbf{Curved (CV):} A radio spectrum is classified as curved when there is a significant deviation from the power-law model, but no peak is seen within our spectral range. Either the reduced $\chi^2$ value is lower for the curved power law model or $|q|\gtrsim0.1$.  
    Usually $\nu_{min}$ is  greater than the calculated peak frequency, $\nu_{\rm peak} + \sigma_{\nu_{\rm peak}}$.

    \item \textbf{Flat (F):} When the $\alpha$ estimated from either power-law or curved-power model is $|\alpha| < 0.5$.
    
    \item \textbf{Inverted (I):} Sources from this class have a steeply rising spectrum, $\alpha>0.5$.
    
    \item \textbf{Upturned (U):} 
    Sources that show a concave spectrum are categorized as having upturned spectrum. 
\end{itemize}

Figure~\ref{fig:spshape} shows the distribution of spectral classes for our entire sample with less reliable spectra shown hatched. We find  47 sources with power-law spectra, 38 sources with curved, 58 sources with peaked,
eight sources with flat, one source with inverted spectrum, and three sources with upturned spectra.  Table~\ref{tab:alpha_calc} lists the spectral shape classification of our entire sample.

\begin{figure*}
\centering
\includegraphics[width=0.9\linewidth,clip=true,trim=28cm 8cm 8cm 6cm]{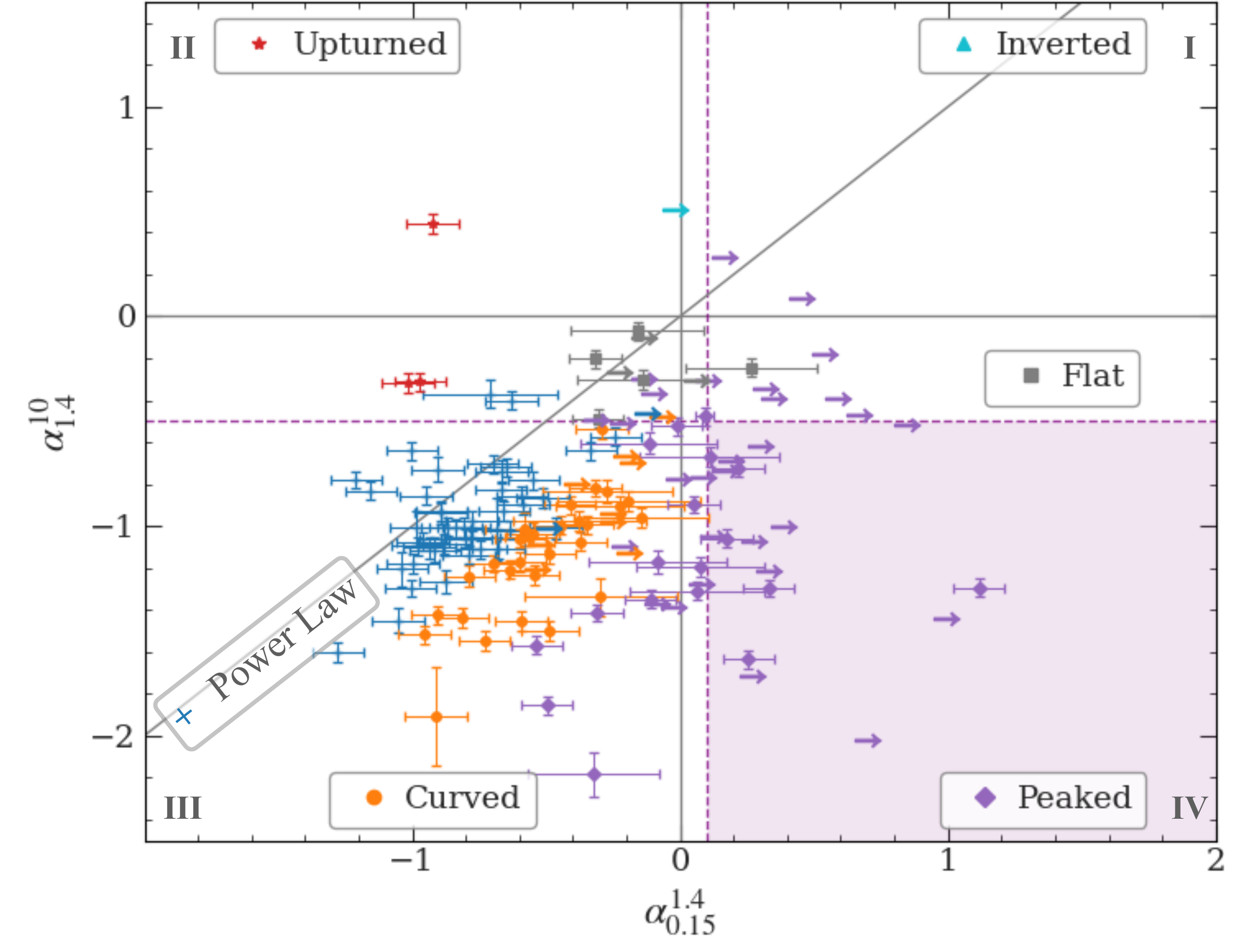}
\caption{The radio color-color plot showing spectral indices, $\alpha_{1.4}^{10}$ and $\alpha_{0.15}^{1.4}$, calculated using flux densities from TGSS (0.15 GHz), NVSS (1.4 GHz), and our VLA 10 GHz survey. 
Different symbols represent different radio spectral shape classes described in Section~\ref{sec:sedclass}. These include: Blue cross: Power law; Orange circle: Curved; Purple diamond: Peaked; Gray square: Flat; Light blue triangle: Inverted; and Red star: Upturned. The horizontal arrows are lower limits to the value of  $\alpha_{0.15}^{1.4}$. The two dashed lines are taken from \citet{callingham+17} and they define the peaked spectral shapes which lie in the purple shaded region. 
\label{fig:rcolor} 
}
\end{figure*}

\subsubsection{Radio Color-Color Plot}
A common, though simpler, approach to characterising radio spectra is to use flux densities measured at three widely separated frequencies to define two radio colors (see  Figure~\ref{fig:rcolor}). In our case, we choose flux densities at 10 GHz (our survey), 1.4 GHz (NVSS) and 150 MHz (TGSS), which together define two spectral indices: $\alpha_{0.15}^{1.4}$ and $\alpha_{1.4}^{10}$. 
As one expects, most of our classifications lie in the appropriate quadrant. However, there are exceptions. For example, if a peak falls close to the two outer frequencies, it may not fall in the ``peaked'' quadrant. We include this plot in part to provide complementary quantification of spectral shape and in part to allow comparison with other work.



\subsubsection{The Final Sample}\label{sec:final}

\begin{figure*}
    \centering
    \setlength{\fboxsep}{1pt}%
    \setlength{\fboxrule}{1.2pt}%
    \includegraphics[width=0.8\textwidth]{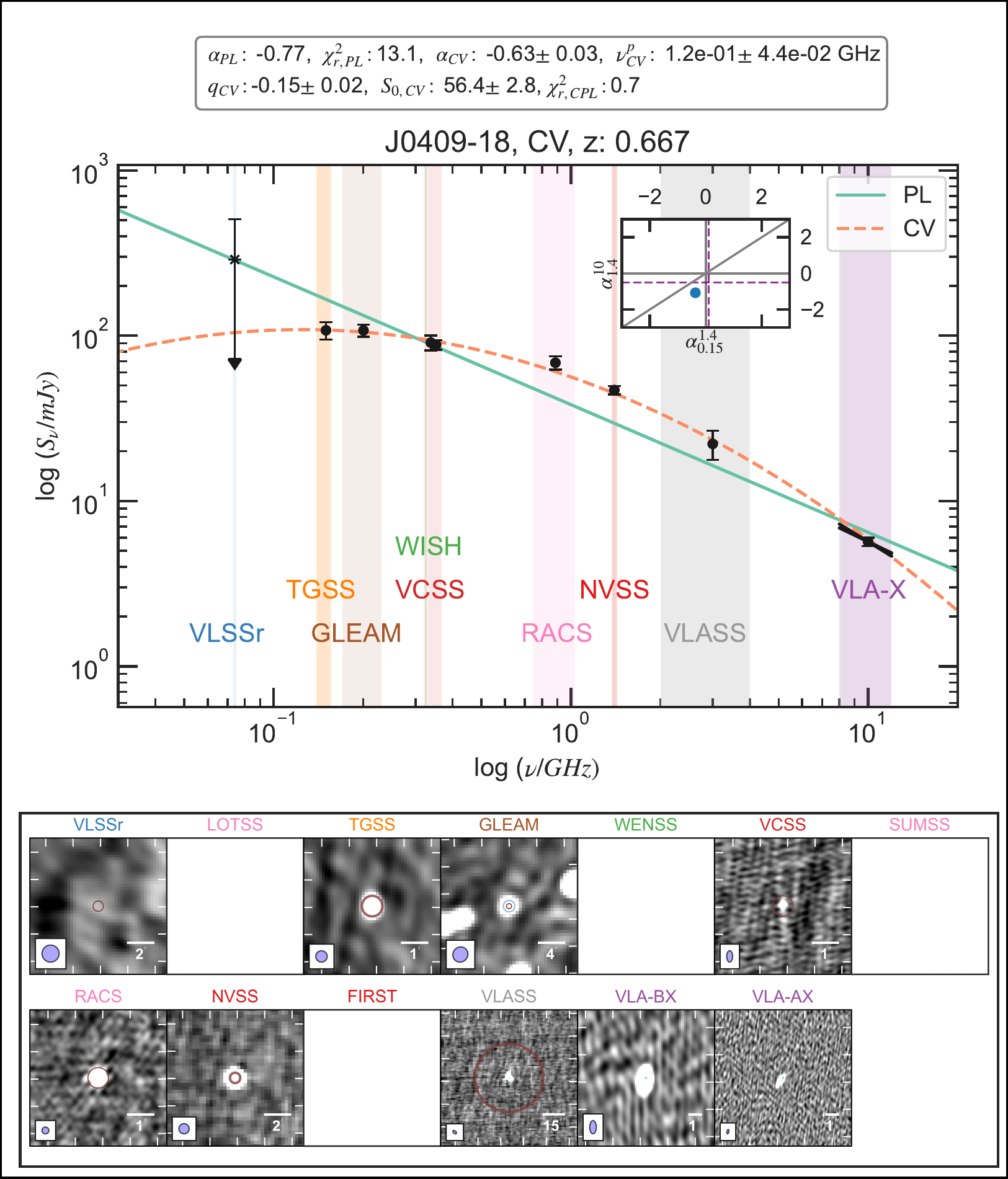} 
    \caption{Example illustrating our approach to classifying spectral shape class and assessing its reliability.  The main top panel shows the radio spectrum of J0409-18. The solid green and dotted orange lines are the  best fit power-law (PL, Equation~\ref{eqn:pl}) and curved parabola (CV, Equation~\ref{eqn:cpl}) solutions, respectively. The text box at the top lists their best-fit  parameters, uncertainties, and reduced $\chi^2$ values. The colored vertical stripes show the spectral range for each radio survey. We also show the in-band spectral index over the entire X ($8-12$) GHz band as indicated by the two crossed lines at the centered at the 10 GHz. See Section~\ref{sec:spectral_fitting} for additional details on the incorporation of bandwidth in our spectral fitting.
    The radio color-color plot is in the inset (see Figure~\ref{fig:rcolor} for reference) with the location of the target shown by the blue dot. The bottom panel shows image cutouts from each survey, with its synthesized beam in the lower-left (purple) and angular scale on the lower-right of each cutout. The red circle is the typical NVSS resolution beam of $45\arcsec$. 
    The cyan symbol is the location of the WISE position and its size shows the $2\sigma$ positional uncertainty of $0.14^{\prime\prime}$. 
     The median $2\sigma$ WISE positional uncertainty for our sample is $\sim 0.25^{\prime\prime}$.}
    \label{fig:sedcheck}
\end{figure*}

Figure~\ref{fig:sedcheck} illustrates the various issues described above that result in our final sample of sources with reliable power law, peaked, or curved spectra. The spectral shape class is assigned based on a comparison of the two fitted spectra (power law or parabola) and the location of the source in the radio color-color plot. Visual inspection of the various survey images allows us to keep 96 out of 155 sources that are unresolved in all surveys. 
Of the remaining 59 sources, we rejected from our final sample 22 due to issues  arising from using multi-telescope data (11 curved, one flat, four peaked, and six power-law; See Section~\ref{sec:resolution}). We kept the remaining 37 which did not seem to suffer from any resolution-dependent effects.  Our final sample now includes a total of 133 sources. 


%
All of the results are tabulated in Table~\ref{tab:alpha_calc} which provides the spectral shape classification (where ``:'' indicates an uncertain classification), the two$-$band spectral indices used in Figure~\ref{fig:rcolor}, and an indication of whether the source is included in the final spectral sample. 


\begin{figure*}[ht!]
\includegraphics[width=\textwidth]{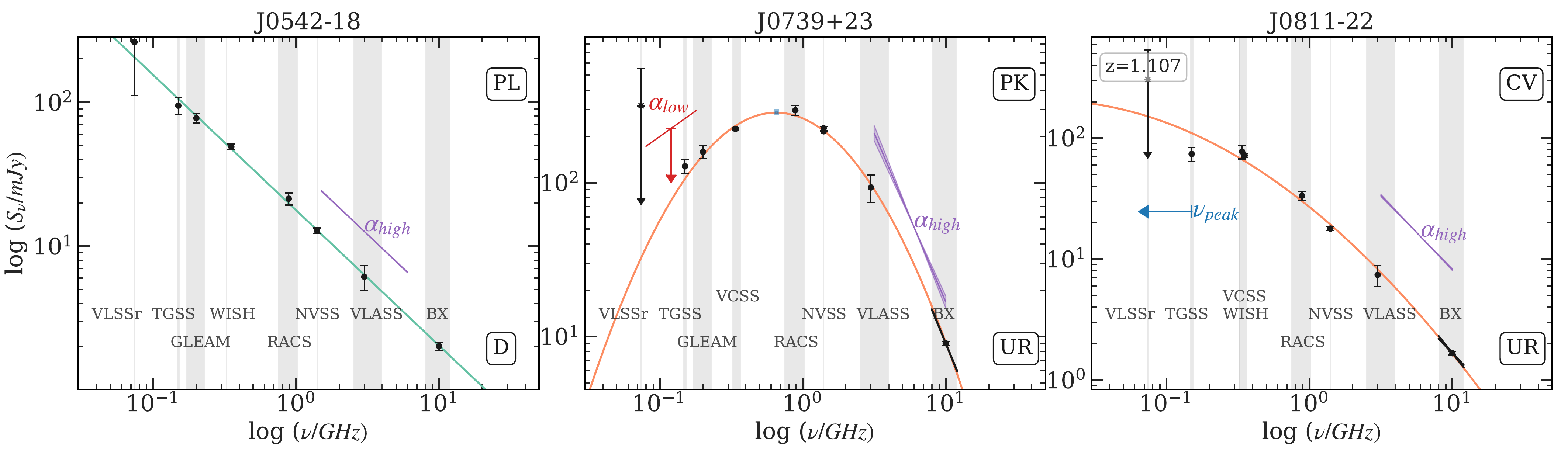}
\caption{Three different spectral shapes illustrating our parameters. The overall format of each panel is described in the caption to Figure~\ref{fig:sedcheck}.  The best-fit power-law (green) or parabola (orange) is shown. Spectral shape class and 10~GHz radio morphology class are given in the upper  and lower right boxes.  In the top left, the spectroscopic redshift is given, when available. \textbf{Left:} Power-law spectrum.  The purple line with its bow-tie is the best-fit power-law and uncertainty.  \textbf{Middle:} Peaked spectrum. The red and purple lines with bow-ties show $\alpha_{\rm low}$ and $\alpha_{\rm high}$, with their uncertainties.  The blue shaded box in the center gives the location of $\nu_{\rm peak}$ and $S_{\rm peak}$ with their $1~\sigma$ uncertainties.
\textbf{Right:} A curved-spectrum. The purple line and bow-tie show $\alpha_{\rm high}$ and its uncertainty estimated from VLASS, NVSS, and our 10 GHz data. The blue arrow denotes the upper limit to $\nu_{\rm peak}$.
}\label{fig:sedparam}
\end{figure*}

\begin{figure*}[ht!]
    \centering
    \includegraphics[width=\textwidth]{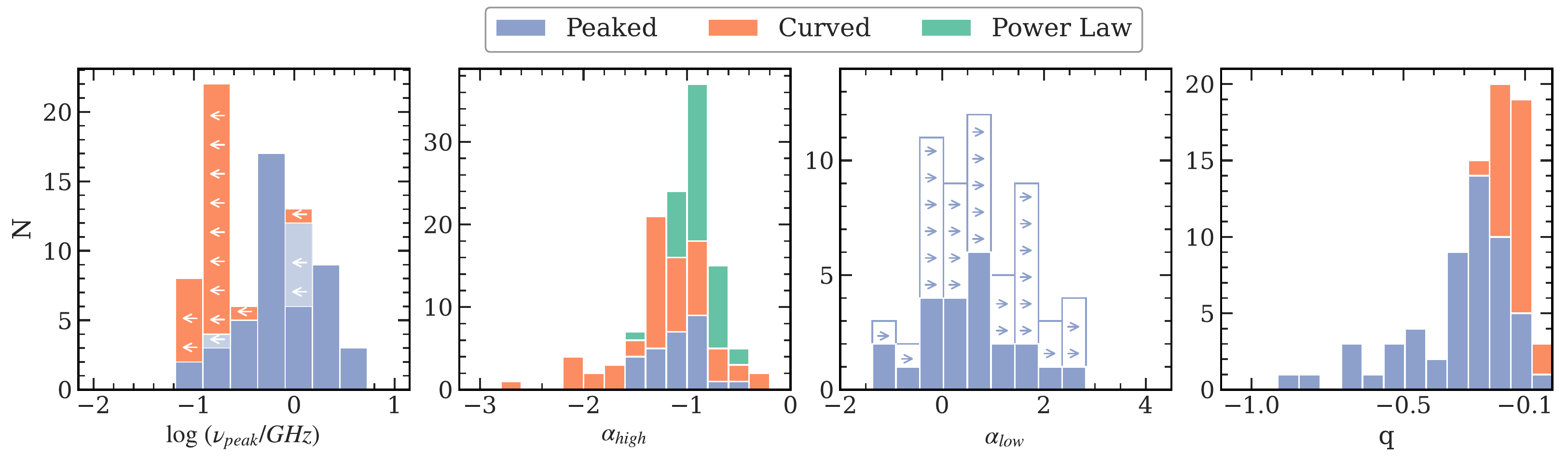}
    \caption{Distribution of radio spectral shape parameters (see caption to Figure~\ref{fig:sedparam}). Arrows indicate limits. In the first panel, faded blue bars with arrows show peaked sources that have large uncertainty the peak frequency. In the third panel, $\alpha$ limits refer to peaked sources with flux density upper limits for one or more measurements below the peak frequency. }
    \label{fig:paramd}
\end{figure*}

\subsection{Spectral Shape Parameters}\label{sec:sedparam}
Although we use a straight line and parabolic fits to help classify the spectral shape, we measure additional parameters that are more commonly used to discuss the physical properties of the radio source. These parameters include: the location of the peak frequency, $\nu_{\rm peak}$; spectral indices, $\alpha_{\rm high}$ and $\alpha_{\rm low}$, at frequencies greater and less than $\nu_{\rm peak}$; and the $q$ parameter from Equation~\ref{eqn:cpl} to indicate the width of the peak or degree of curvature. These are now defined in slightly more detail. 
\begin{itemize}
    \item $\alpha_{\rm high}$: This spectral index is thought to characterize the optically thin part of the underlying synchrotron emission. It is most useful for steep power-law, curved, and peaked sources. For peaked sources, we estimate $\alpha_{\rm high}$ using all the flux density  measurements available at  $\nu > \nu_{\rm peak}$. For curved sources, we use flux densities at $\nu\gtrsim1$~GHz. For power-law spectra, $\alpha_{\rm high}$ is simply the best-fit value of $\alpha$. 
    
    \item $\alpha_{\rm low}$: 
    We only estimate $\alpha_{\rm low}$ for sources with peaked and inverted spectra. Since spectral turnover is likely to result from absorption, the value of $\alpha_{\rm low}$ can help distinguish between different kinds of absorption, such as free-free absorption (FFA) or synchrotron self-absorption (SSA).  We estimate its value using all flux density measurements at $\nu < \nu_{\rm peak}$. 
    
    \item $\nu_{\rm peak}$: For peaked spectra, we calculate $\nu_{\rm peak}$ using Equation~\ref{eqn:cpl_nup}, with an uncertainty  estimated by propagation of errors.  For curved spectra, we use $\nu_{min}$ as the upper limit of   $\nu_{\rm peak}$, where $\nu_{min}$ is the lowest frequency observations available. Similarly, for inverted spectra, we take  10~GHz as the lower limit to  $\nu_{\rm peak}$. If the radio spectrum is peaked due to absorption, the value of $\nu_{\rm peak}$ and $S_{\rm peak}$ can give valuable information on the source properties (see Section~\ref{sec:peaked}).
    
    
    \item $q$: This parameter is only given for peaked or curved spectra and is taken directly from the fit of  Equation~\ref{eqn:cpl}. It indicates the width of the peak or the degree of spectral curvature. Its value is likely determined by the nature of the absorption (on low frequency side) and spectral aging (on the high-frequency side) and also on the complexity of the source (e.g., simple vs.\ multiple screen; single vs.\ multiple electron populations).  

\end{itemize}

Figure~\ref{fig:sedparam} illustrates these parameters using three sources with different radio spectral class. 
We list all these parameters and their uncertainties for the final sample  in Table~\ref{tab:alpha_calc}, and Figure~\ref{fig:paramd} shows their distributions.


For the curved sources, peak frequency is $<400$ MHz. The median peak frequency for the peaked sources is 887 MHz with an interquartile range from 450 MHz to 1.6 GHz, though this range reflects, in part, the frequency range used in our study. For sources classified as having curved spectra, the upper limit of peak frequency is  either 150 MHz or 74 MHz depending on whether data from TGSS or VLSSr is available. The distribution of $\alpha_{\rm high}$ (all spectral types) ranges from $-0.3$ to $-2.7$ with a median near $-1.01$. Many sources therefore have somewhat steeper spectra than the canonical index of $-0.7$ for optically thin synchrotron, and we discuss this result further in Section~\ref{sec:steep_alpha}.  The distribution of spectral indices below the peak, $\alpha_{\rm low}$, (peaked sources only) is dominated by lower limits with measured values in the range $0 - 3$. Finally, we find $q$ values in the range $-0.1$ to $-1.0$ with an equivalent range in full-width half maximum (FWHM) of $4.0$ to $0.5$ dex.

\section{Radio Spectral Shape vs. Radio Morphology}\label{sec:shape_morph}
In \citet{patil+20}, we used our VLA images to classify the 10 GHz radio morphology of the sample, at resolutions of $\sim 0.1$\asec\ and $\sim 0.6$\asec\ for the A- and B-array observations, respectively. We now look to see if there is any relation between morphology classifications and spectral shape.  As discussed in Sections~\ref{sec:resolution} and~\ref{sec:final}, we include all PK, CV, and PL sources that are part of the final sample irrespective of their 10 GHz morphology. This then allows us to see if there is any relation between morphological classification at 10 GHz (compact, double, triple etc) and the spectral index of the entire source (PL, CV, PK etc).

\begin{figure}[t]
    \centering
    \includegraphics[width=\linewidth]{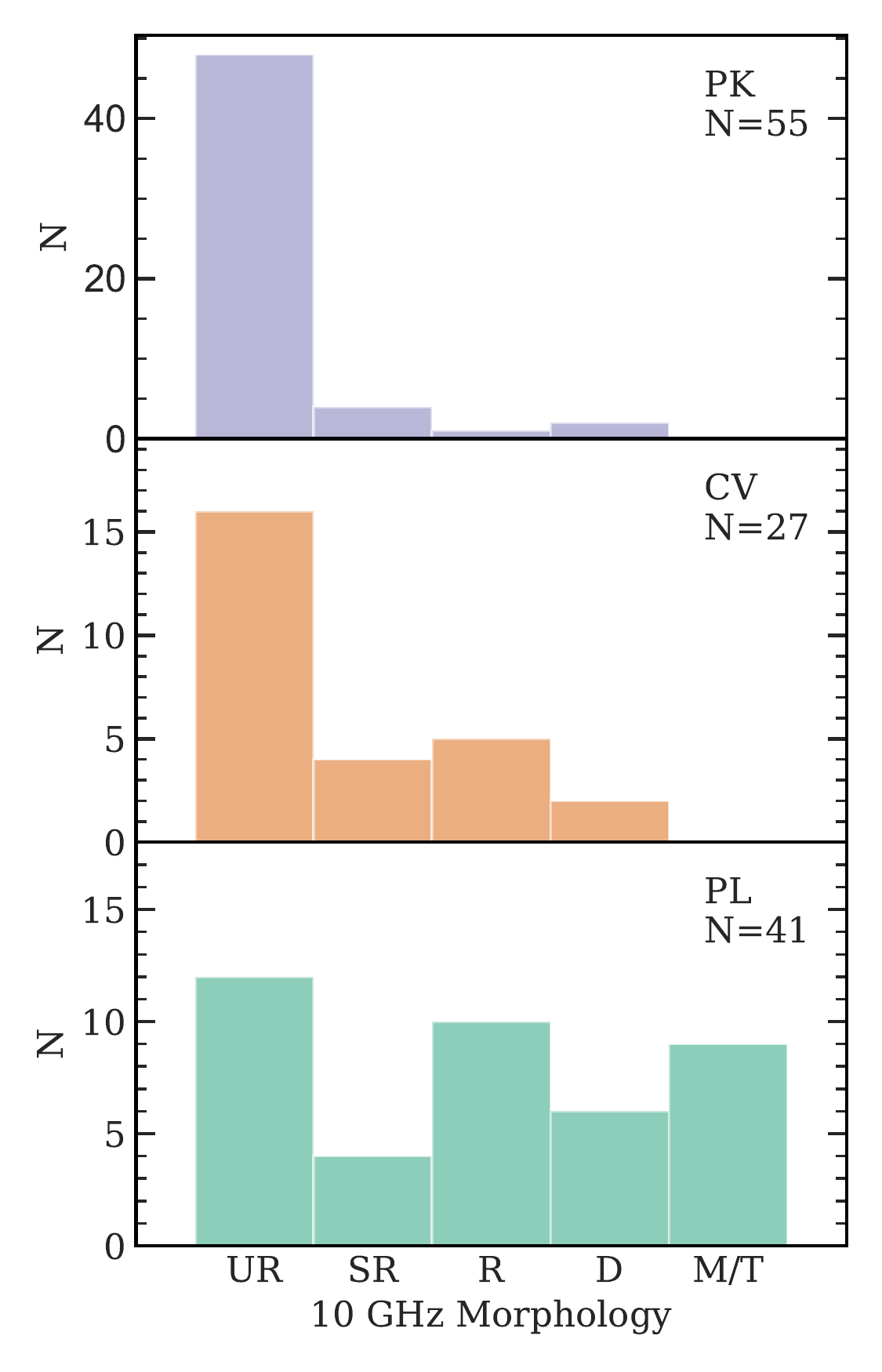}
    \caption{Histograms of source morphologies at 10 GHz for three spectral shape classes, PK, CV, and PL. The morphological categories go from most compact (unresolved; UR) on the left to most extended (multi- or triple-component; M/T) on the right. We find  a tendency for the peaked sources to  have most compact and unresolved morphologies.}
    \label{fig:spectra_vs_morph}
\end{figure}

Figure~\ref{fig:spectra_vs_morph} shows histograms of source morphology, with increasing compactness to the left, for the three radio spectral shape classes, PK, CV, and PL. The radio morphological classes as defined in \citet{patil+20} are unresolved (UR), slightly/marginally resolved (SR), resolved (R), double (D), and multi-component/triple (M/T). 
There is a clear tendency for the sources with peaked spectra to be unresolved.  Indeed, while 87\% of the peaked sources are unresolved, only 29\% of the power law sources are unresolved. 
We ran the Fisher exact test for a $2\times2$ contingency table  and found this difference to be significant at 5.8$\sigma$ (p-value = 5.3$\times10^{-9}$) level.  There is some evidence that the sources classified as ``curved'' (CV) are also preferentially compact, as one would expect if they were also self-absorbed but with peaks at lower frequencies (below our lowest spectral window), suggesting a lower degree of compactness compared to peaked sources (however, see Section \ref{sec:nonpeak} for other explanations of the curved spectra). 

A turnover in the radio spectrum is associated with the presence of an absorption mechanism caused by either thermal gas (FFA) or the synchrotron-emitting plasma (SSA; \citealt{kellarmann+66}). If the source is synchrotron self-absorbed, the turnover frequency is inversely proportional to the emitting region sizes; thus, the peaked spectrum implies very compact emitting regions. Thus, our result in Figure~\ref{fig:spectra_vs_morph} is consistent with the expectations of compact source size and a spectral peak in the range $0.3 - 3$ GHz. In Section~\ref{sec:peaked}, we use synchrotron theory to obtain additional constraints on the source sizes for the peaked sources, finding angular extents that are indeed somewhat below the $\sim 0.2$\asec~resolution  of the VLA A-array observations.



\section{High-frequency Spectral Indices}\label{sec:steep_alpha}



\subsection{Distributions}\label{sec:alpha_distrib}
In Paper II, we presented in-band spectral indices, $\alpha_{IB}$, measured from our $10$ GHz X-band observations using the two WIDAR side-bands at $8-9$ and $11-12$ GHz. We found a rather broad distribution centered near $\alpha_{IB} \sim -1$. Here, we confirm the reliability of these in-band indices, and explore why they are somewhat steeper than the canonical value of $\alpha \sim -0.7$ for radio sources dominated by optically thin synchrotron emission.

\begin{figure}
    \centering
    \includegraphics[ width=\linewidth]{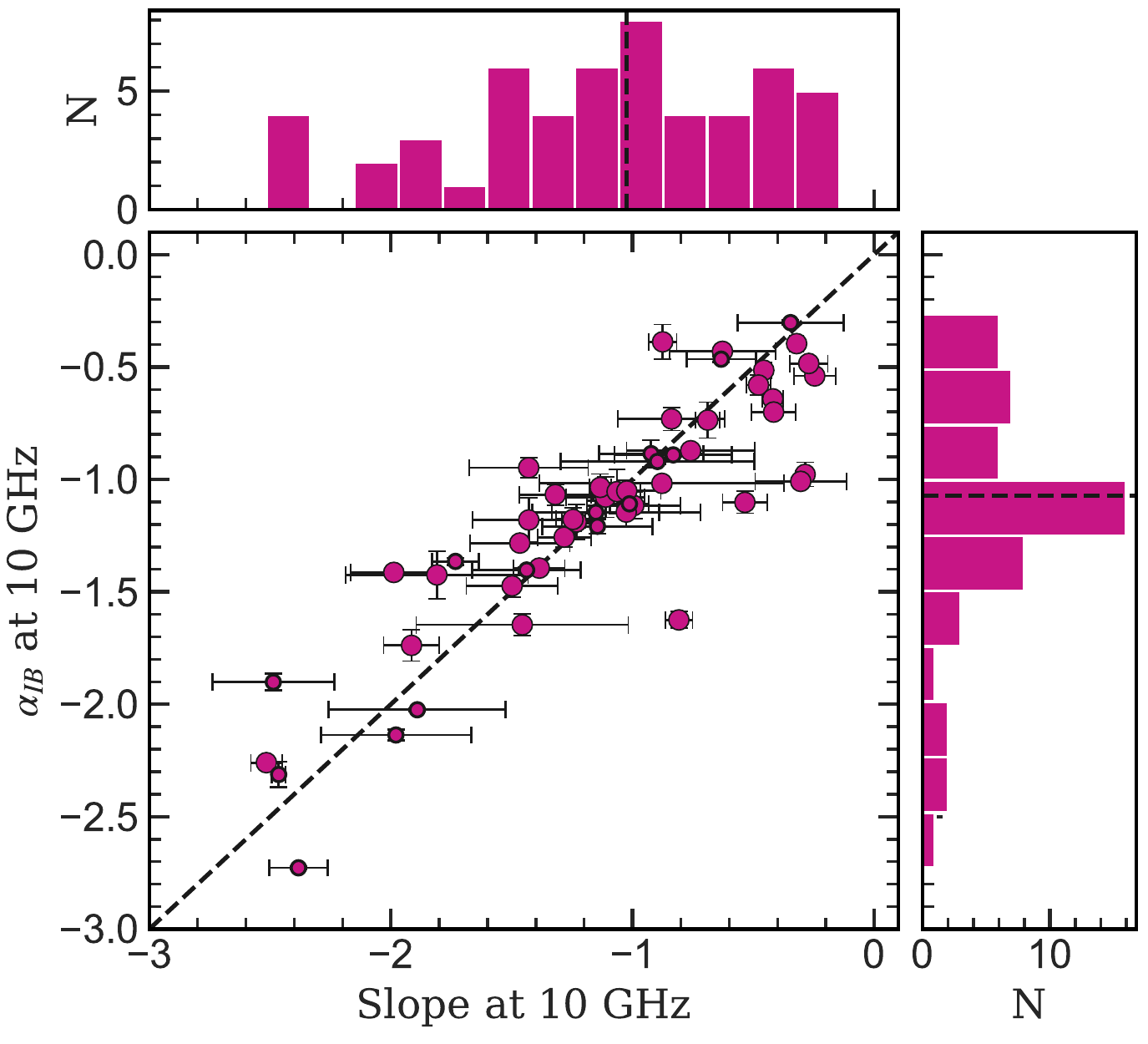}
    \caption{
    Comparison of 10~GHz in-band spectral index, $\alpha_{IB}$, with the slope of the best-fit model at 10 GHz. We only include sources that are unresolved on all scales and have 10 GHz flux density S/N $>$ 70 to ensure a robust estimation of $\alpha_{IB}$ (Paper II, \citealt{ Condon+2015}).  The top and right panels show  distributions of each index and its median value is shown by a dashed line. These two independent distributions confirm the steepness of the high-frequency spectral slopes relative to the canonical $-0.7$.
    }
    \label{fig:alpha_comp}
\end{figure}


Figure~\ref{fig:alpha_comp} compares $\alpha_{IB}$ with the slope at 10 GHz derived from the spectral fit, with distributions of each shown to the right and top. 
We only include sources that are unresolved on all scales and have a 10 GHz detection with S/N of at least 70 to ensure $\alpha_{IB}$ has an uncertainty of 0.1 or less (see Paper II for further discussion). We have also excluded the three sources which used $\alpha_{IB}$ in the spectral fit (see Section~\ref{sec:spectral_fitting}).
Overall, there is excellent  agreement, confirming that our estimates of $\alpha_{IB}$ are reliable. More importantly, both distributions of the 10 GHz index have median values near $-1.0$ which is significantly steeper than the canonical value of $-0.7$. We found similar distributions using the closely related indices $\alpha_{\rm high}$, $\alpha_{1.4}^{10}$, and  $\alpha_{3}^{10}$ which had median values of $ -1.01$, $-0.98$, and $-0.99$, respectively. Any spectral curvature would tend to reduce these indices, so their median values are likely lower limits, if anything.  Taken together, all these measures of the 10 GHz index suggest more negative (steeper) indices in our sample than the standard $-0.7$ value seen for most radio-loud AGN.
Since many of the sources in our sample are also GPS or CSS sources, we now compare the spectral indices of our sample with those of well-known GPS and CSS sources. Such sources presented in \citet{odea+98} and \citet{callingham+17} have flatter spectral indices above the peak frequency with median values of $-0.77$ and $-0.75$, respectively. A non-parametric Kolmogorov$-$Smirnov test confirms that the distributions of $\alpha_{\rm high}$ for our sample and the sample of \citet{callingham+17} are different with 5.8$\sigma$ significance.

In a wide range of astrophysical systems, optically thin synchrotron emission has a spectral index near $\alpha \sim -0.7$. A number of processes, however, can lead to deviations from this value, such as radiative and adiabatic losses from aging electron populations; inverse Compton scattering off a lower energy photon background; and various absorption processes which in turn may depend on the radio source environment. 
We now explore possible reasons for this spectral steepening. 

\subsection{Possible Causes of Spectral Steepening}


\subsubsection{Resolution Effects}
Before considering a physical origin of  spectral steepening, it is important to rule out the possibility that it has arisen due to systematic loss of extended emission in the high-frequency, higher-resolution observations. As discussed in Section~\ref{sec:resolution}, the final sample used in our spectral analysis excludes sources with any indication of extended emission, particularly in the lower-frequency survey images. Specifically, amongst the high-frequency observations, all the sources in our spectral analysis are essentially unresolved. While a definitive statement must await multi-frequency observations at matched resolution, we feel that our approach to defining our final sub-sample is not significantly compromised by resolution effects. 

\subsubsection{K-Correction and Spectral Aging}
Although the spectral indices are measured near 10 GHz, at redshifts of $z\sim1-3$ this corresponds to rest frame $20-40$ GHz. At such high frequencies, the electron lifetimes are quite short ($\sim10^2-10^5$ yr), especially in compact sources with relatively high magnetic field strength. As a result, one expects spectral aging with an associated spectral steepening \citep[e.g.,][]{pacholczyk+70, krolik+91}. Using alternate terminology, the observed spectral steepening due to redshift is an example of K-correction. 

To test this possibility, we looked for a correlation between spectral index, $\alpha_{\rm high}$, and redshift, $z$. We used a sample of  57 sources for which we have spectroscopic redshift and good quality radio spectra. A  Kendall$-\tau$ test (Section~\ref{sec:stats}) finds no correlation ($\tau=0.08$ with $P_{\rm null} = 0.42$). 

For extended classical radio sources, such a correlation between $\alpha$ and $z$ does indeed exist \citep[e.g.,][]{krolik+91, ker+12, morabito+18} and has even been used to find high-redshift radio galaxies \citep[e.g.,][]{ miley+08}. However, the existence of such a correlation in compact radio sources is less strong \citep{ker+12}, and in any case the $\alpha - z$ correlations are quite weak, so it is unclear whether the absence of a correlation in our sample can be taken as evidence against spectral aging. We therefore consider high-frequency radiative losses causing steepening at higher frequencies to be a possible contributing cause to the observed steep spectra near 10 GHz. 

\subsubsection{Inverse Compton Losses off the CMB}

Another  explanation for a correlation between $\alpha$ and $z$ in extended radio sources is inverse Compton Scattering of the relativistic electrons off Cosmic Microwave Background (CMB) photons. Because the CMB photon energy density is proportional to $(1+z)^4$, the losses can become significant at $z>1$. Indeed, it is thought that this gives rise, at least in part, to the class of Ultra Steep Spectrum (USS) sources at high redshift \citep[e.g.,][]{ miley+08}.

As before, the absence of a correlation between $\alpha_{\rm high}$ and $z$ in our sample argues against the importance of this process, though perhaps by itself does not rule it out, for the reasons given in the previous section. A stronger argument against its importance is that the magnetic field energy density, $u_{\rm mag}$, in our radio sources is significantly greater than the energy density in the CMB, $u_{\rm CMB}$. In this situation the cross-section for synchrotron self-absorption dominates, and IC losses off the CMB are of secondary importance. 

The ratio of these two energy densities, which sets the threshold above which inverse Compton scattering off the CMB is important, is:

\begin{equation}
\label{eqn:rcmb}
    R_{CMB} = \frac{u_{CMB}}{u_{mag}}
\end{equation}
The energy density in the CMB radiation field is

\begin{equation}
\label{eqn:ucmb}
     u_{CMB} = aT_0^4(1+z)^4 = 4.22\times10^{-13}(1+z)^4\ \textrm{erg cm}^{-3}
\end{equation}
where $T_0 = 2.725$ K is the CMB temperature at the current epoch, $a$  is the radiation constant, which is equal to $7.565\times10^{-15}$erg cm$^{-3}$ K$^{-4}$.

The energy density in the magnetic field is $u_{\rm mag} = B_{\rm min}^2/8\pi$ where $B_{min}$ is the magnetic field in Gauss calculated from synchrotron theory assuming minimum energy (approximately equipartition) conditions and is given by \citep{miley+80}:  

\begin{equation}
\label{eqn:bmin}
B_{min} \approx 0.0152
  \Bigg[ \frac{a}{f_{rl}}\frac{(1+z)^{4-\alpha}}{\theta_{mas}^3} 
    \frac{S_{mJy}}{\nu_{GHz}^{\alpha}}    \frac{X_{0.5}(\alpha)}{r_{Mpc}}\Bigg]^{2/7},
\end{equation}
where the radio source has a flux densityin $S_{\rm mJy}$ mJy at frequency $\nu_{\rm GHz}$ GHz with spectral form $S_\nu \propto \nu^{\alpha}$ and angular size $\theta_{\rm mas}$ milliarcsec, $z$ is the redshift of the source, and $r_{\rm Mpc}$ is the comoving distance in Mpc (assuming a flat geometry). For the current calculation, we adopt a spectral index $\alpha \simeq -1$, a filling factor for the relativistic plasma $f_{\rm rl} = 1$, and a relative contribution of the ions to the energy $a = 2$.  The function 
$X_{0.5}(\alpha)$ handles integration over a 
frequency range from $\nu_1 = 0.01$ GHz to $\nu_2 = 100$ GHz and is defined as: 

\begin{equation}
    X_{p}(\alpha) = \frac{(\nu^{p+\alpha}_{2} - \nu_{1}^{p+\alpha})}{(p+\alpha)},
\end{equation} 
where p is 0.5 in this case, and for $\alpha = -1$, we have $X_{0.5}(-1) \simeq 20$. Using these values, we find:

\begin{equation}
\label{eqn:umag}
    u_{mag} \approx 7.54\times 10^{-5} \left(\frac{S_{mJy} \nu_{GHz} (1+z)^5}{\theta_{mas}^3 r_{Mpc}}\right)^{4/7}\ {\rm erg\ cm^{-3}.}
\end{equation} 
Finally, given the limits of this kind of analysis, we make use of a good approximation for the comoving distance in Mpc:

\begin{equation}
\label{eqn:rcompc}
    r_{Mpc} \approx 4430 \frac{z}{(1+z/2)(1+0.5q_0z/(1+z))} 
\end{equation}
where $q_0 = -0.55$ is the deceleration parameter, and 4430 Mpc is the Hubble radius associated with $H_0 = 67.7$ km s$^{-1}$ Mpc$^{-1}$. Substituting these into Equation~\ref{eqn:rcmb} we find: 

\begin{equation}
\label{eqn:rcmb2}
    R_{CMB} \approx 1.36\times10^{-6}\ z\ \Bigg[ \frac{\theta_{mas}^3}{S_{mJy}\nu_{GHz}} \Bigg]^{4/7}
\end{equation}
where the combined $z$ dependence is well approximated by $2z$, and this yields the leading $z$ in Equation~\ref{eqn:rcmb2}. 

For our sources, with typical $\theta_{\rm mas} \sim 100$, $S_{\rm mJy} \sim 10$, $\nu_{\rm GHz} = 10$, and $z \sim 1$, we find $R_{\rm CMB} \sim 10^{-4}$ and so inverse Compton scattering off the CMB is unlikely. This is perhaps not surprising, since our sources are compact with significantly higher magnetic fields than the high$-z$, well-extended USS sources, for which $R_{\rm CMB} > 1$.


 
 \subsubsection{Inverse Compton Losses off the AGN Radiation}

While the CMB radiation field is relatively weak in the vicinity of our radio sources, the radiation from the AGN itself may be sufficiently intense that inverse Compton scattering off of those photons might cause spectral steepening. This seems promising, since a luminous MIR source is a key component of our selection criteria, yielding IR luminosities in the range $\log (L_{\rm IR}/L_\odot)\ \sim\ $ 12.4$-$14  (Paper I). Such effects have been proposed before in similar contexts \citep[e.g.,][]{ wilson+87, blundell+95, blundell+99}.



Following a similar approach to the previous section, we consider the ratio of the AGN bolometric photon energy density to the magnetic energy density, averaged over the volume of the radio source: 

\begin{equation}
    R_{bol} = \frac{u_{bol}}{u_{mag}}\ .
\end{equation}
For a spherical region of angular diameter $\theta_{\rm c}$ in radians with bolometric flux $f_{\rm bol}$ in erg s$^{-1}$ cm$^{-2}$, the energy density averaged over the sphere is:

\begin{equation}
\label{eqn:ubol1}
    u_{bol} = 12\  \frac{f_{bol}(1+z)^4}{\theta_c^2\ c} \ \ {\rm erg\ cm^{-3}}
\end{equation} where the factor $(1+z)^4$ tracks the standard relativistic dimming of surface brightness. We choose to specify the bolometric flux using the MIR flux and a bolometric correction factor, $B_{\rm c}$: 

\begin{equation}
\label{eqn:ubol2}
    f_{bol} = \nu_{34}f_{\nu 34}\times B_c\ \ {\rm erg\ s^{-1} cm^{-2}}
\end{equation} where ``34'' here refers to the mean frequency and flux density of the WISE W3 and W4 bands.  Converting $\theta$ to milliarcsec, we find:

\begin{equation}
\label{eqn:ubol3}
    u_{bol}=1.5\times 10^{-6} \frac{(W3+W4)_{mJy} (1+z)^4 B_c}{\theta_{mas}^2}\ \ {\rm erg\  cm^{-3}}.
\end{equation}

Combining Equations~\ref{eqn:umag} and \ref{eqn:ubol3} we find a relatively simple expression for $R_{bol}$:

\begin{equation}\label{eqn:rbol}
    R_{bol}\approx 4.74\frac{z\  (W3+W4)_{mJy}\  B_c}{(\theta_{mas}^{1/2}\  S_{\nu,mJy}\  \nu_{GHz})^{4/7}}
\end{equation} where once again the combined $z$ dependence is well approximated by $2z$, and this yields the leading $z$ in Equation~\ref{eqn:rbol} for $R_{\rm bol}$. An important quality of this relation is the muted dependence on the angular size of the source. Smaller sources have higher photon flux, but, for a given radio flux, a smaller source has higher magnetic field. 

The last parameter of interest is the  bolometric correction factor, $B_{\rm c}$. We estimate this parameter by inspecting the optical$-$IR SEDs for our sample presented in Paper I. These SEDs were fit using the following three components: starlight (constrained by the optical flux);  AGN/torus emission peaking in the MIR; and a colder black body from larger scale starburst or AGN heated dust, (constrained by 345 GHz ALMA flux densities). For our purposes, we only include radiation that is  cospatial with the radio source, so we only consider the compact AGN/torus emission. In a plot of $\log \nu$ vs.\ $\log \nu f_{\nu}$, this component peaks in the MIR and falls either side. An upper limit to $B_{\rm c}$  assumes a flat SED  spanning 2 dex in $\log \nu$, giving  $B_{\rm c} = \ln 100 = 4.6$. A lower limit comes from a typical AGN/torus fit given by Equation 2 in Paper I, for which $B_{\rm c} \simeq 1.7$. We adopt $B_{\rm c} \simeq 2$ recognizing (a) there is likely some emission from a compact nuclear starburst, (b) torus emission may be anisotropic which might increase the emission seen by the radio sources, and (c) the approximations do not warrant further refinement. 

\begin{figure*}[t]
    \centering
    \includegraphics[width=0.7\linewidth]{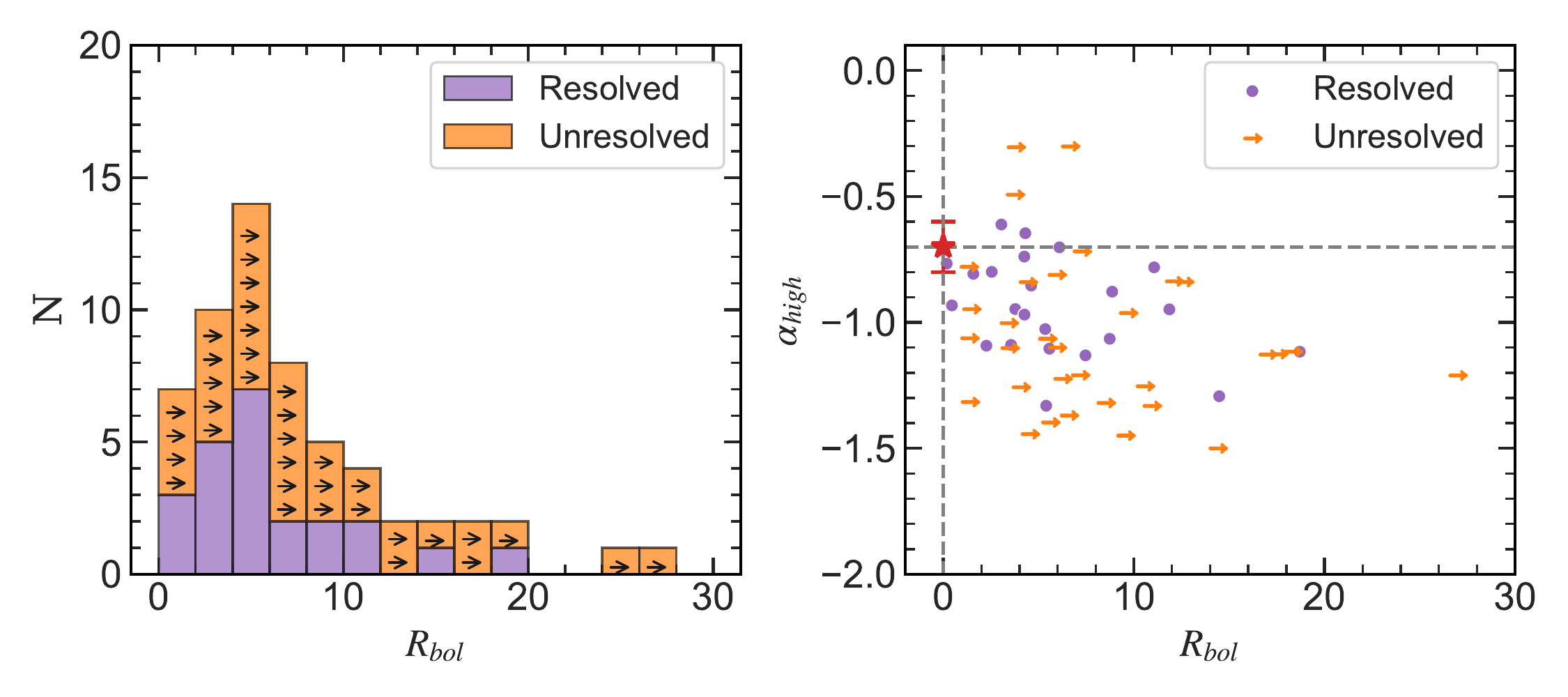}
    \caption{\textbf{Left:} Distribution of $R_{\rm bol}$ for our sub-sample of 110 sources. The unresolved sources yield lower limits and are shown in orange. Sources with resolved 10 GHz morphologies are shown in purple. \textbf{Right:} $\alpha_{\rm high}$ vs.\ $R_{\rm bol}$.  The large red symbol shows the location of canonical classical radio sources with spectral index $\alpha_{\rm high} \sim -0.7$. The horizontal and vertical dashed lines mark the canonical spectral index, $\alpha_{high} \sim-0.7$ and R$_{bol} \sim 0$.}
    \label{fig:r_ratio}
\end{figure*}

Figure~\ref{fig:r_ratio}a shows the distribution of $R_{\rm bol}$ for our sample, with lower limits shown for unresolved sources. Almost all the values are greater than 1.0, with most falling in the range $2-10$, with some as high as $20$. Figure~\ref{fig:r_ratio}b plots $\alpha_{\rm high}$ against $R_{\rm bol}$, with the canonical value of $R_{\rm bol} \simeq 0$ and $\alpha_{\rm high} \simeq -0.7$ for more extended classical radio sources shown as a red star symbol. The data suggest  a weak tendency (a significance of 2.4$\sigma$ for a censored Kendall$-\tau$ test) for sources with larger $R_{\rm bol}$ to have steeper $\alpha_{\rm high}$, though a robust analysis is undermined by the large number of lower limits.  

In summary, it does seem that inverse Compton scattering off a near-nuclear AGN radiation field provides a plausible explanation of the steep high-frequency spectra. In retrospect, this is perhaps not too surprising: for this process to be relevant, one needs high luminosity AGN with compact radio sources. Our sample provides both these -- the WISE-NVSS sample are luminous AGN, and the radio sources are physically small (but not as small as the classical flat spectrum cores).

\subsubsection{Dense Ambient Medium}
The environment in which a radio source develops may also affect its spectral index. For example, the tendency for radio sources at higher redshift (e.g., USS) to have steeper spectral indices has been explained as arising from their development in a much denser ambient medium  \citep{athreya+98, klamer+06, bornancini+10}. Since one of the selection criteria for our sample is a high MIR/optical ratio, the near-nuclear regions are likely to be highly obscured 
(Paper I).  Thus, the radio jets are likely to be interacting with a denser near-nuclear medium and for this reason exhibit steeper spectra.

\section{The MIR$-$Radio Connection}\label{sec:mir_radio}

Since our sample has unusual MIR emission, we consider the possibility that the radio and MIR emission regions are interacting; for example, an expanding radio sources might heat gas and dust which in turn might might confine the expanding radio sources. In this case, the interaction would depend on the degree to which the two emission regions are cospatial. Our radio sources span a few 10s pc to a few kpc, with most near a few hundred pc (Paper II, \citealt{lonsdale+21}). What about the MIR emission region sizes? Paper I  considered several possibilities: an optically thick torus; an AGN heated cocoon; and a compact starburst. All of these options are compact, spanning a few pc to a few hundred pc. A simple estimate of the largest distance considers a clear line of sight from the AGN to the dust and solves for an equilibrium temperature of T$_{\rm dust} \sim 300$ K which generates the  MIR.
The distance, $d_{pc}$, is given by, 
\begin{equation}\label{eqn:Tdust}
    d_{pc}\approx 16(44)\   L_{13}^{\frac{1}{2}}\  T_{300}^{-3}\ \ {\rm pc}
\end{equation} where $L_{13}$ is the AGN bolometric luminosity in units of $10^{13}\ L_{\odot}$, $T_{300}$ is the dust temperature in units of 300~K, and we take the dust optical absorption and near-IR emission coefficients for silicate (graphite) grains from \citet[Chapter 6.4]{ryden_pogge_2021}.
Hence, we confirm that the MIR emission region must indeed be compact, and might be cospatial with the most compact radio sources in our sample (see Paper II).

In what follows, we first analyze the MIR properties of our sample and compare them with closely related samples of compact radio AGN, and then we look for direct correlations between the MIR and radio properties.

\subsection{WISE Colors of Compact Radio AGN}

\begin{figure}[ht]
    \centering
    \includegraphics[width=\linewidth]{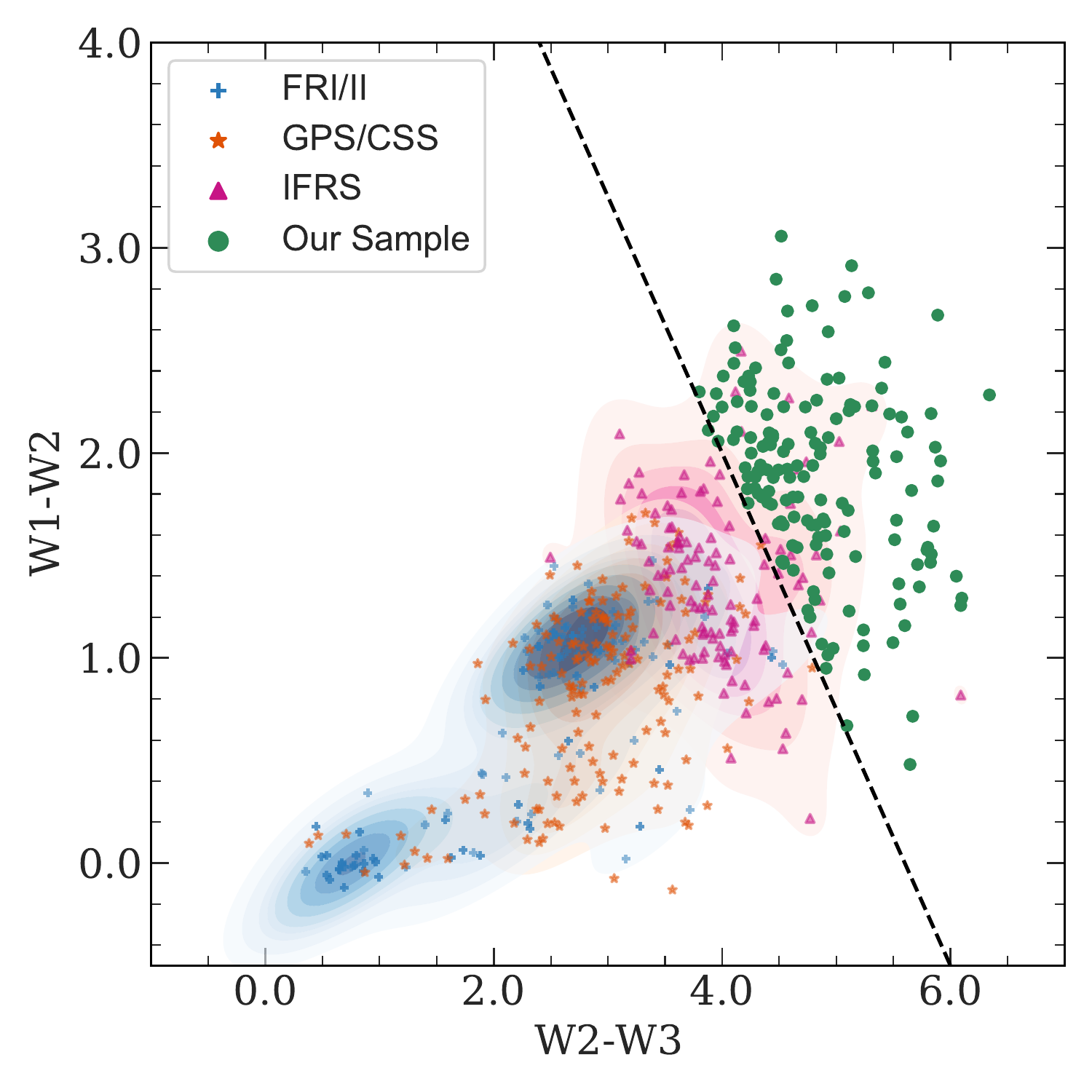}
    \caption{WISE $3.6-4.5\,\mu\textrm{m}$ and $12-22\,\mu \textrm{m}$ color diagram. The green dots are our sources, blue crosses are classical FRI and FRII sources taken from the references given in  \citet[][Figure 2]{an+12}, orange stars are well-known GPS, CSS, and HFP sources from \citet{jeyakumar+16}. We also show a density plot in the background for the FRI/II as well as compact GPS/CSS sources. The magenta triangles are IFRS taken from \citet{collier+14}. Most of FRI/II and compact radio sources are bluer than our sample, although the IFRS sources show some overlap with our sources.  }
    \label{fig:wisecols}
\end{figure}

Figure~\ref{fig:wisecols} shows WISE colors (W1$-$W2 vs.\! W2$-$W3) for our sample (green) and samples of FR\ I and FR\ II radio sources compiled by \citet[][ blue]{an+12}, and GPS, CSS and HFP sources from \citet[][ orange]{jeyakumar+16}. 
Very few FRI/II and GPS/CSS sources have WISE colors as red as our sample. 
A similar result was seen by \citet{chhetri+20}, who found that compact radio AGN have bluer WISE colors than AGN selected by optical/IR techniques. Thus, although our sample was pre-selected to have very red WISE colors, other classes of object with similarly compact radio sources (CCS/GPS/HFP) do not share these MIR colors. This is consistent with our findings in Papers I and II that our sources are rarer and likely caught in the short-lived evolutionary phase when dense gas has arrived near the nucleus during a merger event and the AGN has only recently turned on. 

Also included in Figure~\ref{fig:wisecols} are the  Infrared Faint Radio Sources (IFRS) which have relatively faint MIR flux density (W1 $<30\,\mu$Jy) and high radio/MIR flux density ratios ($S_{20\textrm{cm}}/S_{3.6\mu\textrm{m}}>500$)  
\citep[e.g.,][]{norris+06, middleberg+08, collier+14}.
The IFRS radio sources are similar to ours: often compact with either steep or peaked radio spectra \citep{herzog+16}. As can be seen from Figure~\ref{fig:wisecols} they lie in the region of obscured AGN with WISE colors that are redder than most other classes of AGN but bluer than our sources, though with some overlap. It is possible, given their similar radio properties and MIR colors, that they are physically related to our sources, although further work is needed to explore the relationship between these two classes of AGN. 



\subsection{MIR and Radio Parameters}

Table~\ref{tab:corr_pars} lists parameters 
that characterize the MIR and radio emission. They include for the radio emission and source: measures of linear extent, luminosity, pressure, shape of radio spectra and, from Paper II, model dependent estimates for the dynamical age, expansion velocity and ambient density. For the MIR parameters, we consider the $6\mu\textrm{m}$ luminosity, the MIR colors, and an MIR energy density estimated using the radio source size and the $6\mu\textrm{m}$ luminosity. For completeness, we also include the $870 \mu\textrm{m}$ luminosity from Paper I, where available.


\subsection{Correlation Analysis}\label{sec:stats}
%
%
%
%

To test for the presence of a significant correlation between MIR and radio parameters, we use the censored Kendall$-\tau$ correlation test \citep{isobe+86}. 



We find that 
most MIR-radio parameter pairs are not correlated, while some are correlated for uninteresting reasons. For example, MIR and radio luminosity correlate because of a common dependence on redshift, and MIR energy density correlates with radio source size because the size is used to define the MIR energy density. 

Ultimately, we find no significant correlations that are not of the kind just described. 
In particular, there are no differences in the distribution of MIR parameters between the three spectral shape classes: peaked, curved or power law. 
This is consistent with an MIR emitting region that is more compact than the radio source, such as an optically thick torus. Alternatively, if the MIR emitting gas is cospatial with the radio source, then it does not affect the radio spectrum.    

\begin{table}
    \centering
    \begin{tabular}{|c|c|}
    \hline
    Selection Criteria &  Parameters\\ \hline
         10~GHz Continuum & LLS, $L_{1.4\,\rm GHz}$ , $P_{\rm lobe}$, $t_{\rm dyn}$, $n_{\rm a}$\\
         Radio Spectra & $\alpha_{\rm high}$, $\alpha_{\rm low}$, q, $\nu_{\rm peak}$, $\nu_{\rm peak, rf}$ \\
         MIR & $L_{6\mu m}$, W1-W2, W2-W3, $u_{\rm IR}$\\
         FIR & $S_{870\mu \textrm{m}}$\\\hline
    \end{tabular}
    \caption{List of Parameters used in Correlation Analysis; LLS: Largest Linear Size in kpc from 10~GHz radio morphology; $L_{1.4\,\textrm{GHz}}$: 1.4 GHz luminosity; $P_{\rm lobe}$: Radio source pressures from \citet{patil+20}; $t_{\rm dyn}, n_{\rm a}$: Dynamical age and ambient density calculated using a self-similar lobe expansion model \citep{patil+20}. See Section~\ref{sec:sedparam} for the explanation of the radio spectral parameters.  $L_{6\,\mu \textrm{m}}$: 6 $\mu\textrm{m}$ luminosity; W1$-$W2, W2$-$W3: WISE colors using WISE 3.4 $\mu\textrm{m}$ (W1), 4.6 $\mu\textrm{m}$ (W2), and 12 $\mu\textrm{m}$ (W3) bands; $u_{\rm IR}$: IR energy density calculated using Equation~\ref{eqn:ubol1}; $S_{870\mu\textrm{m}}$: ALMA Band 7 flux density  presented in Paper  I. }\label{tab:corr_pars}
\end{table}

\begin{figure}
\centering
\includegraphics[width=\linewidth]{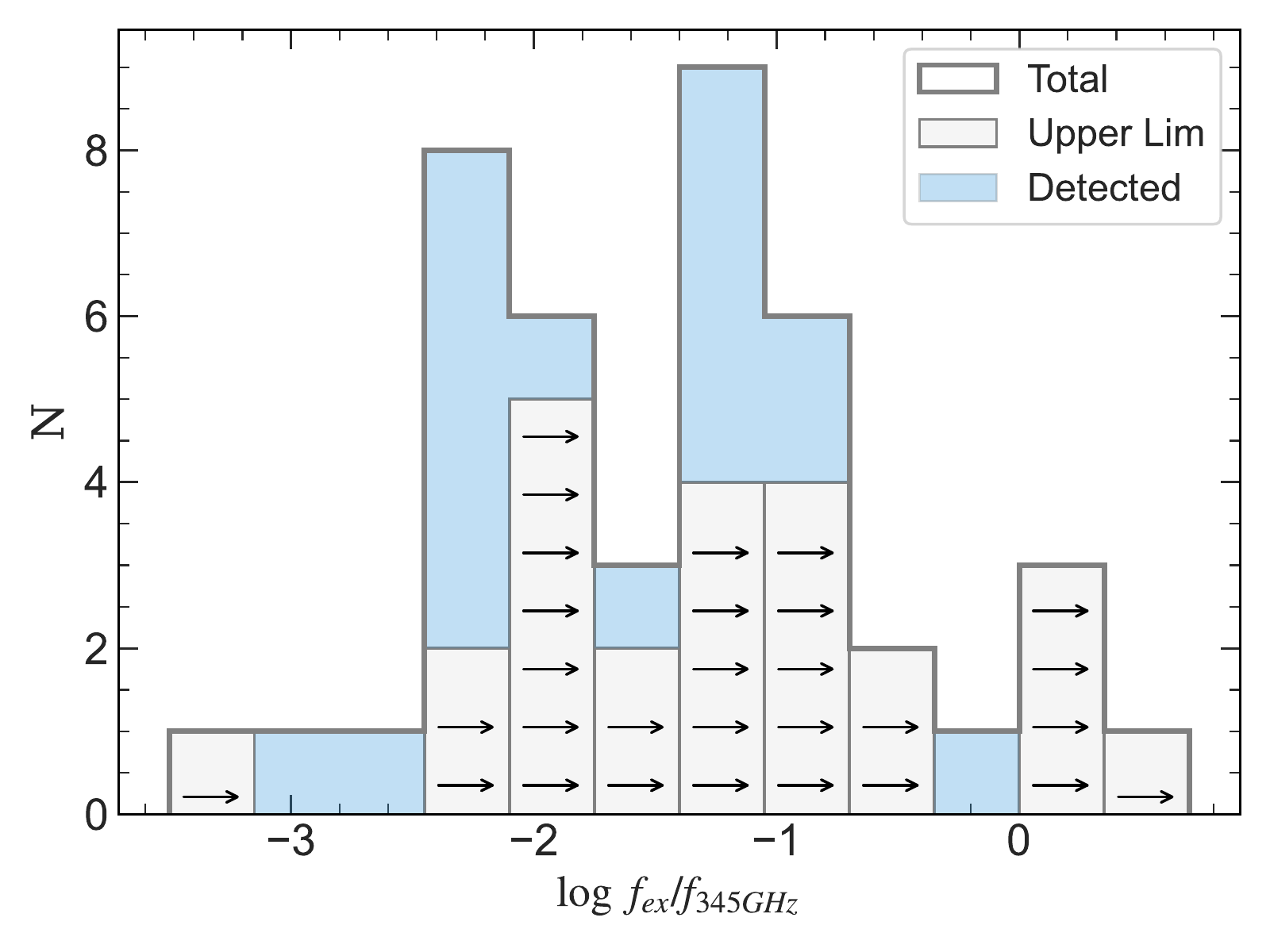}
\caption{Ratio of extrapolated synchrotron flux density to the observed ALMA flux density  at 345~GHz (870 $\mu m$).  The blue histogram includes sources detected at 870$\mu m$. The light grey histogram with arrows includes non-detections at 870 $\mu\textrm{m}$. We use $3\sigma$ as the upper limit on the ALMA flux density to estimate the ratio. 
}\label{fig:fsyn}
\end{figure}

\section{Origin of the ALMA 870~$\mu$\lowercase{m} Emission} \label{sec:alma_emission}

In Paper I, ALMA data at 345 GHz (870 $\mu\textrm{m}$) were presented for 49 southern sources taken from the main WISE-NVSS sample, detecting 26 of them above $3\sigma$. In that paper, it was assumed that the 345 GHz emission was dominated by thermal emission from warm dust, with a plausible origin for heating coming either from star formation or an AGN. With our new radio data, we are able to check an alternate possibility: namely whether the 345 GHz emission could arise from a high-frequency synchrotron extension of the radio source. 


To assess the synchrotron contribution to the ALMA emission, we use our radio spectral fits to extrapolate to 345 GHz and ask how that extrapolation, $f_{\rm ex}$, compares to the measured 345 GHz flux density, $f_{345{\rm GHz}}$, using the ratio $f_{ex}/f_{\rm 345GHz}$. A conservative approach is to adopt a power law rather than a parabolic fit that would include spectral steepening to higher frequencies. For the power law index, we use  $\alpha_{\rm high}$ if available, but if not, we take $\alpha_{3}^{10}$ and failing that $\alpha_{1.4}^{10}$. Thus, our flux density ratios are  conservative in the sense that the true synchrotron flux density at 345 GHz is either close to $f_{\rm ex}$ or lower. 
Figure~\ref{fig:fsyn} shows the distribution of $\log (f_{\rm ex} / f_{\rm 345GHz})$ with arrows showing lower limits for the 22 ALMA non-detections. 

Clearly, all sources detected by ALMA have $f_{\rm ex} / f_{345 \rm GHz} < 1$ confirming that non-thermal synchrotron from the radio source does {\textit{not}} contribute significantly to the 345 GHz ALMA flux density, which are likely the long wavelength extension of the thermal dust component. A stronger conclusion is not possible because of the large number of lower limits, with at least four sources with $f_{\rm ex} / f_{345 \rm GHz} \gtrsim 1$ (J0354$-$33, J0519$-$08, J0823$-$06, and J1308$-$34).

\section{The Linear Size vs. Turnover Frequency}\label{sec:lin_vs_to}

\begin{figure}
    \centering
    \includegraphics[width=\linewidth]{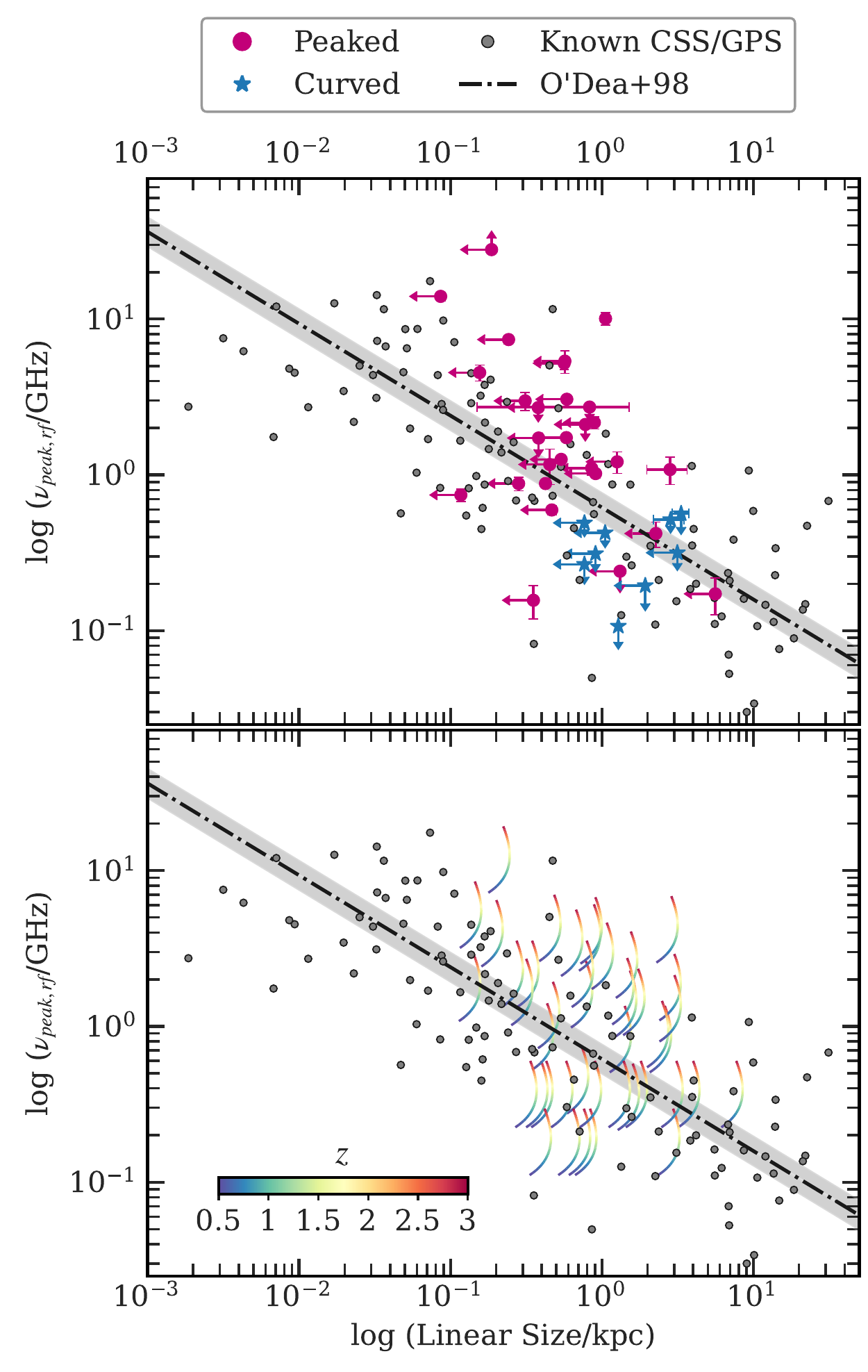}
    \caption{The linear size vs. rest-frame turnover frequency  relation. Purple circles and blue stars in the top panel are 28 peaked (pink) and 10 curved (blue) sources from our sample with known $z$. Small gray circles in both panels are known GPS, CSS, and HFP sources compiled by \citet{jeyakumar+16}. The dashed-dotted line is the best-fit relation from \citet{odea+98}, with the grey band indicating the $1\,\sigma$ errors on the best-fit. The lower panel includes 44 peaked or curved sources with no redshift, which are shown as tracks  for $0.5 < z < 3$.}
    \label{fig:lvsnu}
\end{figure}

Figure~\ref{fig:lvsnu} shows the well-known linear size (LS) vs. rest-frame turnover frequency ($\nu_{\rm peak, rf}$) relation for various samples of CSS, GPS, and HFP sources (grey points) compiled by \citet{jeyakumar+16}, together with a fit to GPS sources taken from \citet{odea+97} (see also  \citet{falcke+04} and \citet{orienti+14}). The upper diagram includes the 38 peaked and curved sources with redshifts, with size limits indicated for unresolved sources, and peak frequency limits for the curved sources taken to be the lowest measured frequency. The remaining 44 sources with no redshift are shown on the lower diagram as tracks that span $0.5 < z < 3$ which is the full range seen in our sample.


Most of our peaked spectrum sources (purple circles) are within the scatter seen for the HFP, GPS, and CSS sources.  Although some peaked sources lie above the relation, many have upper limits on linear size (they are unresolved) and so could be consistent with the underlying relation.  Four extreme outliers fall at least 3.6$\sigma \approx 0.8$ away from the relation given by \citet{odea+98}. The curved spectrum sources (blue stars) have a smaller scatter and a few with upper limits to $\nu_{\rm peak, rf}$ may lie significantly off the relation, although the overall uncertainties may account for some of these outliers. Overall, we find our sources to be consistent with the \citet{odea+98} relation. It is also worth noting that some discrepancies have been reported in recent studies \citep[e.g.,][]{coppejans+16, collier+18, keim+19}. In a high-resolution study of  low-luminosity  ($L_{1.4{\rm GHz}}<10^{27}$ WHz$^{-1}$) GPS/CSS sources by \citet{collier+18}, two of their five sources fell significantly away from the relation. As the primary relation was derived using luminous peaked sources, lower-luminosity peaked sources may not follow this interesting relation.

The empirical $\nu_{\rm pk,rf} - $LS relation may help shed light on the turnover mechanism and radio source evolution \citep{odea+97}.  SSA provides a natural explanation for the dependence of the turnover frequency on the emitting lobe size, which in turn scales with the source total linear extent  \citep[e.g.,][]{odea+97, jeyakumar+16}. 
Alternatively, FFA resulted in the $\nu_{\rm pk,rf}-$LS relation in the analytic models of \citet{bicknell+97} and the subsequent hydrodynamical simulations of \citet{bicknell+18}. These models do, however, require a dense inhomogeneous ionized medium external to the source which does not seem applicable for the general population of young radio sources many of which lie in early type hosts. Furthermore, GPS/CSS sources with independent evidence for FFA can lie far from the  $\nu_{\rm pk,rf}-$LS relation \citep[e.g.,][]{keim+19}. For our sources, however, a dense ionized gas is likely to be present, suggesting FFA may be present, and this may account for some of the sources lying far from the relation.  
Unfortunately, further improvement in Figure~\ref{fig:lvsnu} for our sample must await observations with higher angular resolution, wider spectral coverage, and higher redshift completeness. 


We defer to the next section a more detailed discussion of whether SSA or FFA causes the turnover in the peaked sources.


\section{The Peaked Sources}
\label{sec:peaked}

One of the important results of this study is that a significant fraction of our sources show peaked radio spectra ($37$\% PK) or curved spectra ($25$\% CV), suggesting absorption of low-frequency emission. Furthermore, as discussed in Section~\ref{sec:shape_morph}, a high fraction of these are spatially unresolved in our VLA imaging ($83$\% PK; $53$\% CV), suggesting compact sources. In this section, we discuss the nature of the absorption and try to use it to constrain the properties of the emitting region. We consider in turn the two well-known absorption mechanisms: Synchrotron Self-Absorption and Free-Free Absorption. 

The explanation of spectral turnover in AGN is still debated \citep[e.g.,][]{tingay+03, tingay+15, callingham+15}. The standard approach is to compare the low-frequency spectral index with the idealized treatment: SSA generates $\alpha \lesssim 2.5$ while FFA generates a steeper (exponential) index. Using our values of $\alpha_{low}$, we find almost all to be less steep than $2.5$ consistent with SSA. This is indeed the most common interpretation for the turnovers seen in, for example, the GPS or HFP sources \citep[see ][and references therein]{odea+20}. However, our spectral sampling is too sparse to yield a robust value for $\alpha_{low}$. Furthermore, more detailed observations of more local sources rarely conform to  the idealized SSA or FFA spectral shapes \citep[e.g.,][]{callingham+15}.  For these reasons, we choose to explore both possibilities, and try to learn more about the radio source and its environment assuming first SSA (Section~\ref{sec:bsolve}) and second FFA (Section~\ref{sec:FFA}). 

\subsection{Synchrotron Self-Absorption  (SSA)}\label{sec:SSA}
%
%
%
%
Synchrotron self absorption, SSA, occurs when relativistic electrons absorb their own synchrotron emission \citep{slish+63, kellarmann+66}. The simplest model considers a homogeneous population of relativistic electrons and yields an inverted power law with spectral index $\alpha_{\rm thick} = +2.5$.  Typically,  shallower $\alpha_{\rm thick}$ values are observed, likely due to contribution from multiple electron populations  \citep[e.g.,][]{odea+98, orienti+14, callingham+17}. SSA has been the favored interpretation of peaked sources in many studies \citep[e.g.,][]{snellen+00, devries+09, jeyakumar+16, orienti+16}. Furthermore,  it can explain the global properties of the GPS/CSS population, e.g., the observed linear size vs.\ turnover relation \citep{odea+97} and provides magnetic field estimates consistent with the equipartition fields estimated from the optically thin part of the spectrum \citep{orienti+08}. The consensus, then, is that SSA will always be present to some degree in radio-emitting plasma \citep{fanti+09, orienti+16, odea+20}.  In what follows we continue to assume that the peaked spectra result from SSA. In Section~\ref{sec:FFA}, we explore the alternate possibility that the peaked spectra result from free-free absorption (FFA).  

\subsection{Deriving Emitting Region Properties}\label{sec:bsolve}

Under the assumption that the peaked sources are synchrotron self-absorbed, we can use standard synchrotron theory to derive important properties of the emitting region. For a source at redshift $z$ with angular size $\theta_{\rm mas}$ [mas], whose self-absorbed spectrum peaks at observed frequency $\nu_{\rm peak}$ [GHz] with flux density of $S_{\rm peak}$ [mJy], the magnetic field is given by \citep{kellermann+81}:

\begin{equation}
  \label{eqn:bssa}
  B_{SSA} = 23 \ \frac{\theta_{mas}^4}{S_{peak}^2} \frac{\nu_{peak}^5}{1+z} \ \ {\rm [Gauss]}.
\end{equation}

\begin{figure}[ht]
    \centering
    \includegraphics[width=\linewidth]{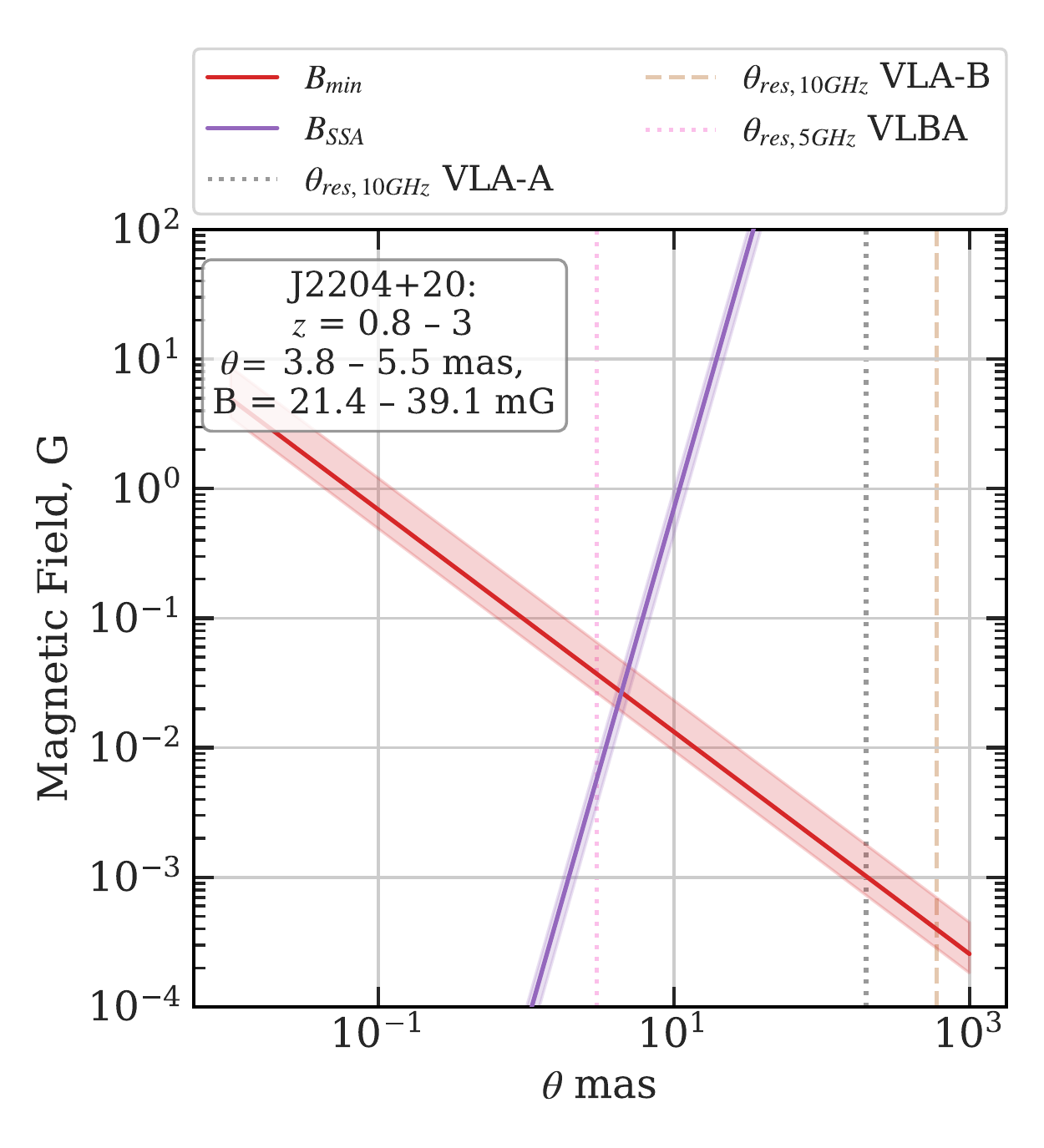}
    \caption{  
    An example of applying the SSA (purple) and $B_{min}$ (red) relations to the source J2204+20, yielding independent solutions for emitting region sizes and magnetic fields.
    The  band  widths  reflect  uncertainties  in  the  spectral measurements  and  the  redshift  range  of  $0.8< z <3$ (J2204+20 has no measured redshift).
    The intersection suggests sizes of $\theta_{\rm mas} = 3.9-5.5$ mas and $B = 21-38$ mG. The gray dotted  and brown dashed lines at $\theta_{\rm mas} \sim 200\ \&\ 600$ show the limiting angular resolution of our VLA 10 GHz A- and B-array observations. The pink dotted line at 3 mas shows the  angular resolution of 5 GHz VLBA observations. }
    \label{fig:bsolve}
\end{figure}

To illustrate, Figure~\ref{fig:bsolve} shows the steeply rising purple line of $\theta_{\rm mas}$ and $B_{SSA}$ values that satisfy Equation~\ref{eqn:bssa} for J2204+20, using values of $S_{\rm peak}$ and $\nu_{\rm peak}$ taken from Table~\ref{tab:alpha_calc}. Uncertainties in the measurements, and redshift range $0.5 < z < 3$ (as this source does not have a measured redshift) yield the band that surrounds the line. 

However, this same emitting region is also generating the power-law emission on the optically thin high-frequency side of the peak, which yields independent constraints on $\theta_{\rm mas}$ and $B_{\rm min}$ from the minimum energy (approximately equipartition) condition, as given by Equation~\ref{eqn:bmin}. These constraints are also shown on Figure~\ref{fig:bsolve} as the more gently decreasing red line, with its band for uncertainties. Also shown are two vertical lines at $\theta_{\rm mas} = 200$ (gray dotted) $\&\ 600$ (brown dashed) which indicate the resolution of our 10 GHz VLA A- and B-array observations. Recall from Section~\ref{sec:shape_morph} that most peaked sources, including this one, are unresolved (UR), suggesting their sources lie to the left of these vertical lines. 

Since the SSA and $B_{\rm min}$ analyses apply to the \textit{same} emitting region, then they should yield the same magnetic field and source size. Hence, we set: 

\begin{equation}
  \label{eqn:b_equal}
  B_{SSA} = B_{min}
\end{equation}
and then use Equations~\ref{eqn:bmin} and \ref{eqn:bssa} to find the source size and magnetic field. For J2204+20 this condition occurs where the two lines cross in  Figure~\ref{fig:bsolve}, yielding a source size of $\theta_{\rm mas} \approx 5$ mas and magnetic field $B_{SSA} = B_{min} \approx 130$ mG. For the specific case of $\alpha = -1$, there is a simple relation for the source size in mas: 

\begin{equation}
  \label{eqn:bminbssa}
  \theta_{mas} = 0.38\ (1.2 + z/4)\ S_{peak}^{0.41}\ \nu_{peak}^{-1.03}(S_{mJy}\nu_{GHz})^{1/17}
\end{equation}
where the total redshift dependence is well approximated by (1.2 + z/4). The magnetic field is then given by inserting $\theta_{\rm mas}$ into Equation~\ref{eqn:bssa}. 

There is some evidence to support the assumption that $B_{\rm SSA} \approx B_{\rm min}$. \cite{orienti+08} measure angular source sizes and spectral peaks in a sample of HFP and GPS sources and are able to separately evaluate $B_{\rm SSA}$ and $B_{\rm min}$, finding broad agreement. Recognizing that the agreement may be only approximate in individual cases, we nevertheless proceed to use the condition to estimate source sizes and magnetic field strengths.

Table~\ref{tab:bsoltab} gives the angular source sizes and magnetic fields for the 46 sources with peaked spectra. For the 24 sources with measured redshifts, Table~\ref{tab:bsoltab} also gives the source size in pc, the region pressure and total energy, as well as radiative cooling timescale given by the total energy divided by the radio luminosity. The table also includes synchrotron  electron lifetimes at 10 GHz. 


\begin{figure*}[t]
    \centering
    \includegraphics[width=0.65\linewidth]{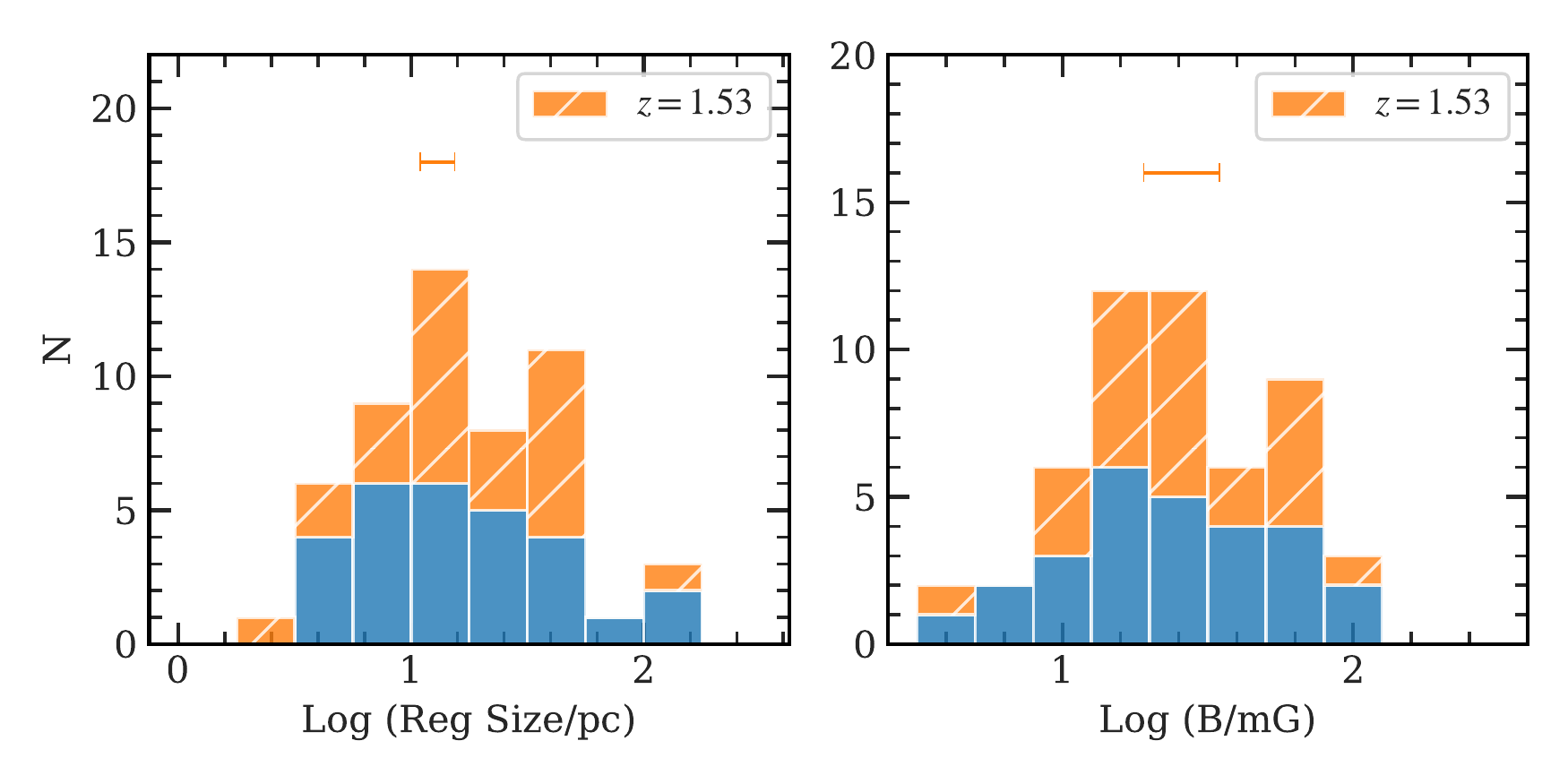}
    \caption{ Distributions of physical sizes of emitting regions (left) and magnetic fields (right), derived assuming $B_{\rm SSA} = B_{\rm min}$. Sources with redshifts are in blue, while sources without are in hatched orange, and assume the median redshift $z = 1.53$ (the redshift dependence is weak, with the error bar showing values for $0.8 < z < 3$).  }
    \label{fig:bmin_hist}
\end{figure*}

Figure~\ref{fig:bmin_hist} shows the distribution of component sizes and magnetic fields estimated using the method described above. We also plot sources without redshifts (hatched) using the sample median redshift, $z_{\rm med} = 1.53$ (the redshift dependence is weak, with values shifting less than one bin width for $0.8 < z < 3$).  Overall, the distributions of sources with and without redshift are quite similar.  Typical region sizes are $3-100$ pc,  magnetic fields are $6-100$ mG, and pressures are $10^{-6} - 10^{-3}$ dyne cm$^{-2}$. These are comparable to the region sizes and pressures found for the luminous HFP sources, estimated using long-baseline observations \citep[e.g.,][]{orienti+08, orienti+14}. Thus, although only indicative, our sample's physical properties derived here are similar to those seen in young radio sources.

These estimates are well below the resolution of our VLA 10 GHz observations and need to be checked using direct VLBA milli-arcsec observations, which  will be presented in a future paper \citep{lonsdale+21}. Independent evidence for our approach comes from existing VLBI observations, where measurements of source size and  turnover frequency yield similar values of magnetic field under SSA and minimum energy assumptions   \citep[e.g.,][]{readhead+94, orienti+08b}. 

\begin{figure*}
    \centering
    \includegraphics[width=\textwidth]{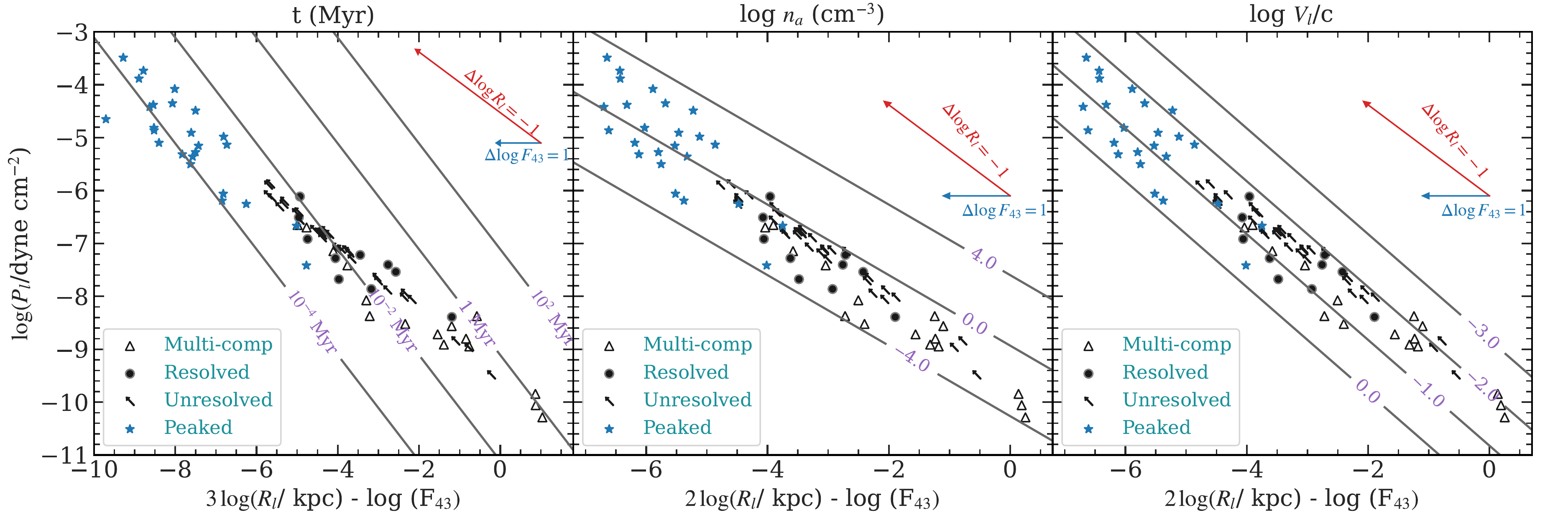}
    \caption{Updated adiabatic lobe expansion model from Paper II (Equations~A6-A8 in that Paper).  The panels isolate  source age ($t_{\scriptsize{\textrm{Myr}}}$), ambient particle density ($n_{\rm a}$), and lobe expansion speed ($V_{\rm l}/c$) using observed parameters $R_{\rm l}$, $p_{\rm l}$ and $F_{43}$ as described in Paper II. The blue stars are unresolved peaked sources with known redshifts with $R_{\rm l}$ and $p_{\rm l}$ estimated assuming $B_{SSA} = B_{min}$  (see Section~\ref{sec:bsolve}). The arrows, triangles, and filled circles are same as Figure~12 in Paper II.  Open triangles are individual resolved lobe components for double or triple sources; filled circles are partially resolved sources; arrows are unresolved sources. Red and blue vectors show the effect of a decrease in source size by one dex and increase in jet power by one dex. }
    \label{fig:new_lobe_model}
\end{figure*}

\subsection{Revisiting The Lobe Expansion Model}

In Paper II, we presented a simple ``bubble model'', which provides an analytic solution for a radio lobe expanding into an ambient medium driven by energy input from a jet. We can now include our new values for the lobe pressures and sizes for the peaked and unresolved sources, assuming SSA and equipartition as described in Section~\ref{sec:bsolve}.  Figure~\ref{fig:new_lobe_model} shows as blue stars the new values for the peaked sources, while the arrows show unresolved non-peaked sources with the linear extent taken as the limit to lobe sizes.
The peaked sources have younger ages ($40-2000$ years with a median of $215$ years), relatively high ISM densities ($0.02-1000$ cm$^{-3}$ with a median of $2$ cm$^{-3}$), and modest to high expansion speeds ($0.01c-0.5c$ with a median of $0.07c$). This is consistent with the expectations that the peaked sources are likely to have the youngest and most compact regions expanding into a dense ISM. The range of ages now approaches the new class of peaked radio sources that have just turned on in the past $10-20$ years \citep{nyland+20, wolowska+21}.

While these values for source ages, expansions speeds and ambient densities arise from application of this simple expanding bubble model, they are nevertheless within the ranges seen in much more detailed numerical simulations, such as those of \citet{mukherjee+16, mukherjee+20}.

\subsection{Timescales}
\label{sec:timescales}

Since one of the primary themes of this work is to identify young radio sources, it is appropriate to consider some direct estimates of relevant timescales. We will discuss three somewhat different timescales (see Table~\ref{tab:bsoltab}). 

First, the lobe expansion model presented in Paper II yields a dynamical time,  $t_{\rm dyn}$, for the jet to inflate the lobe to its observed size and pressure. The model assumes energy and momentum conservation with only adiabatic losses arising from lobe expansion into an ambient medium, and a jet power estimated from the radio source power. For our resolved and slightly resolved sources we find  $3 < \log (t_{\rm dyn}/\rm{year}) < 6$. Our unresolved sources give only upper limits, but we can now include them if we use the source sizes and pressures derived in Section~\ref{sec:bsolve}. Perhaps not surprisingly, we find shorter timescales, $1 < \log (t_{\rm dyn}/{\rm year}) < 3$. When one recalls that some of the source sizes we derive are only a few parsecs, these short timescales are not unreasonable. Indeed the shortest derived timescales begin to approach those of the new class of extremely young radio sources that have ``turned on'' in the last $\sim25$ years between the FIRST survey in the 1990s and the recent VLASS survey \citep{nyland+20}. These  new radio sources are similar to ours, with compact ($< 0.1\arcsec$) peaked spectra (peak frequencies $1-10$ GHz observed, $10-50$ GHz rest). Thus the youngest of our sources may form part of a continuum that reaches down to truly newborn jets and lobes. 

Our second timescale is a radiative cooling timescale: 

\begin{equation}
    \label{eqn:rad_time}
    t_{rad} = E_{lobe}/L_{rad}
\end{equation}
where $E_{\rm lobe}$ is the total energy stored in the lobes and $L_{\rm rad}$ is the radio luminosity, $\nu L_\nu$, at 10 GHz.  This timescale is simply the time to drain the lobe of its energy via radio emission, assuming no other energy gains or losses. Evaluating $E_{\rm lobe}$ and $L_{\rm rad}$ for the unresolved peaked sources gives cooling times of $3 < \log (t_{\rm rad}/{\rm year}) < 4.8$. Overall, $t_{\rm rad}$ is significantly longer than the time to create the lobes, $t_{\rm dyn}$. This is important because it justifies excluding radiative losses in the simple dynamical model of lobe inflation that only considered adiabatic losses. In practice, of course, as long as the jet remains active, the lobes will continue to inflate and the radiative losses will not compete with the energy input from the jet. 

Our third timescale, $t_{\rm syn}$, gives the cooling time of the relativistic electrons that generate synchrotron emission near $\nu_{\rm GHz}$: 

\begin{equation}\label{eqn:syn_time}
    t_{syn} = 5.0 \times 10^4\, P_{-8}^{-3/4}\, \nu_{GHz}^{-1/2}\ \  \textrm{yr}
\end{equation}
where $P_{-8}$ is the equipartition pressure in units of 10$^{-8}$ dyne cm$^{-2}$. For our unresolved peaked sources, the synchrotron lifetimes at 10 GHz are $1.8 < \log t_{\rm syn} < 4$ years. While this range overlaps the range in $t_{\rm dyn}$, object by object we find $t_{\rm syn} > t_{\rm dyn}$ typically by factors $\sim 20-50$. We conclude that the electron lifetimes at 10 GHz are significantly longer than the lobe inflation dynamical times, $t_{\rm dyn}$, suggesting that spectral steepening at high-frequencies is not occurring for these compact sources. As discussed in Section~\ref{sec:steep_alpha}, the fact that the high-frequency spectra are steeper than normal, $\alpha_{\rm high} \approx -1.0$, can be explained by inverse Compton scattering off the intense AGN radiation field.

\subsection{Free-Free Absorption (FFA)}\label{sec:FFA}

An alternate possible cause of the peaked radio spectra is free-free absorption (FFA) by electrons in a thermal gas either interior or exterior to the radio-emitting plasma. In its simplest form, FFA generates a steep, exponentially truncated, low frequency spectrum, though more complex geometries can generate a wider variety of optically thick spectra \citep[e.g.,][]{bicknell+97, tingay+03, callingham+15, callingham+17, mhaskey+19}. For our sample, the distribution of $\alpha_{low}$ (third panel in Figure~\ref{fig:spshape}) shows that only three sources have measured indices steeper than the SSA limit of $+2.5$, disfavoring simple FFA for the majority of our sample. However, with only one or two low frequency observations, our ability to reliably measure a low frequency index is limited, and furthermore $\sim60\%$ of our peaked sources only have lower limits for $\alpha_{\rm low}$. We therefore continue to examine simple models that aim to ascertain whether FFA is a credible possibility. 

First, the standard relation for the optical depth to FFA at frequency $\nu_{\rm GHz}$ from a uniform gas with electron density $n_e$ cm$^{-3}$, temperature $T_4\times10^4$ K and depth $l_{\rm kpc}$ kpc, is \citep{condon+16}:

\begin{equation}
\label{eqn:ff_tau}
    \tau_{f\!f} \approx 3.3\times10^{-4}\ T_4^{-1.35}\ \nu_{\rm GHz}^{-2.1}\ \int n_e^2\ dl_{\rm kpc}\ .
\end{equation} 
Assuming a uniform medium, and setting $\tau_{f\!f} \approx 1$ at the peak frequency $\nu_{\rm pk,GHz}$ we obtain the following constraint on the path length and electron density:

\begin{equation}
\label{eqn:ff_tau1}
    n_e^2\ l_{\rm kpc} \approx 3000\ T_4^{1.35}\ \nu_{\rm pk,GHz}^{2.1}.
\end{equation}

If we now focus on the fact that our sample is deeply embedded, with extreme  MIR/optical flux density ratios and red WISE colors, then a plausible assumption is that the dust along the line of sight to the nucleus becomes optically thin somewhere between W2 and W3 (observed) or $\sim5\mu$m (rest) or, in magnitudes, $A_{\rm M}\sim 1$. While the UV extinction in AGN at high-redshift is quite uncertain, the NIR-MIR extinction seems better behaved \citep[e.g.,][]{hirashita+19} so we use a standard   reddening curve \citep[][]{wang+19} with $A_{\rm M}/A_{\rm V} \approx 0.023$ and dust-to-gas ratio $A_{\rm V} \approx 5.3\times10^{-22} N_H$, where $N_H$ is the total hydrogen column density in cm$^{-2}$. If we further assume this column is ionized then we find:

\begin{equation}
\label{eqn:nuc_column}
    n_e\ l_{\rm kpc} \approx 26\ A_{\rm M}.
\end{equation}
Combining equations \ref{eqn:ff_tau1} and \ref{eqn:nuc_column}, we find plausible values for the ionized gas density,  region size, and mass (assuming the region to be spherical):

\begin{equation}
\label{eqn:combined}
\begin{split}
n_e\ \approx 117\ A_{\rm M}^{-1}\ T_4^{1.35}\ \nu_{\rm pk,GHz}^{2.1} \ {\rm cm}^{-3},\\
l_{\rm kpc}\ \approx 0.22\ A_{\rm M}^{2}\ T_4^{-1.35}\ \nu_{\rm pk,GHz}^{-2.1}\ {\rm kpc},\\
M_{\rm ion}\ \approx 1.4\times10^8\ A_{\rm M}^{5}\ T_4^{-2.7}\ \nu_{\rm pk,GHz}^{-4.2} \rm M_{\odot}.
\end{split}
\end{equation}

To check the assumption that the gas could be ionized by the AGN, we consider the standard Stromgren condition that ionization balances recombination: 

\begin{equation}
\label{eqn:stromgren}
    (4\pi/ 3)\ d_S^3\ n_e^2\ \alpha_B \approx Q_{\rm ion} \equiv f_{\rm ion}\ L_{\rm bol}\ /\ h\nu_0.
\end{equation} 
where $\alpha_{B} = 2.6\times10^{-13}\  T_4^{-0.8}$ cm$^3$ s$^{-1}$ is the case B recombination coefficient, $Q_{\rm ion}$ is the number of ionizing photons per second, $h\nu_0 = 13.6$ eV, and $f_{\rm ion} \equiv h\nu_0\  Q_{\rm ion} / L_{\rm bol}$ approximates the ionizing fraction and is $\sim 0.1-0.2$ for typical AGN SEDs. Expressing $d_S$ in kpc and the bolometric luminosity in units of $10^{13} L_{\odot}$, we find:

\begin{equation}
\label{eqn:stromgren_constraint}
    n_e^2\ d_{\rm S,kpc}^3 \approx 55,000\  f_{\rm ion} L_{\rm bol,13}\ T_4^{0.8}.
\end{equation} 
Finally, combining Equations~\ref{eqn:nuc_column}, \ref{eqn:combined} and \ref{eqn:stromgren_constraint} we find:

\begin{equation}
\label{eqn:dist_ratio}
    \frac{d_{\rm S,kpc}}{l_{\rm kpc}} \approx 7.15\ A_{\rm M}^{-4/3}\ T_4^{0.72}\ \nu_{\rm pk,GHz}^{0.7}\ f_{\rm ion}^{1/3}\ L_{\rm bol,13}^{1/3}.
\end{equation} 
For $f_{\rm ion}\sim0.1-0.2$ and $A_M\sim1-$few, we have $d_{\rm S,kpc}\sim l_{\rm kpc}$ and we confirm that a high column-density that is optically thick into the NIR can be photoionized by an AGN with  bolometric luminosity typical of our sources. We conclude that our deeply embedded AGN are sufficiently luminous that enough of the high-column nuclear material can be ionized to generate free-free absorption peaks in the $0.1 - 10$ GHz frequency range.  

An alternate scenario that might give rise to free-free absorption is to consider the adiabatic lobe expansion model introduced in Paper II. The model is based on energy and momentum conservation as the lobe expands into an ISM of density $n_a$, creating a shell surrounding the lobe of density $n_a / 3\delta$ where the shell's thickness is $\Delta R_{\rm shell} = \delta \times R_{\rm lobe}$. If the shell is ionized, then the emission measure across it is $\int n_e^2 dr = n_a^2 R_{\rm lobe} / 9\delta$. If we now substitute this for the factor $n_e^2\ l_{kpc}$ in Equations~\ref{eqn:ff_tau} and \ref{eqn:ff_tau1} then we get an equation for the free-free turnover frequency:

\begin{equation}
\label{eqn:peak_shell}
    \nu_{\rm pk,GHz}^{2.1} \approx 3.7\times10^{-5}\ T_4^{-1.35}\ n_a^2\ R_{\rm lobe,kpc}\ \delta^{-1}.
\end{equation} 
This strongly suggests that the swept up gas is {\it not} optically thick to free-free absorption, since combining values of $n_a$ taken from the middle panel of Figure~\ref{fig:new_lobe_model} with values of lobe radius $0.01 < R_{\rm lobe,kpc} < 10$ and shell thickness $0.01 < \delta < 0.1$ still cannot compete with the prefactor $3.7\times10^{-5}$, so the implied radio turn-over frequencies are well below the GHz range. 

In summary, we find that the lobe expansion model cannot yield shells that are free-free thick at GHz frequencies. However, the large column densities implied by the embedded nature of our sources could be free-free thick at GHz frequencies. This latter condition also requires the high column-density to be ionized by the central AGN, which  seems possible since our sources are of high bolometric luminosity. 

\subsection{Further discussion of FFA {\it vs} SSA}
\label{sec:SSA_vs_FFA}

Although we have shown that the deeply embedded nature of our sample is consistent with the high column-densities necessary for FFA, there are some problems with this particular idea. 

First, if the radio source lies interior to the MIR emitting region then we might expect both the presence of a peak and the frequency of the peak to correlate with possible tracers of the nuclear column density, such as MIR color (W1-W2 or W2-W3), MIR \& FIR luminosity ($L_{6\mu m}$, $L_{870\mu m}$), or MIR energy density ($u_{IR}, R_{bol}$). However, as discussed in Section~\ref{sec:mir_radio}, no such correlations are seen in our data, and in particular the distributions of these MIR parameters are, statistically, identical for the three spectral classes: PL, CV, and PK. This suggests either the obscuring columns lie interior to the radio source, or FFA from those columns isn't significant. 

Second, the only property that is clearly different between these spectral groups is radio source size (see Figure~\ref{fig:spectra_vs_morph}), where peaked sources are preferentially unresolved compared to curved or power law sources. This association, while not conclusive, does weigh in favor of SSA.


Third, our peaked radio sources are very similar to typical GPS, CSS and HFP sources, with similar sizes, magnetic fields, and turnover frequencies (see Paper II). However, these other samples are {\it not} deeply embedded. It seems unlikely, therefore, that one can  transfer our arguments for FFA given in \S\ref{sec:FFA} to these other sources. The natural preference for a {\it single} explanation for radio peaks in all these sources cannot, therefore, invoke the embedded nature of our sources, because many other peaked sources are not embedded. This argument therefore weighs against FFA that is directly associated with the high MIR columns found in our sources.

\section{Non-Peaked Sources}\label{sec:nonpeak}

About 25\% of our sources have spectra classified as ``Curved''. It is currently unclear whether their spectral shapes have a similar origin to the peaked sources (but with lower frequency turnover, $\lesssim100$ MHz), or whether their spectral shape has a different origin. The degree of curvature, quantified by the $q$ parameter, is significantly smaller than the peaked spectra (see fourth panel in Figure~\ref{fig:paramd}), and may therefore not be easily explained by the standard SSA or FFA processes that tend to yield stronger curvature with larger (more negative) $q$. In their study of extended classical radio sources, \citet{duffy+12} interpret their curved spectra as arising from non-power law (log normal) electron energy distributions. While this remains a possibility for our curved sources, there are no obvious differences between our curved and peaked sources --- both are dominated by compact emission (Figure~\ref{fig:spectra_vs_morph}) and both have similar high-frequency spectral indices ($\alpha_{high}$,  Figure~\ref{fig:paramd}), suggesting that at these energies the electron energy distributions are similar and neither group has experienced significant radiative losses. We therefore remain undecided as to the nature of the ``Curved'' sources.

In addition to the peaked, curved and power-law sources, thirteen have more unusual spectral shapes: eight are flat (F), four are upturned (U), and one is inverted (I). While we cannot be certain what underlies these different spectral shapes, possibilities include superposition of multiple components, restarted sources \citep[e.g.,][]{callingham+17}, blazar or variable sources, patchy opacity \citep[e.g.,][]{bicknell+18}, or absorbed sources with $\nu_{pk}$ $\gtrsim20$ GHz \citep[e.g.,][]{nyland+20}. The fact that these unusual spectra are so few suggests that our sample is fairly homogeneous with similar properties subject to similar physical processes. 

\section{Summary and Conclusion}\label{sec:sed_conclusion}
We continue our analysis of the sample of 167 heavily obscured, luminous quasars in the redshift range $0.4 <z <2.8$ identified by  \citet[][Paper I]{lonsdale+15}.  Our sample selection identified optically faint but MIR-bright quasars with a luminous ($L_{1.4 \rm GHz}\gtrsim10^{25}$ W Hz$^{-1}$) radio source that is unresolved by NVSS ($<45$\asec). In Paper II, we presented 10 GHz VLA images of 155 ($93\%$) of the sample, which reveals that 57\% are compact even on sub-arcsecond scales ($\lesssim1.7$ kpc at $z\sim2$). In the present study, we add archival data to our 10 GHz flux densities to construct radio spectra spanning 150 MHz to 10 GHz. We use these spectra to further analyze the nature of the radio sources. Our main results are as follows: 

\begin{enumerate}
    \item In this sample of 155 sources, 58 (37\%) have peaked spectra while 38 ($25\%$) have curved spectra that might turn over below our lowest frequency (150 MHz). Thus, up to $62\%$ of our sample have spectra that could place them in the well-known classes of CSS, GPS and HFP sources. 
    
    \item Our sources have somewhat steeper high-frequency (optically thin) spectra than the canonical $\alpha \approx -0.7$ value for the general population of AGN radio sources. Of the 119 sources with well-measured high-frequency spectral indices, about $80\%$ have
    $\alpha_{\rm high} < -0.8$ with a median value near $-1.0$. 
    If the high bolometric radiation field is roughly cospatial with the radio source, we find its energy density to be factors of $2-20$ greater than the magnetic energy density in the radio source. This suggests that inverse Compton scattering of the relativistic electrons off the MIR photons may explain the steeper high-frequency spectra. Alternatively, it has been suggested that jet interaction with a dense ambient medium, which is likely present in our sources, may also generate steep spectra. 
    
    \item We looked for correlations between radio spectral parameters and several MIR and radio continuum parameters. Other than mutually dependant variables, no new statistically significant two-parameter correlations were found. Although the radio properties of our sample are similar to samples of CSS/GPS sources, their MIR colors are quite different. This is consistent with our sources being caught in a short-lived stage possibly following a gas-rich merger. 
    
    \item Extrapolating our 10 GHz flux densities to 345 GHz confirms that the synchrotron contribution to the measured 345 GHz ALMA flux densities is typically $<10\%$, confirming that thermal dust emission dominates at sub-mm wavelengths. This result supports our estimates of physical properties such as AGN luminosity and star formation rates presented in Paper I. 
    
    \item Most of our peaked and curved sources lie close to the linear size vs.\ turnover frequency relation seen for other CSS/GPS/HFP samples. Synchrotron self-absorption (SSA) is likely to be dominant in those sources, or free-free absorption (FFA) under particular circumstances such as the presence of a dense ambient medium \citep[e.g.,][]{bicknell+18}. 
    
    \item Assuming the same magnetic field underlies SSA at low frequencies and equipartition at higher frequencies, we invoke the condition $B_{\rm SSA} \approx B_{\rm min}$ to derive emitting region sizes and magnetic fields for the peaked sources. We find the emitting regions have sizes in the range $3-100$ pc and magnetic fields in the range $6-100$ mG, similar to values found in luminous GPS/HFP sources. 
    We reapply the dynamical model from Paper II using the new values of source size and magnetic field and find even younger ages. The ages of the youngest sources now approach the ages of a recently discovered class of compact peaked sources that have turned on in the last $10-20$ years \citep{nyland+20}.  

    \item The deeply embedded nature of our sources, together with their high AGN ionizing luminosity, raises the possibility that free-free absorption may also play a role in causing the peaked radio spectra. We show this 
    is possible assuming the regions become optically thin to dust in the MIR, which is consistent with the extremely red WISE colors. We confirm that these columns can also be ionized by the high AGN luminosity. Conversely, the shells of ionized gas swept up by the expanding radio lobes cannot themselves provide sufficient column density for FFA. 
    
    \item Two arguments weigh against FFA associated with the high MIR column densities. First, there are no correlations between MIR properties and radio spectral properties, including the spectral shape classes of peaked, curved or power law.  Second, our radio sources are very similar to other classes of peaked radio source (GPS, HFP) but these other sources are not deeply embedded. If a single mechanism creates spectral peaks it cannot depend on the high column densities associated with deeply embedded AGN. 
    
    \item Roughy $25\%$ of our sample have gently curved spectra across our spectral window. We are unable to ascertain whether these sources are simply peaked sources with peaks below $150$ MHz, or whether they are optically thin sources with non-power law electron energy distributions (as suggested by \citealt[][]{duffy+12}).  
\end{enumerate}

Overall, we find many sources that have peaked or curved radio spectra indicating compact emission regions likely arising from recently triggered radio jets. 
Further work on this important sample of young and deeply embedded radio sources is ongoing. VLBA observations can help probe the most compact radio source components, while deeper, multi-frequency VLA, VLBA and e-MERLIN observations can yield spectral index maps that further reveal the nature and evolution of the radio sources.

\section*{Acknowledgments}
We thank the anonymous referee for many helpful suggestions that have significantly improved the paper. Pallavi Patil is a Jansky Fellow of the National Radio Astronomy Observatory.
The National Radio Astronomy Observatory is a facility of the National Science Foundation operated under cooperative agreement by Associated Universities, Inc. Support for this work was  provided by the NSF through the Grote Reber Fellowship Program administered by Associated Universities, Inc./National Radio Astronomy Observatory. 
Basic research in radio astronomy at the U.S. Naval Research Laboratory is supported by 6.1 Base Funding.  MK was supported by by the National Research Foundation of Korea (NRF) grant funded by the Korea government (MSIT, No. 2020R1A2C4001753).
This publication makes use of data products from the Wide-field Infrared Survey Explorer, which is a joint project of the University of California, Los Angeles, and the Jet Propulsion Laboratory/California Institute of Technology, funded by the National Aeronautics and Space Administration. This
research made use of NASA’s Astrophysics Data System, the VizieR
catalog access tool, CDS, Strasbourg, France. This research has made use of the NASA/IPAC Extragalactic Database (NED),
which is operated by the Jet Propulsion Laboratory, California Institute of Technology,
under contract with the National Aeronautics and Space Administration.
The authors have made use of SciPy \citep{2020SciPy}, {\sc Astropy}, a community-developed core {\sc Python} package for Astronomy \citet{astropy+13}, PANDAS, a data analysis and manipulation python module \citep{pandas}, and NumPy \citep{numpy}. This research made use of APLpy, an open-source plotting package for {\sc Python} hosted at \url{http://aplpy.github.com}. This research made use of matplotlib, a {\sc Python} library for publication quality graphics \citep{hunter+07}.


\facilities{VLA, ALMA, WISE}

\software{CASA \citep{mcmullin+07};\,
         {\sc Astropy} \citep{astropy+13};\,
          MATPLOTLIB \citep{hunter+07};\,
          TOPCAT \citep{topcat};\,
          seaborn \citep{seaborn}}

\bibliography{jvla_SED_v1.bib}

\appendix



\section{Tables and Figures}
\subsection{Radio Spectra of the Entire Sample}
Figure~\ref{fig:sed} shows radio spectra of our entire sample. The figure caption is the same as that of  Figures~\ref{fig:sedcheck} and~\ref{fig:sedparam}. 
\begin{figure*}[htpb!]
\centering
\includegraphics[clip=true, width=\textwidth]{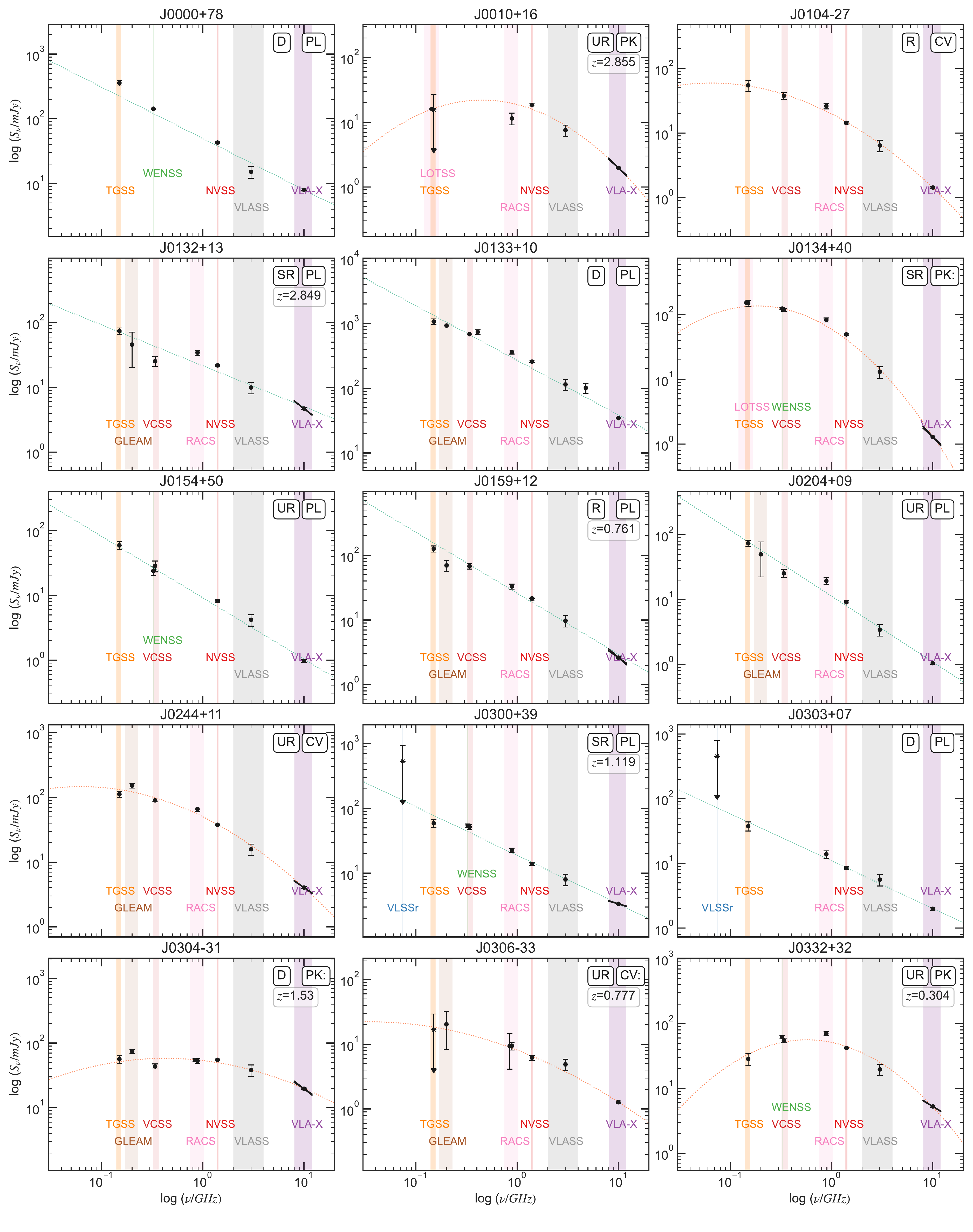}
\caption{Radio spectra of our entire sample. See Figures~\ref{fig:sedcheck} and~\ref{fig:sedparam} for  the caption. }\label{fig:sed}
\renewcommand{\thefigure}{\arabic{figure} (Cont.)}
\addtocounter{figure}{-1}
\end{figure*}
\begin{figure*}
\centering
\includegraphics[clip=true, width=\textwidth]{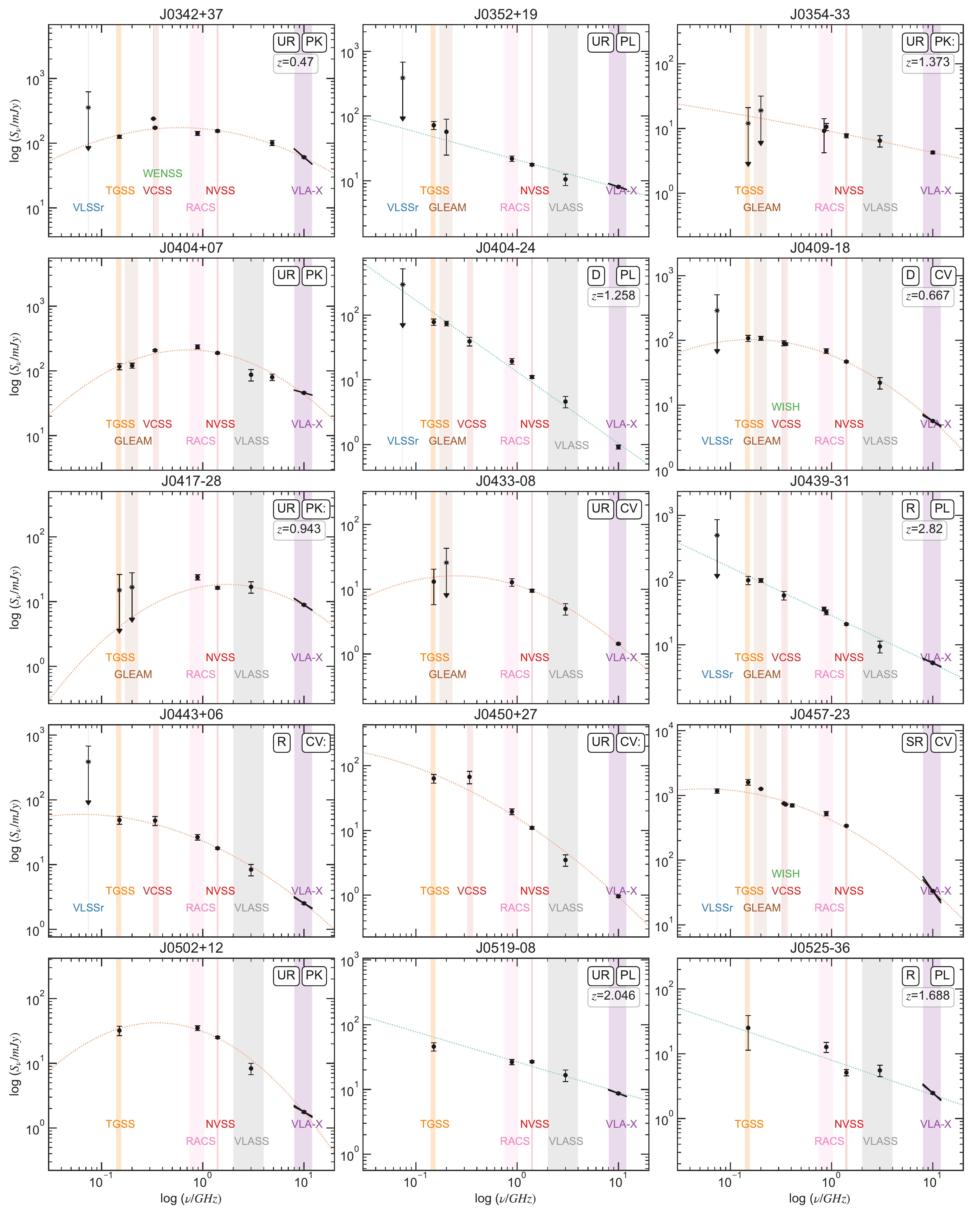}\captcont{\it Continued}
\end{figure*}
\begin{figure*}[htpb!]
\centering
\includegraphics[clip=true,  width=\textwidth]{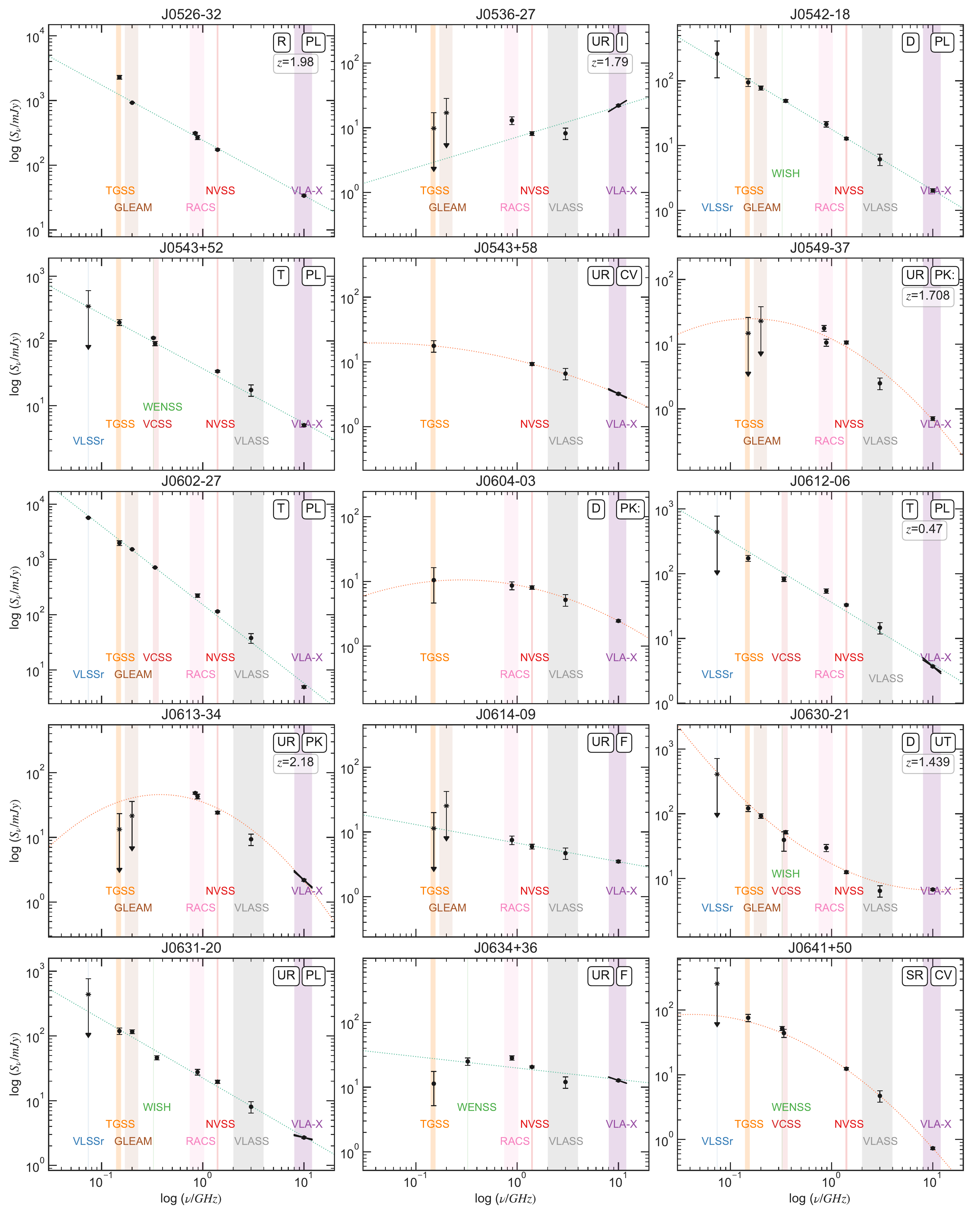}\captcont{\it Continued} 
\end{figure*}
\begin{figure*}[htpb!]
\centering
\includegraphics[clip=true,  width=\textwidth]{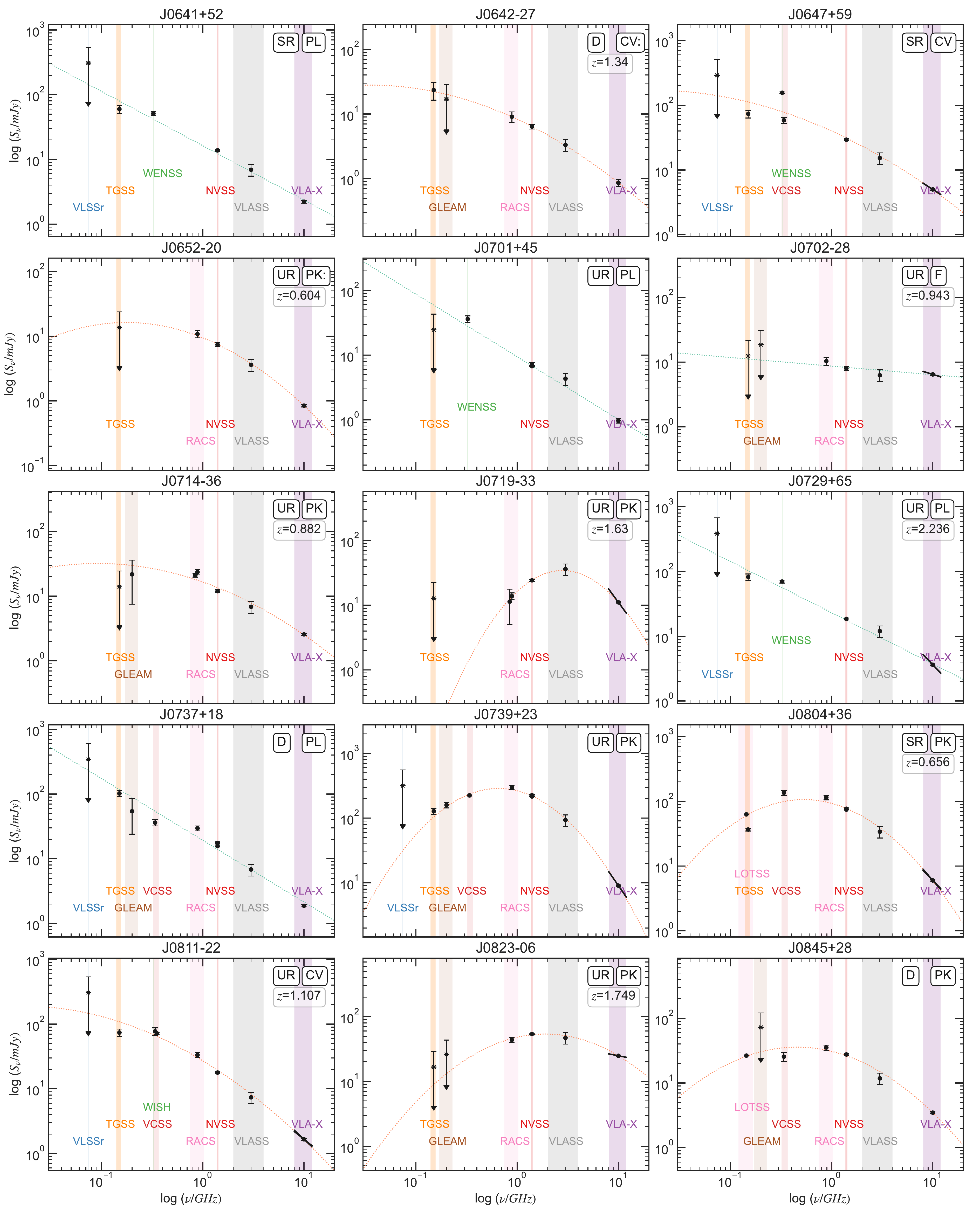}\captcont{\it Continued} 
\end{figure*}
\begin{figure*}[htpb!]
\centering
\includegraphics[clip=true,  width=\textwidth]{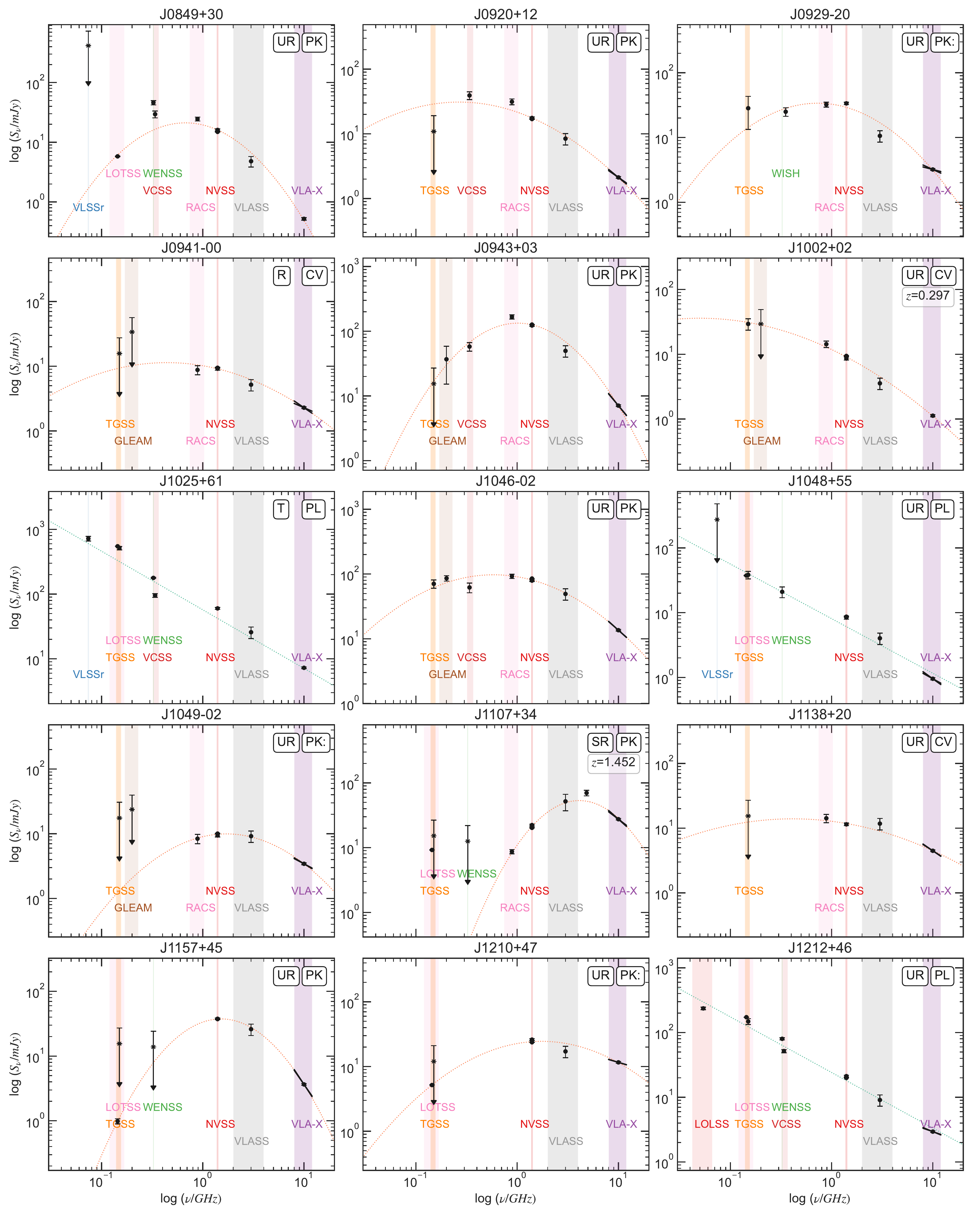}\captcont{\it Continued} 
\end{figure*}
\begin{figure*}[htpb!]
\centering
\includegraphics[clip=true,  width=\textwidth]{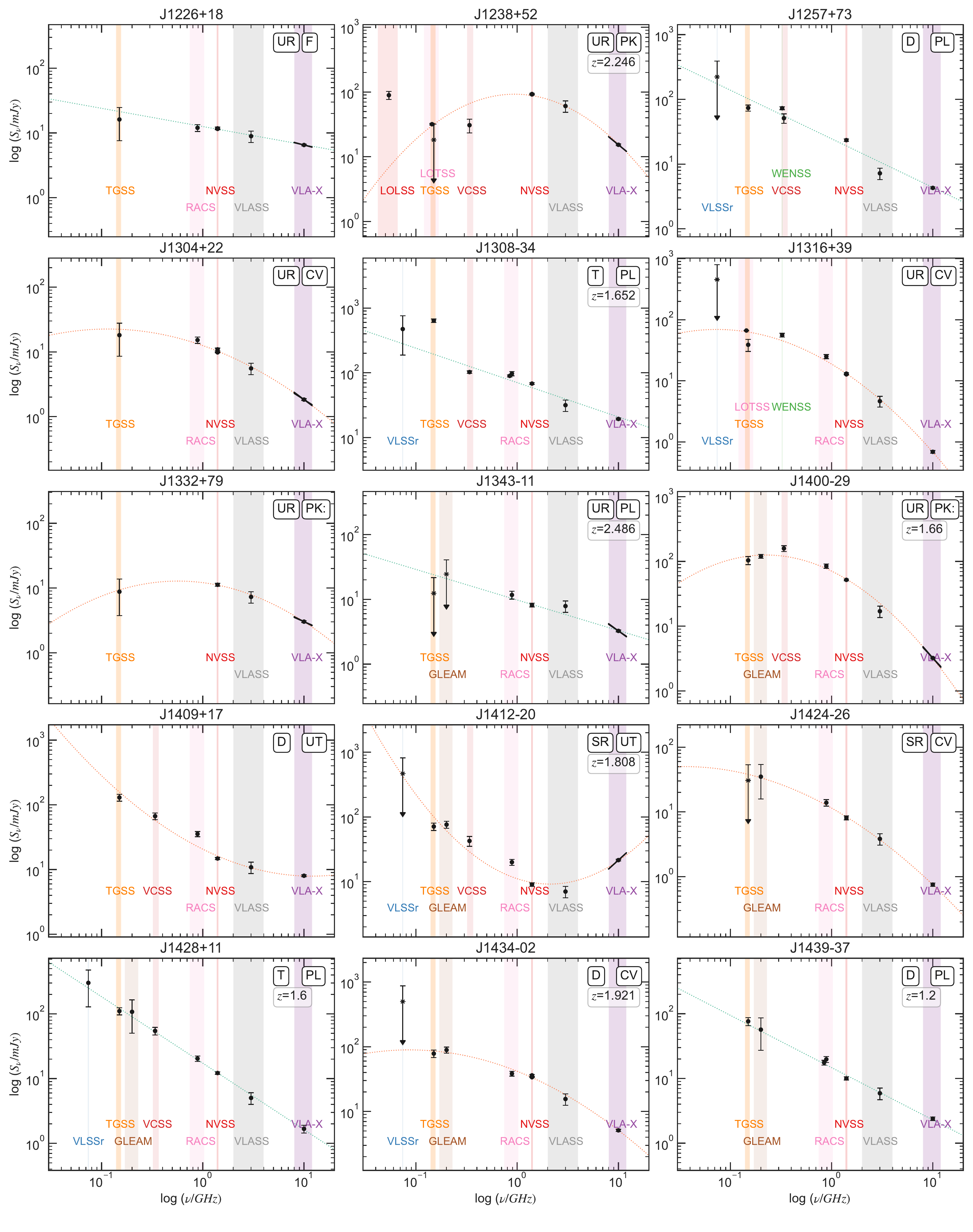}\captcont{\it Continued} 
\end{figure*}
\begin{figure*}[htpb!]
\centering
\includegraphics[clip=true,  width=\textwidth]{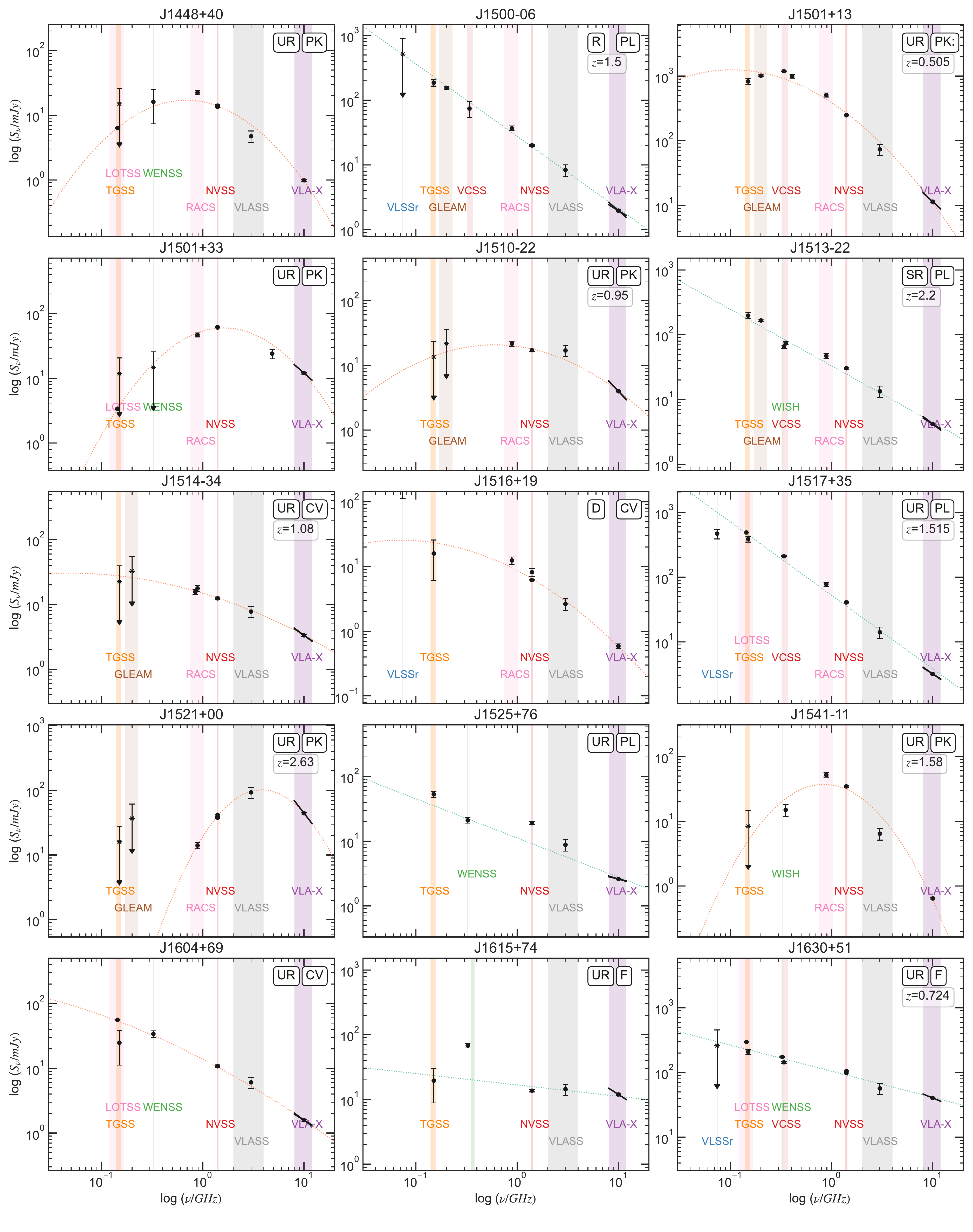}\captcont{\it Continued} 
\end{figure*}
\begin{figure*}[htpb!]
\centering
\includegraphics[clip=true,  width=\textwidth]{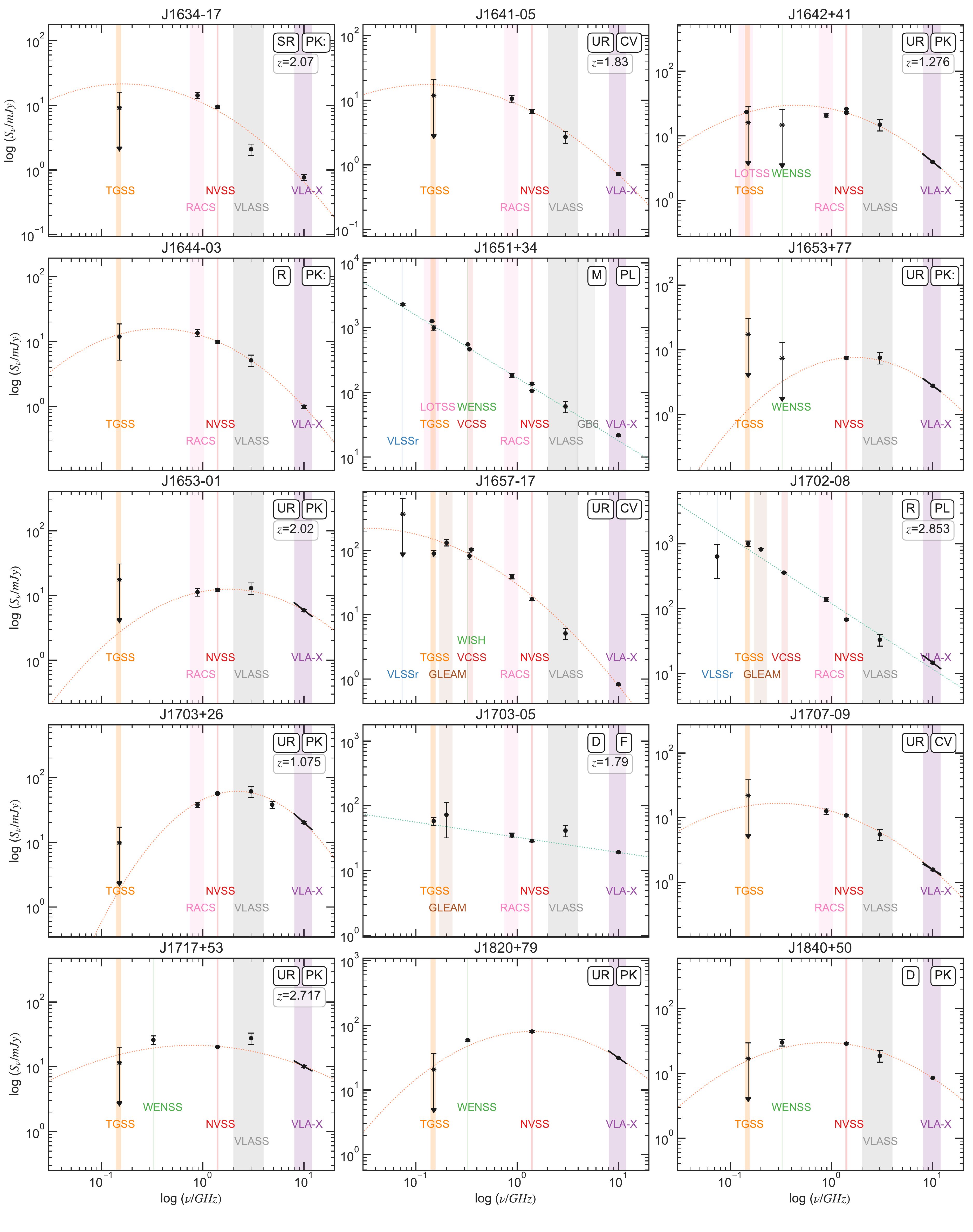}\captcont{\it Continued} 
\end{figure*}
\begin{figure*}[htpb!]
\centering
\includegraphics[clip=true,  width=\textwidth]{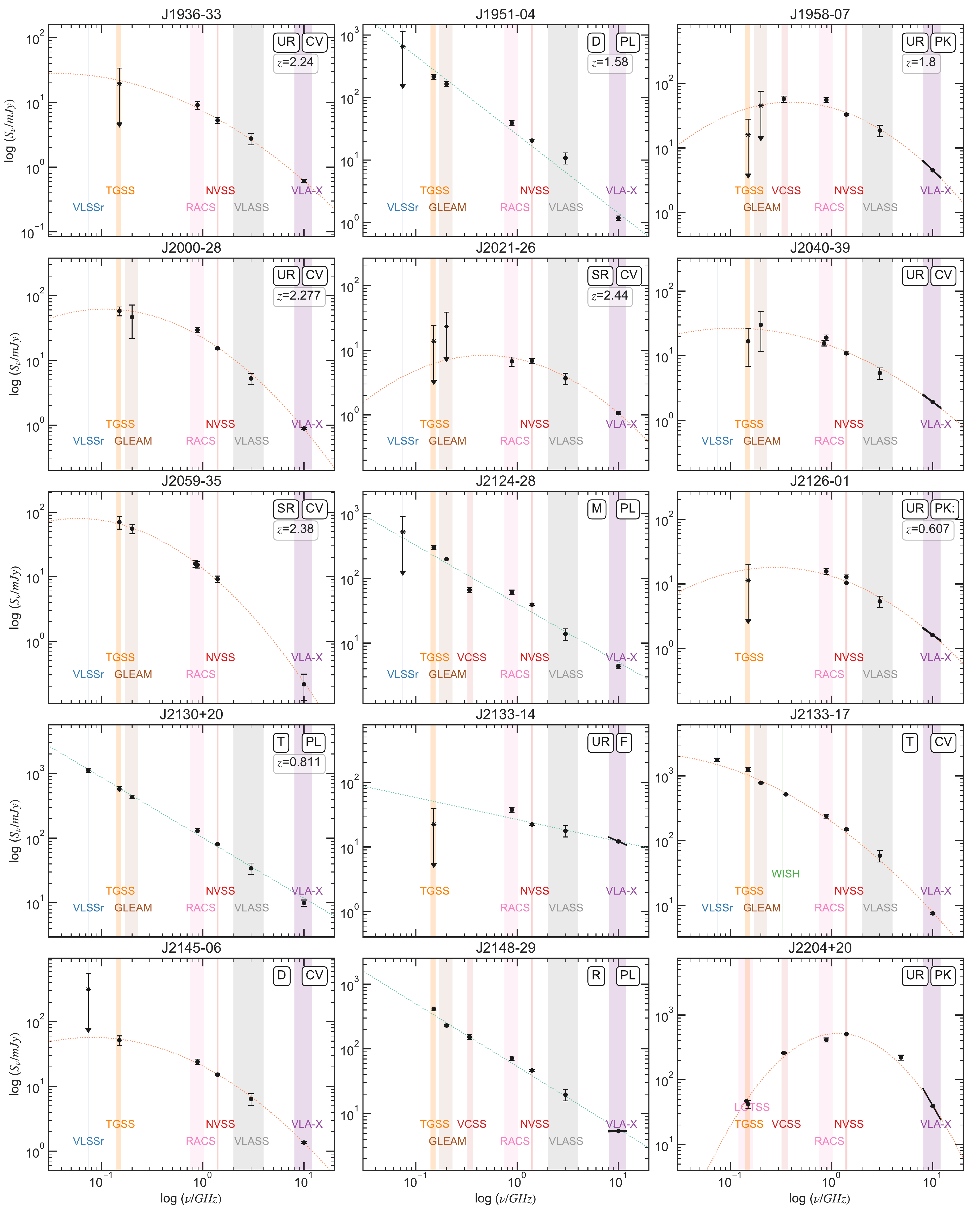}\captcont{\it Continued} 
\end{figure*}
\begin{figure*}[htpb!]
\centering
\includegraphics[clip=true,  width=\textwidth]{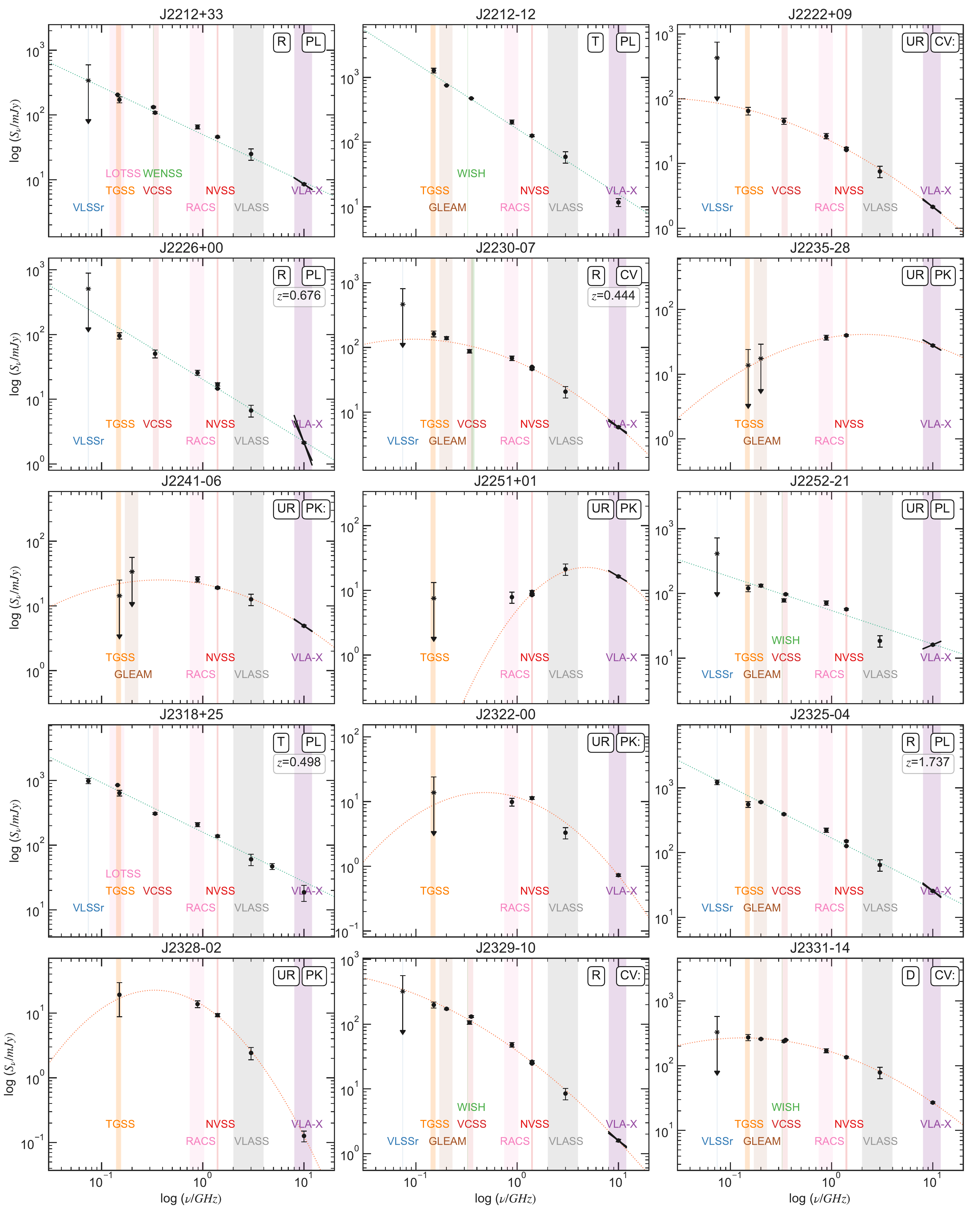}\captcont{\it Continued} 
\end{figure*}

\begin{figure*}[htpb!]
\centering
\includegraphics[clip=true,  width=\textwidth]{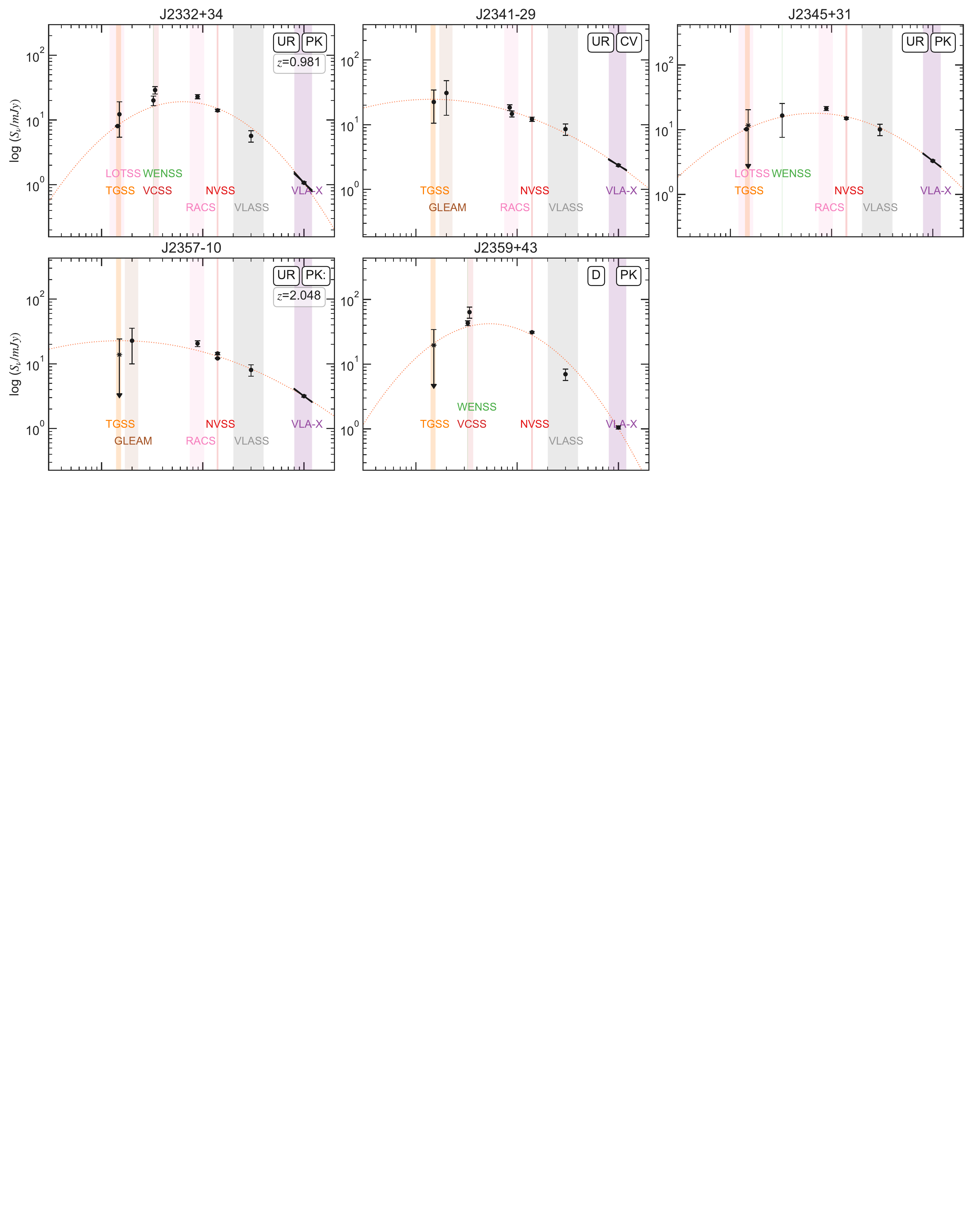}\captcont{\it Continued} 
\end{figure*}

\subsection{Data Tables}
Table~\ref{tab:alpha_calc} provides our spectral shape classification, two-band spectral indices, and  best-fit spectral shape parameters. 
Table~\ref{tab:bsoltab} summarizes the results from our analysis in Section~\ref{sec:bsolve} and Section~\ref{sec:timescales}.  
\input{SED_alpha_measure_v6}

\input{Bsol_tab_v7}

\end{document}

%% file: Survey_tab.tex
\begin{deluxetable*}{ccccccccccccc}
\tablecaption{List of Radio Surveys used in the Construction of  Multiband Radio Spectra\label{tab:surveys}}
\tablehead{
\colhead{Survey} & \colhead{$\nu$} & \colhead{$\Delta\nu$} & \colhead{$\lambda$} & \colhead{$\theta_{res}$} & \colhead{$\sigma_{rms}$} & \colhead{Dec. Range} & \colhead{$\delta_{pos}$} & \colhead{$\theta_{s}$} & \colhead{N} & \colhead{$n_{det}$} & \colhead{$n_{obs}$} &  \colhead{Ref}\\
\colhead{} &  \colhead{GHz} & \colhead{GHz} & \colhead{cm} & \colhead{$^{\prime\prime}$} & \colhead{mJy/beam} & \colhead{deg} &  \colhead{$^{\prime\prime}$}& \colhead{$^{\prime\prime}$} & \colhead{} & \colhead{} & \colhead{} &\colhead{}} 
\colnumbers
\startdata
VLA-A & 10 & 2 & 3 & 0.2 & 0.013 & \nodata & 0.04$^\dagger$ & \nodata & \nodata &  118 & 118 &  1\\
VLA-B & 10 & 2&  3 & 0.6 & 0.013 & \nodata  & 0.1$^\dagger$ &  \nodata & \nodata & 147 & 147 &  1  \\
GB6 & 4.85 & \nodata & 6 & 630 & \nodata  & 0$-$75 & \nodata & 20 & 75,162 & 9 & 9  & 2  \\
VLASS & 3 & 2 & 10 & 2.5 & 0.15 & $>-40$ & 0.5  & \nodata &  5,300,000$^\star$ & 153 & 153 & 3 \\
NVSS & 1.4 & 0.03 & 21 & 45 & 0.45 & $>-40$ & $<\sim1$& 7&  1,773,484 & 155 & 155  & 4  \\
FIRST & 1.4 & 0.03 & 21 & 5 & 0.15 & \nodata & $\lesssim1$ & 5 & 946,432 & 52 & 52  & 5\\
RACS & 0.887  & 0.228 & 35 & 15 & 0.25 & $<41$ & 0.8 &5& 2,800,000 & 133 & 137 & 6\\
SUMSS & 0.843 & 0.002 & 35 & 45 & 6-10 & $<-30$ & 2 & 12 & 211,063 & 15 & 18 & 7\\
MRC & 0.408 & 0.002 & 73  & 92  & 60&  18 $-$ --85  & $\sim$ 5 & 20 &  12,141  &  3 & 3 & 8 \\
VCSS & 0.338 & 0.032 & 88 & 15 & 3 & $>-40$ & \nodata & 5 & \nodata & 96 &  153 & 9\\
WENSS & 0.325 & 0.003 & 92 & 54 & 3.6 & 28$-$76 & 1.5 & 10 & 211,234 & 31 &  49 & 10\\
GLEAM & 0.2 & 0.157 & 150 & 100 & 6-10 & $<30$ &  $\sim$1.6 & 20 & 307,455 & 39 &  110 &  11  \\
TGSS ADR1 & 0.15 & 0.0085 & 200 & 25 & 3.5  & $>-53$& 2& 20 & 623,604 & 86 & 155 & 12\\
LOTSS-DR2 & 0.144 & 0.048 & 208 & 6 & 0.083  & $>28^\ddagger$  & 0.2  & 20 & 4,396,228 & 25 & 25 & 13 \\ 
VLSSr & 0.074 & 0.002 & 405 & 75 & 100 & $>-30$ & \nodata & 20 & 92,965 & 9 & 143 & 14\\
LOLSS &  0.054 & 0.024 & 555 & 47 & 4 & \nodata & 2.5 & 20 &  25,247 & 2 & 3 & 15
\enddata
\tablecomments{Column 1: Name of the Survey; Column 2: Central frequency of the observation in GHz; Column 3: Frequency bandwidth  in GHz; Column 4: Central wavelength of the observation in cm; Column 5: Nominal angular resolution in arcsec; Column 6: 1$\sigma$ rms noise level; Column 7: Declination limit or coverage of the survey; Column 8: Positional accuracy; Column 9: Search radius used in the catalog cross-match with WISE position of the sample; Column 10: Total number of objects detected in the entire survey; Column 11: Total number of cross-matched sources from our sample; Column 12: Total number of our sources within the footprint of each survey; Column 13: Catalog and survey references: 1: \citet{patil+20}; 2: \citet{gregory+96}; 3: \citet{lacy+20}; 4: \citet{condon+98}; 5: \citet{becker+95}; 6: \citet{mcconnell+20}; 7: \citet{mauch+03}; 8: \citet{large+81}; 9: \citet{peters+21}; 10: \citet{rengelink+97}; 11: \citet{hurley+17}; 12: \citet{intema+16};
13: \citet{shimwell+22};  14: \citet{lane+14}; 15: \citet{gasperin+21}. \\
$\dagger$: We adopt a positional accuracy  of $\sim 20\%$ of the synthesized beam. \\
$\star$: The number of sources in the still-ongoing VLASS is only an estimated count of individual source components. \\
$\ddagger$: The latest data release covers two regions  in the sky. Please refer to \citet{shimwell+22} for additional details on the sky coverage.}
\end{deluxetable*}

%% file: SED_alpha_measure_v6.tex
\begin{longrotatetable}
\begin{deluxetable}{ccccccccccccc}
\tabletypesize{\scriptsize}
\tablecaption{Radio Spectral Shape Parameters and Two-band Spectral Indices of our Sample\label{tab:alpha_calc} }
\tablehead{
\colhead{Source} & \colhead{$z$} & \colhead{Morph} &  \colhead{Sp Class} & \colhead{$\alpha^{10}_{1.4}$} & \colhead{$\alpha^{1.4}_{0.15}$} 
& \colhead{$\nu_{peak}$} & \colhead{q} & \colhead{$S_{peak}$} & \colhead{$\alpha_{high}$}  & \colhead{$\alpha_{low}$} &  \colhead{QC} & \colhead{Rejected?}\\
\colhead{} & \colhead{} & \colhead{} & \colhead{} & \colhead{} &  \colhead{} &  \colhead{GHz} & \colhead{} & \colhead{mJy} & \colhead{}  & \colhead{} & \colhead{} & \colhead{}} 
\colnumbers
\startdata
J0000+78 & \nodata & D & PL & $-0.86\pm0.04$ & $-0.95\pm0.10$ & \nodata & $-0.01\pm0.06$ & \nodata & $-0.79\pm0.04$ & \nodata & MP & N \\
J0010+16 & 2.855 & UR & PK & $-1.14\pm0.04$ & \nodata & $0.45\pm0.09$ & $-0.25\pm0.05$ & $21.99\pm4.03$ & $-1.21\pm0.10$ & $-0.18\pm0.12$ &  & N \\
J0104$-$27 & \nodata & R & CV & $-1.17\pm0.05$ & $-0.60\pm0.10$ & $<0.15$ & $-0.15\pm0.03$ & \nodata & $-1.17\pm0.02$ & \nodata &  & N \\
J0132+13 & 2.849 & SR & CV: & $-0.78\pm0.04$ & $-0.55\pm0.10$ & \nodata & $-0.12\pm0.06$ & \nodata & $-0.78\pm0.12$ & \nodata & SB & N \\
J0133+10 & \nodata & D & PL & $-1.02\pm0.04$ & $-0.64\pm0.10$ & \nodata & $-0.09\pm0.01$ & \nodata & $-0.84\pm0.03$ & \nodata & MP & Y \\
J0134+40 & \nodata & SR & CV & $-1.86\pm0.04$ & $-0.50\pm0.10$ & $0.18\pm0.05$ & $-0.29\pm0.04$ & $136.30\pm27.13$ & $-1.83\pm0.06$ & $-0.44\pm0.16$ &  & N \\
J0154+50 & \nodata & UR & PL & $-1.09\pm0.05$ & $-0.88\pm0.10$ & \nodata & $-0.09\pm0.05$ & \nodata & $-0.95\pm0.06$ & \nodata &  & N \\
J0159+12 & 0.761 & R & CV: & $-1.06\pm0.04$ & $-0.80\pm0.10$ & \nodata & $-0.09\pm0.01$ & \nodata & $-0.95\pm0.05$ & \nodata &  & N \\
J0204+09 & \nodata & UR & PL & $-1.10\pm0.05$ & $-0.94\pm0.10$ & \nodata & $-0.07\pm0.04$ & \nodata & $-1.02\pm0.05$ & \nodata &  & N \\
J0244+11 & \nodata & UR & CV & $-1.14\pm0.04$ & $-0.49\pm0.10$ & $<0.15$ & $-0.14\pm0.03$ & \nodata & $-1.10\pm0.02$ & \nodata &  & N \\
J0300+39 & 1.119 & SR & PL & $-0.72\pm0.04$ & $-0.65\pm0.10$ & \nodata & $-0.01\pm0.02$ & \nodata & $-0.70\pm0.05$ & \nodata &  & N \\
J0303+07 & \nodata & D & PL & $-0.74\pm0.05$ & $-0.67\pm0.10$ & \nodata & $-0.02\pm0.02$ & \nodata & $-0.73\pm0.03$ & \nodata & MP & N \\
J0304$-$31 & 1.53 & D & PK & $-0.52\pm0.04$ & $-0.01\pm0.10$ & $0.43\pm0.19$ & $-0.11\pm0.04$ & $58.02\pm8.10$ & $-0.61\pm0.15$ & $-0.59\pm0.49$ &  & N \\
J0306$-$33 & 0.777 & UR & CV: & $-0.80\pm0.06$ & $< -0.39$ & $<0.15$ & $-0.09\pm0.05$ & \nodata & $-0.81\pm0.09$ & \nodata &  & N \\
J0332+32 & 0.304 & UR & PK & $-1.06\pm0.04$ & $0.18\pm0.10$ & $0.57\pm0.07$ & $-0.29\pm0.03$ & $56.35\pm4.84$ & $-1.01\pm0.04$ & $0.78\pm0.15$ &  & N \\
J0342+37 & 0.47 & UR & CV & $-0.48\pm0.04$ & $0.09\pm0.04$ & $0.60\pm0.09$ & $-0.13\pm0.02$ & $173.93\pm12.59$ & $-1.32\pm0.13$ & $0.46\pm0.11$ & SB & N \\
J0352+19 & \nodata & UR & PL & $-0.40\pm0.04$ & $-0.63\pm0.10$ & \nodata & $0.06\pm0.01$ & \nodata & $-0.49\pm0.02$ & \nodata &  & N \\
J0354$-$33 & 1.373 & UR & PK: & $-0.30\pm0.06$ & $< -0.14$ & $<0.89$ & $-0.01\pm0.04$ & \nodata & $-0.30\pm0.01$ & $0.01\pm0.57$ &  & N \\
J0404+07 & \nodata & UR & F & $-0.72\pm0.04$ & $0.22\pm0.10$ & $0.72\pm0.08$ & $-0.23\pm0.03$ & $210.91\pm17.41$ & $-0.43\pm0.00$ & $0.92\pm0.18$ &  & N \\
J0404$-$24 & 1.258 & D & PL & $-1.27\pm0.06$ & $-0.88\pm0.10$ & \nodata & $-0.08\pm0.01$ & \nodata & $-1.10\pm0.04$ & \nodata & E/M & N \\
J0409$-$18 & 0.667 & D & CV & $-1.07\pm0.04$ & $-0.37\pm0.10$ & $<0.34$ & $-0.17\pm0.01$ & \nodata & $-1.07\pm0.02$ & $-0.32\pm0.06$ &  & N \\
J0417$-$28 & 0.943 & UR & PK & $-0.31\pm0.04$ & $< 0.10$ & $<0.89$ & $-0.25\pm0.09$ & \nodata & $-0.95\pm0.19$ & $0.20\pm4.92$ &  & N \\
J0433$-$08 & \nodata & UR & CV & $-0.96\pm0.05$ & $-0.14\pm0.25$ & \nodata & $-0.18\pm0.02$ & \nodata & $-0.96\pm0.02$ & $1.71\pm3.99$ &  & N \\
J0439$-$31 & 2.82 & R & PL & $-0.70\pm0.04$ & $-0.70\pm0.10$ & \nodata & $0.01\pm0.02$ & \nodata & $-0.74\pm0.02$ & \nodata &  & N \\
J0443+06 & \nodata & R & CV: & $-1.00\pm0.04$ & $-0.45\pm0.10$ & $<0.07$ & $-0.12\pm0.02$ & \nodata & $-0.99\pm0.01$ & \nodata &  & N \\
J0450+27 & \nodata & UR & CV: & $-1.24\pm0.05$ & $-0.79\pm0.10$ & $<0.15$ & $-0.10\pm0.04$ & \nodata & $-1.24\pm0.05$ & \nodata &  & N \\
J0457$-$23 & \nodata & SR & - & $-1.18\pm0.04$ & $-0.70\pm0.10$ & $<0.07$ & $-0.13\pm0.02$ & \nodata & $-1.20\pm0.15$ & \nodata & FT & Y \\
J0502+12 & \nodata & UR & CV & $-1.35\pm0.04$ & $-0.11\pm0.10$ & $0.35\pm0.07$ & $-0.28\pm0.04$ & $42.22\pm5.18$ & $-1.29\pm0.10$ & $0.05\pm0.13$ &  & N \\
J0519$-$08 & 2.046 & UR & CV & $-0.57\pm0.04$ & $-0.24\pm0.10$ & \nodata & $-0.07\pm0.03$ & \nodata & $-0.50\pm0.05$ & \nodata &  & N \\
J0525$-$36 & 1.688 & R & PK & $-0.37\pm0.07$ & $-0.71\pm0.25$ & \nodata & $-0.39\pm0.20$ & \nodata & $-0.65\pm0.19$ & \nodata &  & N \\
J0526$-$32 & 1.98 & R & PL & $-0.83\pm0.04$ & $-1.16\pm0.10$ & \nodata & $-0.01\pm0.03$ & \nodata & $-0.85\pm0.03$ & \nodata &  & N \\
J0536$-$27 & 1.79 & UR & I & $0.51\pm0.05$ & $< -0.02$ & $1.06\pm0.68$ & \nodata & $8.29\pm1.72$ & $0.96\pm0.04$ & $0.47\pm0.09$ &  & N \\
J0542$-$18 & \nodata & D & PL & $-0.94\pm0.05$ & $-0.90\pm0.10$ & \nodata & $-0.01\pm0.01$ & \nodata & $-0.94\pm0.01$ & \nodata & MP & N \\
J0543+52 & \nodata & T & PL & $-0.98\pm0.05$ & $-0.78\pm0.10$ & \nodata & $-0.08\pm0.02$ & \nodata & $-0.84\pm0.03$ & \nodata & MP & N \\
J0543+58 & \nodata & UR & CV & $-0.54\pm0.05$ & $-0.29\pm0.10$ & $<0.15$ & $-0.06\pm0.00$ & \nodata & $-0.59\pm0.06$ & \nodata &  & N \\
J0549$-$37 & 1.708 & UR & PK: & $-1.38\pm0.06$ & $< -0.09$ & $0.16$ & $-0.21\pm0.11$ & $24.79$ & $-1.37\pm0.13$ & $1.33\pm4.92$ &  & N \\
J0602$-$27 & \nodata & T & PL & $-1.60\pm0.05$ & $-1.28\pm0.10$ & \nodata & $-0.04\pm0.01$ & \nodata & $-1.42\pm0.03$ & \nodata & E/M & N \\
J0604$-$03 & \nodata & D & PK: & $-0.61\pm0.05$ & $-0.12\pm0.25$ & $0.28\pm0.11$ & $-0.11\pm0.03$ & $10.57\pm1.30$ & $-0.61\pm0.00$ & $-0.11\pm0.33$ & MP & Y \\
J0612$-$06 & 0.47 & T & CV: & $-1.11\pm0.04$ & $-0.74\pm0.10$ & \nodata & $-0.12\pm0.03$ & \nodata & $-0.96\pm0.05$ & \nodata & MP & Y \\
J0613$-$34 & 2.18 & UR & PK: & $-1.22\pm0.05$ & $< 0.33$ & $0.38\pm0.18$ & $-0.29\pm0.09$ & $45.84\pm14.10$ & $-1.26\pm0.05$ & $1.50\pm4.92$ &  & N \\
J0614$-$09 & \nodata & UR & F & $-0.27\pm0.06$ & $< -0.23$ & \nodata & $0.02\pm0.03$ & \nodata & \nodata & \nodata &  & N \\
J0630$-$21 & 1.439 & D & CX & $-0.31\pm0.05$ & $-1.02\pm0.10$ & \nodata & $0.17\pm0.03$ & \nodata & \nodata & \nodata & MP & N \\
J0631$-$20 & \nodata & UR & PL & $-1.01\pm0.05$ & $-0.81\pm0.10$ & \nodata & $-0.02\pm0.04$ & \nodata & $-0.87\pm0.06$ & \nodata &  & N \\
J0634+36 & \nodata & UR & F & $-0.24\pm0.04$ & $0.27\pm0.25$ & \nodata & $-0.08\pm0.06$ & \nodata & \nodata & \nodata &  & N \\
J0641+50 & \nodata & SR & CV & $-1.44\pm0.05$ & $-0.81\pm0.10$ & $<0.07$ & $-0.16\pm0.01$ & \nodata & $-1.44\pm0.03$ & \nodata &  & N \\
J0641+52 & \nodata & SR & PL & $-0.93\pm0.05$ & $-0.66\pm0.10$ & \nodata & $-0.06\pm0.01$ & \nodata & $-0.84\pm0.03$ & \nodata &  & N \\
J0642$-$27 & 1.34 & D & PL & $-1.01\pm0.08$ & $-0.58\pm0.14$ & $<0.15$ & $-0.11\pm0.01$ & \nodata & $-1.01\pm0.04$ & \nodata & E/M & Y \\
J0647+59 & \nodata & SR & CV & $-0.90\pm0.04$ & $-0.41\pm0.10$ & $<0.07$ & $-0.09\pm0.07$ & \nodata & $-0.97\pm0.03$ & \nodata &  & N \\
J0652$-$20 & 0.604 & UR & PK: & $-1.10\pm0.06$ & $< -0.21$ & $<0.15$ & $-0.18\pm0.03$ & \nodata & $-1.10\pm0.03$ & $-0.05\pm0.57$ &  & N \\
J0701+45 & \nodata & UR & CV & $-1.01\pm0.06$ & $< -0.49$ & \nodata & $-0.03\pm0.04$ & \nodata & $-0.97\pm0.05$ & \nodata &  & N \\
J0702$-$28 & 0.943 & UR & F & $-0.11\pm0.05$ & $< -0.14$ & \nodata & $0.05\pm0.04$ & \nodata & \nodata & \nodata &  & N \\
J0714$-$36 & 0.882 & UR & PK & $-0.78\pm0.05$ & $< -0.01$ & $0.09$ & $-0.12\pm0.07$ & $32.00$ & $-0.78\pm0.01$ & \nodata &  & N \\
J0719$-$33 & 1.63 & UR & PK & $-0.40\pm0.04$ & $< 0.35$ & $2.81\pm0.48$ & $-0.70\pm0.07$ & $34.40\pm7.13$ & $-2.12\pm0.30$ & $0.01\pm0.66$ &  & N \\
J0729+65 & 2.236 & UR & PL & $-0.83\pm0.04$ & $-0.67\pm0.10$ & \nodata & $-0.03\pm0.02$ & \nodata & $-1.12\pm0.21$ & \nodata &  & N \\
J0737+18 & \nodata & D & PL & $-1.14\pm0.05$ & $-0.79\pm0.10$ & \nodata & $-0.12\pm0.04$ & \nodata & $-0.95\pm0.06$ & \nodata & MP & Y \\
J0739+23 & \nodata & UR & PK & $-1.64\pm0.04$ & $0.26\pm0.10$ & $0.66\pm0.02$ & $-0.46\pm0.01$ & $286.79\pm7.78$ & $-2.19\pm0.18$ & $0.66\pm0.09$ &  & N \\
J0804+36 & 0.656 & SR & PK & $-1.30\pm0.04$ & $0.33\pm0.10$ & $0.53\pm0.04$ & $-0.34\pm0.03$ & $106.69\pm8.38$ & $-1.33\pm0.10$ & $0.90\pm0.13$ &  & N \\
J0811$-$22 & 1.107 & UR & CV & $-1.21\pm0.04$ & $-0.64\pm0.10$ & $<0.15$ & $-0.12\pm0.03$ & \nodata & $-1.23\pm0.04$ & \nodata &  & N \\
J0823$-$06 & 1.749 & UR & PK: & $-0.39\pm0.04$ & $< 0.59$ & $1.89\pm0.47$ & $-0.28\pm0.04$ & $53.99\pm4.85$ & $-0.31\pm0.01$ & $1.39\pm4.92$ &  & N \\
J0845+28 & \nodata & D & CV & $-1.05\pm0.04$ & \nodata & $0.46\pm0.04$ & $-0.24\pm0.02$ & $35.83\pm2.87$ & $-1.05\pm0.01$ & $-0.04\pm0.11$ & MP & Y \\
J0849+30 & \nodata & UR & CV & $-1.74\pm0.05$ & \nodata & $<0.34$ & $-0.52\pm0.04$ & \nodata & $-1.71\pm0.03$ & $2.11\pm0.18$ &  & N \\
J0920+12 & \nodata & UR & PK & $-1.08\pm0.04$ & $< 0.27$ & $<0.34$ & $-0.21\pm0.05$ & \nodata & $-1.07\pm0.03$ & $1.74\pm1.25$ &  & N \\
J0929$-$20 & \nodata & UR & PK: & $-1.20\pm0.04$ & $0.08\pm0.24$ & $0.73\pm0.21$ & $-0.35\pm0.08$ & $33.88\pm5.37$ & $-1.14\pm0.13$ & $-0.15\pm0.65$ &  & N \\
J0941$-$00 & \nodata & R & CV & $-0.71\pm0.05$ & $< -0.18$ & $<0.15$ & $-0.17\pm0.08$ & \nodata & $-0.73\pm0.02$ & $2.50\pm4.92$ &  & N \\
J0943+03 & \nodata & UR & PK & $-1.45\pm0.04$ & $< 0.99$ & $1.02\pm0.09$ & $-0.56\pm0.04$ & $133.28\pm7.33$ & $-1.73\pm0.16$ & $1.49\pm0.47$ &  & N \\
J1002+02 & 0.297 & UR & CV & $-1.03\pm0.05$ & $-0.55\pm0.10$ & $<0.15$ & $-0.12\pm0.02$ & \nodata & $-1.07\pm0.03$ & \nodata &  & N \\
J1025+61 & \nodata & T & PL & $-1.08\pm0.04$ & $-0.96\pm0.10$ & \nodata & $0.06\pm0.06$ & \nodata & $-0.91\pm0.09$ & \nodata & E/M & N \\
J1046$-$02 & \nodata & UR & PK & $-0.90\pm0.04$ & $0.05\pm0.10$ & $0.59\pm0.07$ & $-0.24\pm0.02$ & $97.42\pm5.53$ & $-1.33\pm0.13$ & $-0.27\pm0.39$ &  & N \\
J1048+55 & \nodata & UR & PL & $-1.11\pm0.05$ & $-0.68\pm0.10$ & \nodata & $-0.11\pm0.02$ & \nodata & $-0.85\pm0.05$ & \nodata &  & N \\
J1049$-$02 & \nodata & UR & PK: & $-0.51\pm0.05$ & $< -0.22$ & $1.71\pm0.74$ & $-0.34\pm0.09$ & $9.94\pm2.26$ & $-0.66\pm0.12$ & $0.89\pm4.92$ &  & N \\
J1107+34 & 1.452 & SR & PK & $0.11\pm0.04$ & \nodata & $4.11\pm0.91$ & $-0.79\pm0.08$ & $53.38\pm20.36$ & $-1.09\pm0.23$ & $1.69\pm0.36$ &  & N \\
J1138+20 & \nodata & UR & PK: & $-0.48\pm0.05$ & $< -0.07$ & $<0.15$ & $-0.11\pm0.09$ & \nodata & $-0.84\pm0.21$ & $0.03\pm0.57$ & E/R & Y \\
J1157+45 & \nodata & UR & PK & $-1.19\pm0.04$ & \nodata & $1.52\pm0.01$ & $-0.66\pm0.00$ & $37.55\pm0.36$ & $-1.96\pm0.54$ & $2.82\pm1.24$ &  & N \\
J1210+47 & \nodata & UR & PK: & $-0.41\pm0.04$ & \nodata & $1.75\pm0.12$ & $-0.25\pm0.02$ & $24.52\pm1.47$ & $-0.46\pm0.02$ & $0.71\pm0.03$ &  & N \\
J1212+46 & \nodata & UR & PL & $-1.00\pm0.04$ & $-0.87\pm0.10$ & \nodata & $-0.03\pm0.02$ & \nodata & $-0.84\pm0.04$ & \nodata &  & N \\
J1226+18 & \nodata & UR & F & $-0.30\pm0.05$ & $-0.14\pm0.24$ & \nodata & $-0.04\pm0.02$ & \nodata & \nodata & \nodata &  & N \\
J1238+52 & 2.246 & UR & PK & $-0.92\pm0.04$ & \nodata & $0.94\pm0.12$ & $-0.32\pm0.05$ & $92.59\pm14.14$ & $-1.26\pm0.07$ & $-0.93\pm0.38$ &  & N \\
J1257+73 & \nodata & D & PL & $-0.86\pm0.05$ & $-0.51\pm0.10$ & \nodata & $-0.05\pm0.04$ & \nodata & $-0.75\pm0.04$ & \nodata & MP & N \\
J1304+22 & \nodata & UR & CV & $-0.91\pm0.05$ & $-0.23\pm0.24$ & $<0.15$ & $-0.13\pm0.04$ & \nodata & $-0.90\pm0.05$ & \nodata &  & N \\
J1308$-$34 & 1.652 & T & PL & $-0.64\pm0.04$ & $-1.00\pm0.10$ & \nodata & $-0.09\pm0.05$ & \nodata & $-0.53\pm0.06$ & \nodata & E/R & Y \\
J1316+39 & \nodata & UR & CV & $-1.50\pm0.05$ & $-0.49\pm0.11$ & $<0.07$ & $-0.19\pm0.02$ & \nodata & $-1.49\pm0.02$ & \nodata &  & N \\
J1332+79 & \nodata & UR & PK: & $-0.67\pm0.05$ & $0.11\pm0.26$ & $0.57\pm0.12$ & $-0.18\pm0.03$ & $12.81\pm0.74$ & $-0.70\pm0.02$ & $0.11\pm0.26$ &  & N \\
J1343$-$11 & 2.486 & UR & CV & $-0.47\pm0.05$ & $< -0.13$ & \nodata & $-0.06\pm0.06$ & \nodata & $-0.72\pm0.14$ & \nodata &  & N \\
J1400$-$29 & 1.66 & UR & CV & $-1.41\pm0.04$ & $-0.31\pm0.10$ & $0.22\pm0.04$ & $-0.26\pm0.03$ & $125.72\pm14.66$ & $-1.50\pm0.10$ & $0.53\pm0.01$ &  & N \\
J1409+17 & \nodata & D & CX & $-0.31\pm0.04$ & $-0.97\pm0.10$ & \nodata & $0.16\pm0.05$ & \nodata & \nodata & \nodata & MP & N \\
J1412$-$20 & 1.808 & SR & CX & $0.44\pm0.05$ & $-0.92\pm0.10$ & \nodata & $0.34\pm0.05$ & \nodata & \nodata & \nodata &  & N \\
J1424$-$26 & \nodata & SR & CV & $-1.21\pm0.06$ & $< -0.54$ & $<0.15$ & $-0.13\pm0.03$ & \nodata & $-1.21\pm0.04$ & \nodata &  & N \\
J1428+11 & 1.6 & T & PL & $-1.01\pm0.08$ & $-0.99\pm0.10$ & \nodata & $-0.00\pm0.01$ & \nodata & $-1.03\pm0.02$ & \nodata & E/M & N \\
J1434$-$02 & 1.921 & D & CV & $-1.00\pm0.04$ & $-0.35\pm0.10$ & $<0.07$ & $-0.13\pm0.02$ & \nodata & $-0.96\pm0.01$ & $0.37\pm0.70$ & MP & Y \\
J1439$-$37 & 1.2 & D & PL & $-0.73\pm0.06$ & $-0.91\pm0.10$ & \nodata & $0.03\pm0.03$ & \nodata & $-0.81\pm0.04$ & \nodata & E/M & N \\
J1448+40 & \nodata & UR & PK & $-1.35\pm0.05$ & \nodata & $0.69\pm0.04$ & $-0.40\pm0.02$ & $17.02\pm1.26$ & $-1.33\pm0.01$ & $0.89\pm0.67$ &  & N \\
J1500$-$06 & 1.5 & R & PL & $-1.18\pm0.04$ & $-1.00\pm0.10$ & \nodata & $-0.04\pm0.01$ & \nodata & $-1.12\pm0.02$ & \nodata &  & N \\
J1501+13 & 0.505 & UR & CV & $-1.57\pm0.04$ & $-0.54\pm0.10$ & $0.10\pm0.04$ & $-0.23\pm0.04$ & $1249.57\pm337.02$ & $-1.45\pm0.04$ & \nodata &  & N \\
J1501+33 & \nodata & UR & PK & $-0.83\pm0.04$ & \nodata & $1.62\pm0.05$ & $-0.49\pm0.02$ & $59.96\pm2.86$ & $-1.33\pm0.12$ & $1.38\pm1.23$ &  & N \\
J1510$-$22 & 0.95 & UR & PK & $-0.74\pm0.04$ & $< 0.17$ & $0.60\pm0.26$ & $-0.21\pm0.07$ & $20.69\pm3.58$ & $-1.32\pm0.38$ & $1.48\pm4.92$ &  & N \\
J1513$-$22 & 2.2 & SR & PL & $-1.01\pm0.04$ & $-0.84\pm0.10$ & \nodata & $-0.04\pm0.05$ & \nodata & $-0.88\pm0.05$ & \nodata &  & N \\
J1514$-$34 & 1.08 & UR & CV: & $-0.67\pm0.05$ & $< -0.21$ & $<0.15$ & $-0.08\pm0.03$ & \nodata & $-0.84\pm0.16$ & \nodata &  & N \\
J1516+19 & \nodata & D & CV & $-1.34\pm0.09$ & $-0.30\pm0.29$ & $<0.15$ & $-0.15\pm0.07$ & \nodata & $-1.20\pm0.06$ & \nodata &  & N \\
J1517+35 & 1.515 & UR & PL & $-1.30\pm0.04$ & $-1.01\pm0.10$ & \nodata & $-0.08\pm0.02$ & \nodata & $-1.13\pm0.04$ & \nodata &  & N \\
J1521+00 & 2.63 & UR & PK & $0.08\pm0.04$ & $< 0.45$ & $3.85\pm0.87$ & $-0.91\pm0.10$ & $101.68\pm43.74$ & $-2.02\pm0.06$ & $2.73\pm4.92$ &  & N \\
J1525+76 & \nodata & UR & PL & $-1.01\pm0.05$ & $-0.46\pm0.10$ & \nodata & $-0.27\pm0.09$ & \nodata & $-0.58\pm0.11$ & \nodata & E/R & Y \\
J1541$-$11 & 1.58 & UR & PK & $-2.03\pm0.05$ & $< 0.70$ & $0.84\pm0.20$ & $-0.67\pm0.14$ & $37.09\pm8.04$ & $-2.03\pm0.03$ & $0.84\pm1.20$ &  & N \\
J1604+69 & \nodata & UR & CV & $-0.98\pm0.05$ & $-0.38\pm0.25$ & $<0.14$ & $-0.06\pm0.03$ & \nodata & $-0.99\pm0.03$ & \nodata &  & N \\
J1615+74 & \nodata & UR & F & $-0.07\pm0.04$ & $-0.16\pm0.25$ & \nodata & $0.11\pm0.10$ & \nodata & \nodata & \nodata &  & N \\
J1630+51 & 0.724 & UR & PL & $-0.49\pm0.04$ & $-0.31\pm0.10$ & \nodata & $0.01\pm0.03$ & \nodata & \nodata & \nodata &  & N \\
J1634$-$17 & 2.07 & SR & CV: & $-1.28\pm0.07$ & $< 0.08$ & $<0.89$ & $-0.21\pm0.15$ & \nodata & $-1.29\pm0.23$ & $0.33\pm0.57$ &  & N \\
J1641$-$05 & 1.83 & UR & CV & $-1.13\pm0.06$ & $< -0.20$ & $<0.15$ & $-0.17\pm0.06$ & \nodata & $-1.13\pm0.01$ & \nodata &  & N \\
J1642+41 & 1.276 & UR & PK & $-0.89\pm0.04$ & \nodata & $0.45\pm0.06$ & $-0.21\pm0.03$ & $29.79\pm2.72$ & $-1.10\pm0.11$ & $-0.98\pm1.23$ &  & N \\
J1644$-$03 & \nodata & R & PK: & $-1.17\pm0.05$ & $-0.09\pm0.26$ & $0.36\pm0.04$ & $-0.25\pm0.02$ & $15.75\pm0.76$ & $-1.17\pm0.05$ & $0.07\pm0.33$ &  & N \\
J1651+34 & \nodata & M & PL & $-0.93\pm0.04$ & $-0.89\pm0.10$ & \nodata & $0.04\pm0.01$ & \nodata & $-0.97\pm0.03$ & \nodata & E/M & N \\
J1653+77 & \nodata & UR & PK: & $-0.50\pm0.05$ & $< -0.32$ & $1.69\pm1.55$ & $-0.32\pm0.15$ & $7.64\pm3.60$ & $-0.83\pm0.24$ & $-1.37\pm1.83$ &  & N \\
J1653$-$01 & 2.02 & UR & PK: & $-0.37\pm0.05$ & $< -0.10$ & $1.78\pm1.00$ & $-0.25\pm0.08$ & $12.53\pm2.85$ & $-1.07\pm0.29$ & $-0.18\pm0.57$ &  & N \\
J1657$-$17 & \nodata & UR & CV & $-1.55\pm0.05$ & $-0.73\pm0.10$ & $<0.15$ & $-0.17\pm0.04$ & \nodata & $-1.55\pm0.01$ & \nodata &  & N \\
J1702$-$08 & 2.853 & R & PL & $-0.78\pm0.04$ & $-1.21\pm0.10$ & \nodata & $0.06\pm0.04$ & \nodata & $-1.13\pm0.04$ & \nodata &  & N \\
J1703+26 & 1.075 & UR & PK & $-0.52\pm0.04$ & $< 0.84$ & $2.18\pm0.53$ & $-0.49\pm0.07$ & $61.30\pm10.12$ & $-1.40\pm0.03$ & $0.84\pm0.57$ &  & N \\
J1703$-$05 & 1.79 & D & F & $-0.20\pm0.04$ & $-0.32\pm0.10$ & \nodata & $0.02\pm0.03$ & \nodata & \nodata & \nodata & MP & Y \\
J1707$-$09 & \nodata & UR & CV & $-0.98\pm0.05$ & $< -0.26$ & $<0.89$ & $-0.19\pm0.05$ & \nodata & $-0.98\pm0.01$ & $-0.24\pm0.57$ &  & N \\
J1717+53 & 2.717 & UR & F & $-0.35\pm0.04$ & $< 0.31$ & $0.80\pm0.47$ & $-0.12\pm0.05$ & $21.48\pm3.47$ & $-0.84\pm0.10$ & $0.96\pm1.31$ &  & N \\
J1820+79 & \nodata & UR & PK & $-0.48\pm0.04$ & $< 0.67$ & $1.40\pm0.04$ & $-0.24\pm0.01$ & $80.21\pm1.38$ & $-1.10\pm0.07$ & $1.25\pm1.30$ &  & N \\
J1840+50 & \nodata & D & PK & $-0.62\pm0.05$ & $< 0.30$ & $0.85\pm0.08$ & $-0.21\pm0.02$ & $29.75\pm1.07$ & $-0.62\pm0.01$ & $0.64\pm1.31$ & MP & Y \\
J1936$-$33 & 2.24 & UR & CV & $-1.10\pm0.07$ & $< -0.52$ & $<0.15$ & $-0.12\pm0.05$ & \nodata & $-1.10\pm0.05$ & \nodata & FT & Y \\
J1951$-$04 & 1.58 & D & PL & $-1.45\pm0.06$ & $-1.05\pm0.10$ & \nodata & $-0.11\pm0.02$ & \nodata & $-1.26\pm0.06$ & \nodata & E/R & Y \\
J1958$-$07 & 1.8 & UR & PK & $-1.01\pm0.04$ & $< 0.38$ & $0.39\pm0.11$ & $-0.23\pm0.05$ & $51.05\pm7.30$ & $-1.21\pm0.17$ & $1.59\pm0.38$ &  & N \\
J2000$-$28 & 2.277 & UR & CV & $-1.45\pm0.05$ & $-0.59\pm0.10$ & $<0.15$ & $-0.21\pm0.03$ & \nodata & $-1.45\pm0.01$ & \nodata &  & N \\
J2021$-$26 & 2.44 & SR & CV & $-0.95\pm0.06$ & $< -0.25$ & $<0.15$ & $-0.22\pm0.08$ & \nodata & $-0.95\pm0.03$ & $1.65\pm4.92$ &  & N \\
J2040$-$39 & \nodata & UR & CV & $-0.88\pm0.05$ & $-0.19\pm0.27$ & $<0.15$ & $-0.13\pm0.05$ & \nodata & $-0.93\pm0.07$ & \nodata & FT & Y \\
J2059$-$35 & 2.38 & SR & CV & $-1.91\pm0.23$ & $-0.91\pm0.11$ & $<0.15$ & $-0.22\pm0.02$ & \nodata & $-1.91$ & \nodata & E/R & Y \\
J2124$-$28 & \nodata & M & PL & $-1.11\pm0.06$ & $-0.92\pm0.10$ & \nodata & $-0.06\pm0.07$ & \nodata & $-0.90\pm0.08$ & \nodata & Mhigh & N \\
J2126$-$01 & 0.607 & UR & PK: & $-1.05\pm0.05$ & $< 0.12$ & $<0.89$ & $-0.19\pm0.08$ & \nodata & $-0.97\pm0.04$ & $0.26\pm0.57$ &  & N \\
J2130+20 & 0.811 & T & PL & $-1.06\pm0.07$ & $-0.88\pm0.10$ & \nodata & $-0.03\pm0.01$ & \nodata & $-0.93\pm0.03$ & \nodata & E/M & N \\
J2133$-$14 & \nodata & UR & PK: & $-0.31\pm0.05$ & $< 0.06$ & $<0.89$ & $-0.03\pm0.11$ & \nodata & \nodata & \nodata &  & N \\
J2133$-$17 & \nodata & T & CV & $-1.52\pm0.04$ & $-0.95\pm0.10$ & $<0.07$ & $-0.13\pm0.02$ & \nodata & $-1.52\pm0.04$ & \nodata & E/R & Y \\
J2145$-$06 & \nodata & D & CV & $-1.23\pm0.05$ & $-0.55\pm0.10$ & $<0.07$ & $-0.16\pm0.02$ & \nodata & $-1.23\pm0.02$ & $-2.38\pm1.45$ & E/R & Y \\
J2148$-$29 & \nodata & R & PL & $-1.10\pm0.04$ & $-0.98\pm0.10$ & \nodata & $-0.07\pm0.01$ & \nodata & $-0.95\pm0.07$ & \nodata &  & N \\
J2204+20 & \nodata & UR & PK & $-1.30\pm0.04$ & $1.12\pm0.10$ & $1.16\pm0.04$ & $-0.55\pm0.03$ & $520.14\pm46.17$ & $-2.72\pm0.10$ & $2.00\pm0.09$ &  & N \\
J2212+33 & \nodata & R & CV & $-0.86\pm0.04$ & $-0.59\pm0.10$ & \nodata & $-0.04\pm0.02$ & \nodata & $-0.85\pm0.06$ & \nodata &  & N \\
J2212$-$12 & \nodata & T & PL & $-1.21\pm0.08$ & $-1.04\pm0.10$ & \nodata & $-0.07\pm0.02$ & \nodata & $-1.01\pm0.04$ & \nodata & E/M & N \\
J2222+09 & \nodata & UR & CV: & $-1.06\pm0.04$ & $-0.60\pm0.10$ & $<0.07$ & $-0.10\pm0.01$ & \nodata & $-1.07\pm0.04$ & \nodata &  & N \\
J2226+00 & 0.676 & R & PL & $-1.06\pm0.05$ & $-0.77\pm0.10$ & \nodata & $-0.04\pm0.02$ & \nodata & $-0.97\pm0.07$ & \nodata &  & N \\
J2230$-$07 & 0.444 & R & CV & $-1.05\pm0.04$ & $-0.56\pm0.10$ & $<0.07$ & $-0.14\pm0.02$ & \nodata & $-1.09\pm0.01$ & $-0.85\pm0.06$ &  & N \\
J2235$-$28 & \nodata & UR & PK & $-0.18\pm0.04$ & $< 0.54$ & $2.13\pm0.21$ & $-0.16\pm0.01$ & $41.16\pm0.98$ & $-0.88\pm0.08$ & $0.67\pm4.92$ &  & N \\
J2241$-$06 & \nodata & UR & PK: & $-0.69\pm0.04$ & $< 0.19$ & $<0.89$ & $-0.15\pm0.04$ & \nodata & $-0.85\pm0.13$ & $2.81\pm4.92$ &  & N \\
J2251+01 & \nodata & UR & PK & $0.28\pm0.05$ & $< 0.16$ & $4.81\pm1.56$ & $-0.61\pm0.07$ & $22.59\pm10.39$ & $-0.91\pm0.12$ & $0.10\pm0.57$ &  & N \\
J2252$-$21 & \nodata & UR & CX & $-0.64\pm0.04$ & $-0.33\pm0.10$ & \nodata & $-0.04\pm0.04$ & \nodata & $0.64\pm0.12$ & \nodata &  & N \\
J2318+25 & 0.498 & T & PL & $-1.02\pm0.15$ & $-0.69\pm0.10$ & \nodata & $-0.01\pm0.04$ & \nodata & $-0.77\pm0.05$ & \nodata & E/M & N \\
J2322$-$00 & \nodata & UR & CV & $-1.39\pm0.05$ & $< -0.03$ & $0.48$ & $-0.32\pm0.20$ & $13.80$ & $-1.39\pm0.04$ & $-0.11\pm0.57$ &  & N \\
J2325$-$04 & 1.737 & R & PL & $-0.90\pm0.04$ & $-0.59\pm0.10$ & \nodata & $-0.02\pm0.01$ & \nodata & $-0.80\pm0.03$ & \nodata &  & N \\
J2328$-$02 & \nodata & UR & PK & $-2.19\pm0.11$ & $-0.32\pm0.25$ & $0.33\pm0.04$ & $-0.45\pm0.04$ & $22.57\pm2.57$ & $-2.16\pm0.12$ & $-0.19\pm0.32$ & E/R & Y \\
J2329$-$10 & \nodata & R & CV: & $-1.43\pm0.05$ & $-0.91\pm0.10$ & $<0.15$ & $-0.11\pm0.02$ & \nodata & $-1.40\pm0.03$ & \nodata &  & N \\
J2331$-$14 & \nodata & D & CV: & $-0.82\pm0.04$ & $-0.32\pm0.10$ & \nodata & $-0.12\pm0.01$ & \nodata & $-0.82\pm0.02$ & $-0.10\pm0.04$ & E/R & Y \\
J2332+34 & 0.981 & UR & PK & $-1.31\pm0.05$ & $0.06\pm0.25$ & $0.63\pm0.04$ & $-0.38\pm0.03$ & $19.14\pm1.92$ & $-1.33\pm0.06$ & $1.14\pm0.36$ &  & N \\
J2341$-$29 & \nodata & UR & CV & $-0.84\pm0.06$ & $-0.27\pm0.24$ & \nodata & $-0.13\pm0.04$ & \nodata & $-0.89\pm0.05$ & \nodata & E/R & Y \\
J2345+31 & \nodata & UR & PK & $-0.77\pm0.04$ & \nodata & $0.68\pm0.04$ & $-0.24\pm0.02$ & $17.97\pm1.17$ & $-0.95\pm0.15$ & $0.34\pm0.67$ &  & N \\
J2357$-$10 & 2.048 & UR & CV & $-0.77\pm0.05$ & $< 0.08$ & $<0.89$ & $-0.11\pm0.07$ & \nodata & $-0.84\pm0.16$ & $2.20\pm3.99$ &  & N \\
J2359+43 & \nodata & D & PK & $-1.72\pm0.05$ & $< 0.27$ & $0.53\pm0.09$ & $-0.43\pm0.07$ & $42.15\pm6.97$ & $-1.72\pm0.05$ & $1.62\pm1.26$ &  & N
\enddata
\tablecomments{Column 1: Source name. Column 2: Spectroscopic redshift. Column 3: 10 GHz Radio morphology as seen in VLA imaging \citep{patil+20}. The morphological codes are UR: Unresolved; SR: Slightly/Marginally resolved; R: Resolved and single component; D: Double; T: Triple; M: Multi-component. Column 4: Spectral shape classes as described in Section~\ref{sec:sedclass}. The notation ``:" in front of the spectral shape indicates uncertainty in determination of the spectral shape. Column 5: $\alpha_{1.4}^{10}$, spectral index calculated between NVSS and VLA 10 GHz data. Column 6: $\alpha_{0.15}^{1.4}$, TGSS-NVSS spectral index. Column 7$-$11: Spectral shape parameters for the curved and peaked spectrum sources. The definition of each parameter is described in Section~\ref{sec:sedparam}. $\nu_{peak}$: The turnover frequency; $q$: best-fit value for the curvature parameter in Equation~\ref{eqn:cpl}; $\alpha_{high}$: High-frequency spectral index calculated using frequencies above  $\nu_{peak}$; $\alpha_{low}$: Low-frequency spectral index calculated using frequencies below $\nu_{peak}$.  Column 12: Quality Code (QC) for individual sources. MP indicates that the source is resolved into multiple but relatively compact sources; SB indicates possible source blending at lower resolution surveys; FT indicates a faint, single source resulting in flux losses and additional steepening of radio spectra (usually at X band); E/R: Extended, multi-component source showing signs of additional spectral steepening at higher frequencies; E/M Extended, multi-component source with no apparent flux losses. Column 13: N indicates source is included in the final sample, where Y  indicates that the given radio spectra is rejected for a variety of reasons as explained in Section ~\ref{sec:spectral_fitting} }
\end{deluxetable}
\end{longrotatetable}

%% file: Bsol_tab_v7.tex
\begin{deluxetable*}{ccccccccc}
\tablecaption{Emitting Region Properties in Peaked Spectrum Sources Assuming $B_{SSA} = B_{min}$\label{tab:bsoltab}}
\tablehead{\colhead{Source} & \colhead{$z$} & \colhead{$\theta_{reg}$} & \colhead{$B$} & \colhead{Reg Size} & \colhead{$P_{reg}$} & \colhead{$\log(t_{dyn})$} &  \colhead{$\log(t_{rad})$} & \colhead{$\log(t_{syn})$} \\
\colhead{} & \colhead{} & \colhead{mas} & \colhead{mG} & \colhead{pc} &  \colhead{$\times 10^{-5} $dyne cm$^{-2}$} & \colhead{yr} & \colhead{yr} & \colhead{yr}
}
\colnumbers
\startdata
J0010+16 & 2.855 & 3.1 & 21.1 & 24.8 & 1.4 & 1.73 & 3.63 & 3.12 \\
J0134+40 & -- & $12.8 - 18.2$ & $3.9 - 7.2$ & $99.1-143.6$ & 0.1 & \nodata & \nodata & \nodata \\
J0304$-$31 & 1.53 & 4.7 & 18.3 & 40.6 & 1.0 & 3.32 & 4.66 & 3.21 \\
J0332+32 & 0.304 & 2.4 & 11.9 & 11.3 & 0.4 & 2.18 & 4.21 & 3.49 \\
J0342+37 & 0.47 & 4.4 & 15.4 & 27.0 & 0.7 & 3.25 & 4.66 & 3.32 \\
J0354$-$33 & 1.373 & 0.9 & 37.9 & 8.2 & 4.5 & 2.69 & 4.15 & 2.74 \\
J0404+07 & -- & $4.1 - 5.8$ & $15.4 - 28.1$ & $31.5-45.6$ & 1.5 & \nodata & \nodata & \nodata \\
J0417$-$28 & 0.943 & 1.3 & 32.4 & 10.5 & 3.2 & 3.12 & 4.43 & 2.84 \\
J0502+12 & -- & $3.9 - 5.5$ & $9.0 - 16.3$ & $30.2-43.7$ & 0.3 & \nodata & \nodata & \nodata \\
J0549$-$37 & 1.708 & 7.8 & 4.6 & 67.6 & 0.1 & 2.07 & 4.29 & 4.12 \\
J0613$-$34 & 2.18 & 4.6 & 12.5 & 39.1 & 0.5 & 1.97 & 3.92 & 3.46 \\
J0652$-$20 & 0.604 & 4.8 & 4.2 & 33.4 & 0.1 & 2.60 & 4.75 & 4.16 \\
J0714$-$36 & 0.882 & 13.6 & 2.6 & 108.5 & 0.1 & 3.43 & 5.31 & 4.47 \\
J0719$-$33 & 1.63 & 0.5 & 102.5 & 4.6 & 32.6 & 2.34 & 3.65 & 2.09 \\
J0739+23 & -- & $5.2 - 7.3$ & $13.2 - 24.1$ & $39.8-57.7$ & 0.5 & \nodata & \nodata & \nodata \\
J0804+36 & 0.656 & 3.7 & 10.1 & 26.8 & 0.3 & 1.98 & 4.08 & 3.60 \\
J0823$-$06 & 1.749 & 1.0 & 77.2 & 8.9 & 18.5 & 2.59 & 3.85 & 2.27 \\
J0849+30 & -- & $3.4 - 4.9$ & $8.9 - 16.1$ & $26.4-38.3$ & 0.2 & \nodata & \nodata & \nodata \\
J0920+12 & -- & $3.9 - 5.5$ & $8.2 - 14.9$ & $29.9-43.3$ & 0.3 & \nodata & \nodata & \nodata \\
J0929$-$20 & -- & $1.7 - 2.4$ & $19.7 - 36.0$ & $13.3-19.2$ & 1.4 & \nodata & \nodata & \nodata \\
J0943+03 & -- & $2.3 - 3.3$ & $22.1 - 40.3$ & $17.8-25.9$ & 1.6 & \nodata & \nodata & \nodata \\
J1046$-$02 & -- & $3.5 - 5.0$ & $13.7 - 25.0$ & $27.0-39.1$ & 1.0 & \nodata & \nodata & \nodata \\
J1049$-$02 & -- & $0.4 - 0.6$ & $48.0 - 87.4$ & $3.1-4.5$ & 17.4 & \nodata & \nodata & \nodata \\
J1107+34 & 1.452 & 0.4 & 144.2 & 3.8 & 64.4 & \nodata & 4.04 & 1.87 \\
J1157+45 & -- & $0.8 - 1.2$ & $37.4 - 68.1$ & $6.5-9.5$ & 5.7 & \nodata & \nodata & \nodata \\
J1210+47 & -- & $0.6 - 0.9$ & $45.1 - 82.2$ & $4.6-6.7$ & 16.9 & \nodata & \nodata & \nodata \\
J1238+52 & 2.246 & 2.8 & 35.1 & 23.2 & 3.8 & 2.06 & 3.69 & 2.79 \\
J1332+79 & -- & $1.4 - 2.0$ & $17.3 - 31.6$ & $10.7-15.5$ & 2.0 & \nodata & \nodata & \nodata \\
J1400$-$29 & 1.66 & 11.5 & 5.3 & 99.6 & 0.1 & 2.23 & 4.31 & 4.02 \\
J1448+40 & -- & $1.3 - 1.9$ & $19.5 - 35.5$ & $10.1-14.6$ & 1.3 & \nodata & \nodata & \nodata \\
J1501+13 & 0.505 & 55.3 & 1.1 & 349.3 & 0.1 & 2.92 & 5.24 & 5.04 \\
J1501+33 & -- & $1.1 - 1.5$ & $52.3 - 95.3$ & $8.3-12.0$ & 7.9 & \nodata & \nodata & \nodata \\
J1510$-$22 & 0.95 & 1.8 & 20.0 & 14.3 & 1.2 & 2.59 & 4.25 & 3.15 \\
J1521+00 & 2.63 & 0.8 & 165.5 & 6.1 & 84.9 & \nodata & 4.09 & 1.78 \\
J1541$-$11 & 1.58 & 1.6 & 16.8 & 13.7 & 0.9 & \nodata & 2.68 & 3.27 \\
J1634$-$17 & 2.07 & 1.1 & 26.9 & 9.2 & 2.2 & 0.75 & 3.01 & 2.96 \\
J1642+41 & 1.276 & 2.9 & 15.1 & 25.1 & 0.7 & 2.53 & 4.27 & 3.34 \\
J1644$-$03 & -- & $2.4 - 3.4$ & $10.5 - 19.2$ & $18.4-26.6$ & 0.5 & \nodata & \nodata & \nodata \\
J1653$-$01 & 2.02 & $0.4 - 0.5$ & $55.2 - 100.6$ & $2.9-4.1$ & 26.1 & nan & nan & nan \\
J1653+77 & -- & 0.5 & 78.7 & 4.7 & 19.2 & \nodata & \nodata & \nodata \\
J1703+26 & 1.075 & 0.8 & 65.0 & 6.8 & 13.1 & 2.33 & 3.74 & 2.39 \\
J1717+53 & 2.717 & 1.8 & 51.9 & 15.0 & 8.3 & 3.02 & 4.25 & 2.53 \\
J1820+79 & -- & $1.3 - 1.9$ & $31.9 - 58.1$ & $10.1-14.7$ & 7.9 & \nodata & \nodata & \nodata \\
J1958$-$07 & 1.8 & 4.6 & 13.1 & 39.6 & 0.5 & 2.33 & 4.15 & 3.43 \\
J2126$-$01 & 0.607 & 0.9 & 22.2 & 6.3 & 1.5 & 1.77 & 3.79 & 3.08 \\
J2204+20 & -- & $3.8 - 5.5$ & $21.4 - 39.1$ & $29.7-43.1$ & 1.7 & \nodata & \nodata & \nodata \\
J2235$-$28 & -- & $0.6 - 0.9$ & $49.4 - 90.0$ & $4.8-7.0$ & 24.7 & \nodata & \nodata & \nodata \\
J2241$-$06 & -- & $1.2 - 1.7$ & $22.7 - 41.3$ & $9.3-13.5$ & 3.2 & \nodata & \nodata & \nodata \\
J2251+01 & -- & $0.2 - 0.3$ & $89.4 - 162.9$ & $1.5-2.2$ & 125.4 & \nodata & \nodata & \nodata \\
J2322$-$00 & -- & $1.7 - 2.4$ & $14.6 - 26.6$ & $13.1-19.0$ & 0.7 & \nodata & \nodata & \nodata \\
J2332+34 & 0.981 & 1.5 & 16.1 & 12.2 & 0.8 & 1.61 & 3.75 & 3.30 \\
J2345+31 & -- & $1.4 - 1.9$ & $19.2 - 34.9$ & $10.5-15.2$ & 2.2 & \nodata & \nodata & \nodata \\
J2357$-$10 & 2.048 & 1.4 & 36.7 & 11.9 & 4.2 & 2.19 & 3.82 & 2.76 \\
J2359+43 & -- & $2.6 - 3.7$ & $13.6 - 24.8$ & $19.9-28.8$ & 0.5 & \nodata & \nodata & \nodata
\enddata
\tablecomments{Column 1: Source name. Column 2: Redshift. Column 3: Estimate of emitting region angular size in mas (see Section~\ref{sec:bsolve}). Column 4: Estimate of magnetic field in Gauss. Column 5: Physical size of the region. Column 6: Source pressure from magnetic field in Column 4. Column 7: Dynamical time calculated using the lobe expansion model. Column 8: Radiative cooling time (Equation~\ref{eqn:rad_time}). Column 9: Synchrotron electron lifetime at 10 GHz (Equation~\ref{eqn:syn_time}). }
\end{deluxetable*}